\documentclass[final]{geophysics}
\usepackage[utf8]{inputenc}
\usepackage{amsmath,amssymb,amsfonts,amsbsy,latexsym,amsthm}
\usepackage[english]{babel}
\usepackage{lineno}
\pdfoutput=1
\usepackage{verbatim}
\usepackage{pgf,tikz}
\usepackage{color}
\usepackage{graphicx}
\usepackage{subfigure}
\usepackage{grffile} 
\usepackage{epstopdf}
\usepackage{algorithm}
\usepackage{algorithmic}
\usepackage{diagbox}
\usepackage{multirow}
\usepackage{chngpage}
\usepackage{indentfirst}
\usepackage{doi}
\usepackage{ulem}
\usepackage{setspace}

\usepackage{marginnote}

\usepackage{url}
\usepackage{hyperref}
\hypersetup{
colorlinks,%
citecolor=blue,%
filecolor=blue,%
linkcolor=blue,%
urlcolor=blue
}
\renewcommand{\sout}[1]{\unskip}
\newcommand{\yc}[1]{\textcolor{black}{#1}}

\title{Deep-learning  inversion:  a next generation seismic velocity-model building {\yc{method}}}
\author
{Fangshu Yang$^{1}$, Jianwei Ma$^1$ \\
$^{1}$Center of Geophysics, Department of Mathematics and Artificial Intelligence Laboratory, Harbin Institute of Technology, Harbin,  China, E-mail: yfs2016@hit.edu.cn; jma@hit.edu.cn
}

\begin{document}

\maketitle

\begin{abstract}

Seismic velocity is one of the most important parameters used in seismic exploration. Accurate velocity models are key prerequisites for reverse-time migration and other high-resolution seismic imaging techniques. Such velocity information has traditionally been derived by tomography or full-waveform inversion (FWI), which are time consuming and computationally expensive, and they rely heavily on human interaction and quality control.  We investigate a novel method based on the supervised deep fully convolutional neural network (FCN) for velocity-model building (VMB) directly from raw seismograms. Unlike the conventional inversion method based on physical models, the supervised deep-learning  methods are based on big-data training  rather than prior-knowledge  assumptions. {{\sout{about the subsurface that are required in many traditional inversion methods.}}} During the training stage, the network establishes a nonlinear projection from the multi-shot seismic data to the corresponding velocity models. During the prediction stage, the trained network can be used to estimate the velocity models from the new input seismic data. One key characteristic of the deep-learning method is that it can automatically extract multi-layer useful features  {{\sout{for velocity model building}}}  without {\yc{the need for}} human-curated activities and initial velocity setup.  The data-driven  method usually {{\sout{takes}}} {\yc{requires}} more time during the training stage,  {{\sout{and uses less time (only seconds) for  actual predictions}}} {\yc{and actual predictions take less time, with only seconds needed.}}  Therefore, the computational time of geophysical inversions, {\yc{including real-time inversions,}} can be dramatically reduced  once  a good generalized network is built.  By using numerical experiments {\yc{on synthetic models}}, {\color{blue}{\sout{we show promising performances using our proposed method for velocity model building}}} {\yc{the promising performances of our proposed method are shown}}  in comparison with conventional FWI even when the input data {{\sout{was}}} {\yc{are in more realistic scenarios.}} {{\sout{contaminated with noises or when the amplitudes have  different magnitude.}}} Discussions on the deep-learning methods, training dataset, lack of low frequencies, and advantages and disadvantages of the new method are also provided.

\end{abstract}

\newpage

\section{Introduction}

Currently, velocity-model building (VMB) is an essential step in seismic exploration because it is used during the entire course of seismic exploration including seismic data acquisition, processing, and interpretation.  Accurate subsurface-image reconstruction from surface seismic  wavefields requires precise knowledge of the local propagation velocities between the recording location and the image location at depth. Good velocity models are prerequisites for reverse-time migration \citep{Baysal1983} and other seismic imaging techniques \citep{Biondi2006}. Estimated velocity models can also be used as initial models to recursively generate high-resolution velocity
models with optimization algorithms \citep{Tarantola2005}. Many of the explored techniques such as migration velocity analysis \citep{AlYahya1989}, tomography \citep{Chiao2001}, and full-waveform inversion (FWI) \citep{Tarantola1984,Mora1987,Virieux2009} share the same purpose of building more accurate velocity models.

Traditional tomography methods \citep{Woodward2008}, including reflection tomography, tuning-ray tomography, and diving-wave tomography \citep{Stefani1995}, are widely used {\yc{for migration of seismic reflection data in building three-dimensional (3D) subsurface velocity models.}} {{\sout{in building 3D subsurface velocity models for migration of seismic reflection data. They}}} {\yc{Such methods}} have worked sufficiently well in most cases. Seismic inversion is performed by means of wave inversion of a simple prior model of {\yc{the}} subsurface, and by using a back propagation loop to infer {\yc{the}} subsurface geological structures \citep{Tarantola1984}. Typically, FWI is a data-fitting procedure {{\sout{that aims at}}} {\yc{used in}} reconstructing high-resolution velocity models  of the subsurface, as well as other parameters that govern wave propagation, from the full information contained in seismic data \citep{Virieux2009,Operto2013}. In FWI, active seismic sources are used to generate seismic waves, and  geophones are placed on the surface to record the measurements. An inverse problem is formulated to {{\sout{connect}}} {\yc{combine}} the measurements {{\sout{and}}} {\yc{with}} the governing physics equations to obtain the model parameters. Numerical optimization techniques are utilized to solve for the velocity models. {{\sout{FWI is highly effective if the initial model is {{\sout{fairly}}} {\yc{suitably}} accurate, but it suffers from cycle skipping if the predicted data from a initial model differs from the acquired data by more than half a period.}}} {\yc{When a suitably accurate initial model is provided, FWI is highly effective for obtaining a velocity structure through iterative updates.}} Efforts have been made recently to overcome the limitations in FWI.  Even though these conventional methods have shown great success in many applications,  they can be limited in some situations owing to a lack of low-frequency components as well as computational inefficiency, subjective human factors, and other issues.  Additionally,  iterative refinement is  expensive {\yc{when used}} in the entire workflow. {\yc{Thus,}} a robust, efficient, and accurate velocity-estimation method is  needed to address these problems. {{\sout{However, so far,  building a practical method for velocity models is still challenging and open because of the above-mentioned bottlenecks.}}}

Machine learning (ML) {\yc{is a field of artificial intelligence that uses statistical techniques to give computer systems the ability to ``learn" from big data.}}  ML has shown its strength in many fields including image recognition, recommendation systems \citep{Bobadilla2013}, spam filters \citep{Androutsopoulos2000}, fraud alerts \citep{Ravisankar2011}, and other applications. Furthermore, ML  has a long history of applications in geophysics. Nonlinear intelligent inverse technologies have been applied since the mid-1980s.  \cite{Roth1994} first presented an application of neural networks to invert from the time-domain {{\sout{of seismic amplitude responses}}} {\yc{seismic data}} to a depth profile of acoustic {{\sout{velocity}}} {\yc{velocities}}. They used pairs of synthetic shot {{\sout{gather}}} {\yc{gathers}} (i.e., a set of seismograms obtained from a single source) and corresponding one-{{\sout{dimension}}}{\yc{dimensional (1D)}} velocity models to train a multi-layer feed-forward neural network with the goal of predicting {\yc{the}}  velocities from new recorded data. They showed that the trained network can produce {{\sout{high resolution}}} {\yc{high-resolution}} approximations to the solutions of the inverse problem. In addition,  their method  can invert the geophysical parameters in the presence of white noise.  \cite{Nath1999} used neural networks for cross-well traveltime tomography. After training the network with synthetic data, the velocities  can be topographically estimated by the trained network with the new cross-well data. In recent years, most  ML-based methods have  focused mainly on pattern recognition in seismic attributes \citep{Zeng2004,Zhao2015b} and facies classifications in well logs \citep{Hall2016}. In the work of \cite{Guillen2015a}, {{\sout{they}}} {\yc{the authors}}  proposed a novel workflow to detect salt {{\sout{body}}} {\yc{bodies}}  based on seismic attributes in a supervised learning method. An {{\sout{machine learning}}} {\yc{ML}} algorithm (i.e., Extremely Random Trees Ensemble) was used to train {\yc{the mapping}} for automatically identifying salt regions. They  concluded that {{\sout{machine learning}}} {\yc{ML}} is a promising mechanism for classifying salt bodies when the selected training dataset has a sufficient  capacity for describing the complex decision boundaries. {{\sout{In 2017,}}}  \cite{Jia2017} used ML with  supported vector regression  \citep{Cortes1995} for seismic data interpolation. Unlike the conventional methods, no assumptions are imposed on  ML-based interpolation problems. On the basis of the above work, \cite{Jia2017a} {{\sout{developed a criterion}}} {\yc{proposed a method}} based on the Monte Carlo method \citep{Yu2016} for  intelligent reduction of training sets. In that study, representative patches of seismic data were selected to train the method for efficient reconstructions.

Deep learning (DL) \citep{LeCun2015a,Goodfellow2016},  a new branch of machine learning,  has drawn widespread interest by showing outstanding performance for recognition and classification \citep{Greenspan2016} in image and speech processing. Recently, \cite{Zhang2014} proposed to use of a kernel regularized least-squares   method \citep{Evgeniou2000} for fault detection from seismic records. The authors used toy velocity models to generate seismic records and set the records and the velocity models as inputs and labels in the training set. The numerical experiments showed that this method  obtained meaningful results.
\cite{Wang2018} developed a salt-detection technique from  raw multi-shot gathers by utilizing a fully convolutional neural network (FCN). The testing performance  showed that salt detections is much faster and efficient by this method than traditional migration and interpretation. \cite{Lewis2017}  {\yc{investigated a combination of DL and FWI to improve the performance for salt inversion. In that study, the network was trained to generate useful prior models for FWI by learning features relevant to earth model building from a seismic image. The authors tested this methodology by generating a probability map of salt bodies in the migrated image and incorporating it in the FWI objective function. The test results showed that this method is promising in enabling  automated salt body reconstruction using FWI.}}  In the work of \cite{Araya-Polo2017a},  a deep neural network (DNN)-based statistical model was used to automatically predict faults directly from synthetic two-dimensional (2D) seismic data. Inspired by this concept, \cite{Araya-Polo2018} proposed an approach for VMB. One key element of this DL tomography is the use of a feature based on semblance that predigests the velocity information. Extracted features are obtained before the training process and are used as DNN inputs to train the network. {{\sout{In 2018,}}}  \cite{Mosser2018}  used a generative adversarial network  \citep{Goodfellow2014} with cycle-constraints  \citep{Zhu2017b} to perform seismic inversion by formulating this problem as a domain-transfer problem. The mapping between the post-stack seismic traces and {{\sout{p-wave}}} {\yc{P-wave}} velocity models was approximated through this learning method. {{\sout{Due to the P-wave velocity models and the seismic amplitudes  are represented as a function of depth, rather than depth and time, respectively, this approach lends itself to perform stratigraphic inversion, where a pre-existing velocity model is used to perform time-to-depth conversion of the seismic amplitude.}}} {\yc{Before training the network, the seismic traces were transformed from the time domain to the depth domain based on the velocity models. Thus, the inputs and the outputs for training were in same domain.}} Most research has focused on identifying features and attributes in migrated images; few studies have discussed VMB or velocity inversion.

Multi-layer neural networks are computational learning architectures that propagate the input data across a sequence of linear operators and simple non-linearities. In this system, a deep convolutional neural network (CNN),  proposed by \cite{LeCun2010},  is implemented with linear convolutions followed by non-linear activation functions.  A strong motivation to use FCN stems from the universal approximation theorem \citep{Hornik1991,Csaji2001}, which states that a feed-forward network with a single hidden layer containing a finite number of neurons can approximate any continuous function on compact subsets under a mild assumption on the activation functions. Additionally, FCN assumes that we learn representative features by convolutional kernels in a data-driven fashion to extract features automatically. Compared with DNN, FCN exhibits structures with fewer parameters to explain multi-layer perceptions while still providing good results \citep{Burger2012}.

{{\sout{What we propose in this paper is}}} {\yc{In this study, we proposed}} the use of FCN to reconstruct  subsurface parameters, i.e., P-wave velocity model{{\sout{ building}}}, directly from  raw {{\sout{input of the}}} seismic data, instead of performing a local-based inversion with respect to {\yc{the}} subsurface represented through a grid. This method is an alternative formulation to  conventional FWI {\yc{and  includes two processes.}} {{\sout{in which the DL techniques are applied to a specific seismic workflow.}}}  {{\sout{In}}} {\yc{During the}} training process, multi-shot gathers are fed into the network together, and the network {{\sout{will}}} effectively {{\sout{approximate}}} {\yc{approximates}} the non-linear mapping between {\yc{the}} data and {\yc{the}} corresponding velocity model. {{\sout{In}}} {\yc{During the}} prediction process, the trained network can be saved to obtain unknown geomorphological structures only with new seismic data. Compared with traditional methods, {{\sout{no}}} {\yc{less}} human intervention or {\yc{no}} initial velocity models are involved {{\sout{in}}} {\yc{throughout}} the  process. Although the training process is expensive, the cost of the prediction stage by the network is negligible once the training is completed. {{\sout{since the wave forward propagation is not involved when we use the network to predict.}}} {{\sout{Meanwhile,}}} {\yc{Alternatively,}}
{{\sout{DL-based}}} {\yc{our proposed}} method {{\sout{may}}}  provides a possible {{\sout{road}}} {\yc{method}} for velocity inversion when the seismic data {{\sout{losses}}} {\yc{are in more realistic cases such as in the presence of noise and with a lack of low frequencies.}} {{\sout{The numerical applications of our method mainly focuses on velocity inversion, that is to evaluate the velocity values  by decreasing the mean square error between the predicted velocities and corresponding ground-truth velocities. Additionally, geometric shape/structure detection is also an important challenge we are faced with. In order to further show promising abilities of DL applied for geological application, we also presented numerical experiments related to detecting  geometric and positional information of particular subsurface  structures, which just be taken as a specific example of our seismic inversion.}}}  {{\sout{From numerical point of view, most}}} {\yc{Moreover,}} numerical experiments {{\sout{performed on various data}}} are used to demonstrate the applicability and feasibilities of our method.

This paper is organized as follows. {{\sout{The second part gives}}} {\yc{In the method section, we present}} a brief introduction to the basic inversion problem, the concepts of  FCN,  {\yc{the mathematical framework, and the special architecture of the network.}} {{\sout{After that, the methodology is presented, including the DL-based mathematical model, data preparation for the training set, and the architecture of the deep-learning-based velocity-model building (DLVMB) network.}}}  {{\sout{The fourth part}}} {{\sout{ shows two aspects of numerical experiments on 2D isotropic acoustic models with a uniform and constant density and Society of Exploration Geophysics (SEG) dataset that includes the DLVMB method for p-wave velocity inversion and a specific example, that is detecting geological structures}}} {\yc{In the results section, we firstly show the dataset design. Two types of velocity models are discussed: a simulated dataset generated by the authors and an open experimental dataset of the Society of Exploration Geophysics (SEG).  In addition, we compare the testing performance of our proposed method with that of conventional methods (i.e., FWI). In the discussion section, we present several open questions related to the utilization of our method for geophysical application.  In the conclusion section, summaries of this study are presented, and future work is outlined.}} All of the acronyms used in this paper  are listed in Table \ref{tab1}.

\section{FCN-based inversion method}
\label{sec2}

\subsection{Basic inversion problem}
The constant density 2D acoustic wave equation is expressed as
\begin{equation}\label{e1}
\left.
\begin{array}{lcl}
\frac{1}{v^{2}(x,z)}\frac{\partial^{2}{u(x,z,t)}}{\partial{{t^{2}}}}=\nabla^{2}{u(x,z,t)}+s(x,z,t),&{}
\end{array}
\right.
\end{equation}
where $(x,z)$ denotes the spatial location, $t$ represents time, $ v(x,z)$ is the velocity of the longitudinal wave at the corresponding location, $u(x,z,t)$ is the wave amplitude, $\nabla^{2}(\cdot)=\frac{\partial^{2}(\cdot)}{\partial{x^{2}}}+\frac{\partial^{2}(\cdot)}{\partial{z^{2}}}$ represents the Laplace operator, and $s(x,z,t)$ is the source signal.

Equation \ref{e1} is usually given by
\begin{equation}\label{e2}
\left.
\begin{array}{lcl}
u=H(v),&{}
\end{array}
\right.
\end{equation}
where {{\sout{$u\in Y$ is the wave field, and $v\in X$ denotes the model parameters that can be  velocity, density, and others}}} the operator $H(\cdot)$ maps $v$ to $u$, and is usually nonlinear.

The classical inversion methods aim at minimizing the following objective function:
\begin{equation}\label{e3}
\left.
\begin{array}{lcl}
\bar{v}=\arg\min\limits_{v}f(v)=\arg\min\limits_{v}\frac{1}{2}\|H(v)-d\|^{2}_{2},&{}
\end{array}
\right.
\end{equation}
where $d$ denotes the measured seismic data, $\|\cdot\|_{2}$ is the $l_{2}$ norm, and $f(\cdot)$ represents the data-fidelity residual.

In many applications, the engine for solving the above equation is to develop a fast and reasonably accurate inverse operator $H^{-1}$. An adjoint-state method \citep{Plessix2006} is used to compute the gradient $g(v)=\nabla f(v)$,  and iterative optimization algorithms are used to minimize the objective function. Owing to  the nonlinear properties of the operator $H$ and the imperfection of the surveys $d$, it is difficult to obtain precise subsurface models. Therefore,  minimizing the above equation is generally an ill-posed problem, and the solutions are non-unique and unstable. If  $d$ contains full-waveform information, the above equation presents an FWI.

\subsection{A review of the FCN}
Many DL algorithms are built with CNNs and  provide state-of-the-art performance in challenging inverse problems such as image reconstruction \citep{Schlemper2017}, super-resolution \citep{Dong2016}, X-ray-computed tomography \citep{Jin2017}, and compressive sensing \citep{Adler2017}. They are also studied as neuro-physiological models of vision \citep{Anselmi2016}.

\begin{figure*}
  \centering
  \includegraphics[width=1.0\textwidth]{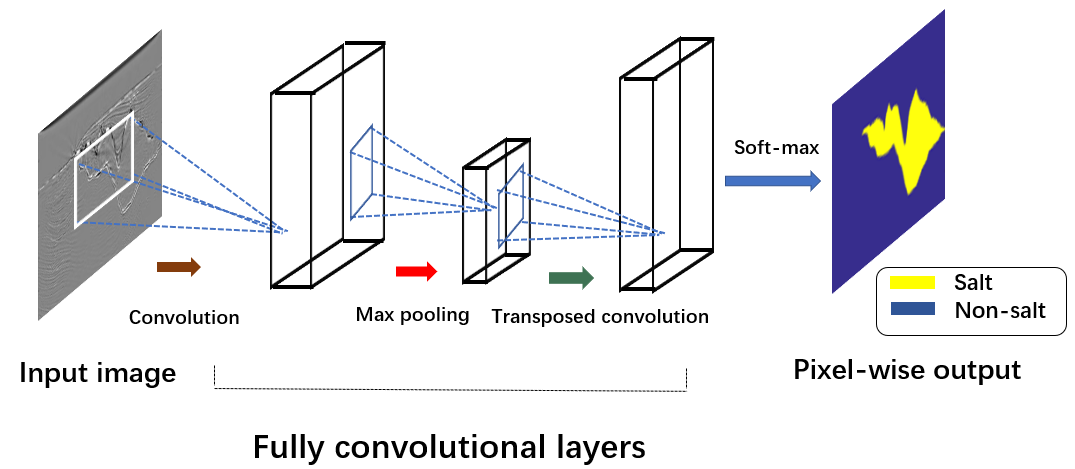}
  \caption{Sketch of a simple fully convolutional neural network (FCN) with {{\sout{a few convolutional layers}}} {{a convolutional layer, a pooling layer and a transposed convolutional layer. Migrated data were adopted as the input, and the pixel-wise output includes  salt and non-salt parts}}. }
  \label{fig2}
\end{figure*}

The FCN, proposed by \cite{Long2015} in the context of image and semantic segmentation, changes the fully connected layers of the CNN into convolutional layers  to achieve end-to-end learning. Figure \ref{fig2} shows a sketch of a simple FCN {{\sout{with a few convolutional layers}}}. In this {{\sout{simple}}} example, {\yc{migrated seismic data are used as input, which is followed by a convolutional layer. Then, a pooling layer is inserted in the middle. After application of max-pooling, the sizes of the feature maps  change to the previous one-half. Afterward, the transposed convolutional operation is applied to enlarge the size of the output to be the same as that of the input. Ultimately, we used a soft-max function to obtain the expected label, which is a label  indicates which pixels belong to the salt structure in the migrated data.}} This FCN method can be described as follows:
\begin{equation}\label{e5}
\left.
\begin{array}{lcl}
y=Net(x;\Theta)=S$($K_{2}*(M(R(K_{1}*x+b_{1})))+b_{2}$)$,&{}
\end{array}
\right.
\end{equation}
where $Net(\cdot)$ denotes an FCN-based network and also indicates the nonlinear mapping of the network, and $x,y$ denotes the inputs and outputs of the network, respectively. $\Theta=\{ K_{1},K_{2},b_{1},b_{2}\}$ is the set of parameters to be learned, including the convolutional weights ($K_{1}$ and $K_{2}$) and the bias ($b_{1}$ and $b_{2}$). $R(\cdot)$ introduces the nonlinear active function, such as the rectified linear unit  \citep{Dahl2013}, sigmoid, or exponential linear unit  \citep{Clevert2015}. $M(\cdot)$ denotes the subsampling function (e.g., max-pooling, average pooling).  $*$ is the convolutional operation, and {\yc{$S(\cdot)$ represents the soft-max function.}}

\subsection{{{\sout{DLVMB}}}  Mathematical framework}
\begin{figure*}
  \centering
  \includegraphics[width=1.0\textwidth]{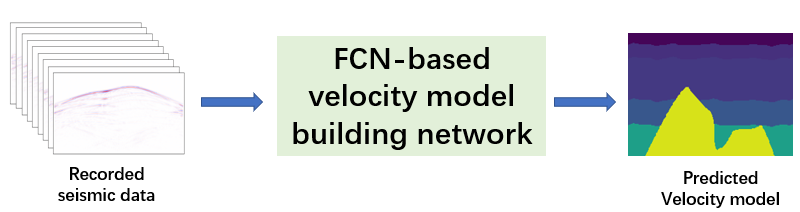}
  \caption{Schematic diagram depicting the velocity-model prediction from recorded seismic data by the {{\sout{deep-learning-based}}} {{fully convolutional neural network}}.}
  \label{fig3}
\end{figure*}

\begin{figure*}
  \centering
  \includegraphics[scale=1.0]{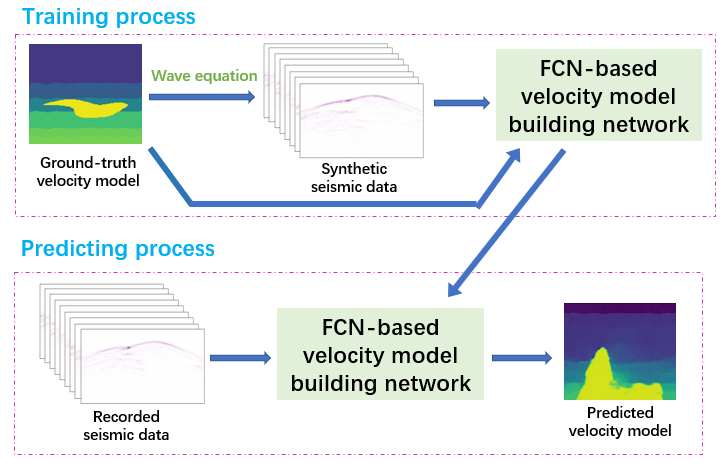}
  \caption{Flow chart of the FCN-based inversion process.}
  \label{fig4}
\end{figure*}

With the goal of  estimating velocity models using  seismic data as inputs directly, the network needs to project seismic data from the data domain $(x,t)$ to the model domain $(x,z)$, as shown in Figure \ref{fig3}. The basic concept of the {{\sout{DLVMB}}} proposed method is to establish the map between inputs and outputs, which can be expressed as
\begin{equation}\label{e9}
\left.
\begin{array}{lcl}
\widetilde{v}=Net(d;\Theta),&{}
\end{array}
\right.
\end{equation}
where $d$ is the raw unmigrated seismic data, and $\widetilde{v}$ denotes the {{\sout{p-wave}}} {\yc{P-wave}} velocity model predicted by the network. Our method contains two stages: the training process and the prediction process, as shown in Figure \ref{fig4}. Before the training stage, many velocity models are generated and are used as outputs.  The supervised network needs pairs of datasets. Therefore, the acoustic wave equation is applied as a forward model to generate the synthetic seismic data, which are used as inputs. Following the initial computation, the input--output pairs, which are named ${\{d_{n},v_{n}\}}_{n=1}^{N}$, are {{\sout{inputted}}} {\yc{input}} to the network for learning the  mapping.

During the training stage, the network learns to fit a nonlinear function from the input seismic data to the corresponding ground-truth velocity model. Therefore, the network learns by solving the optimization problem as
\begin{equation}\label{e10}
\left.
\begin{array}{lcl}
\widehat{\Theta}=\arg\min\limits_{\Theta}\frac{1}{mN}\sum\limits_{n=1}^{N}L(v_{n},Net(d_{n};\Theta)),&{}
\end{array}
\right.
\end{equation}
where $m$ represents the total number of pixels in one velocity model, and $L(\cdot)$ is a measure of the error between ground-truth values $v_{n}$ and prediction values $\widetilde{v}_{n}$. In our numerical experiments, {{\sout{the cross-entropy function (equation \ref{e8}) for detecting the structures and}}} the $l_{2}$ norm is applied for {{\sout{predicting the velocity values}}} {\yc{measuring the discrepancy}}.

For updating the learned parameters $\Theta$, the optimization problem can be solved by using back propagation and stochastic gradient-descent algorithms (SGD) \citep{Shamir2013}. The number of training datasets is large, and the numerical computation of the gradient $\triangledown _{\Theta}L(d;\Theta)$ is not feasible based on our GPU memory. Therefore, to approximate the gradient, the mini-batch size $h$ was applied for calculating $L_{h}$, i.e., the error between the prediction values and the corresponding ground-truth values of a small subset of the whole training dataset,  in each iteration. This  led to the following optimization problem:
\begin{equation}\label{e11}
\left.
\begin{array}{lcl}
\widehat{\Theta}=\arg\min\limits_{\Theta}\frac{1}{mh}L_{h}=\arg\min\limits_{\Theta}\frac{1}{mh}\sum\limits_{n=1}^{h}\|v_{n}-Net(d_{n};\Theta)\|_{2}^{2},&{}
\end{array}
\right.
\end{equation}
Here, the ground-truth velocity models $v_{n}$ are given during the training process but are unknown during testing. {\yc{One epoch is defined when an entire training dataset is passed forward and backward through the neural network once. The training dataset is first shuffled into a random order and is subsequently chosen sequentially in mini-batches to ensure one pass.}} It should be noted that the loss function is different from that (equation \ref{e3}) in  FWI, in which the loss measures the squared difference between the observed and simulated seismograms. In our case, we used the Adam algorithm \citep{Kingma2014}, i.e., a deformation of the conventional SGD algorithm. The parameters are iteratively updated as follows:
\begin{equation}
\left.
\begin{array}{lcl}
\Theta_{t+1}=\Theta_{t}-\delta g(\frac{1}{mh}\triangledown_{\Theta} L_{h}(d_{n};\Theta;v_{n})),&{}
\end{array}
\right.
\end{equation}
where  $\delta$ is the  positive {{\sout{learning rate}}} {\yc{step size}}, and $g(\frac{1}{mh}\triangledown_{\Theta} L_{h}(d_{n};\Theta;v_{n}))$  denotes a function. {{\sout{with respect to $\triangledown_{\Theta} L$ (see Adam algorithm)}}} {\yc{This algorithm is straightforward to implement, computationally efficient,  and well suited for large problems in terms of data or parameters.}}

The network is built once the training process is completed. During the prediction stage, other unknown velocity models are obtained by the available learned network. In our work, the input seismic data for prediction is also synthetic seismic traces. In a real situation, however, the input  is field data.  The method can be calculated by algorithm \ref{alg1}.

\begin{algorithm}[!htbp]
\caption{\small  FCN-based inversion method}
\textbf{Input:} $\{d_{n}\}_{n=1}^{N}$: seismic data, $\{v_{n}\}_{n=1}^{N}$: velocity models, $T$: epochs, $lr$: learning rate of network, $h$: batch size, $num$ : number of training sets \\
\textbf{Given notation:} $\ast$ : 2D convolution with channels including zero-padding, $\ast_{\uparrow}$ : 2D deconvolution (transposed convolution), $R(\cdot)$ :  rectified linear unit, $B(\cdot)$ : batch normalization, $M(\cdot)$ : max-pooling, $C(\cdot)$ : copy and concatenate,  $\Theta=\{K, b\}$ : learnable parameters, $L$: loss function, $Adam$: SGD algorithm\\
\textbf{Initialize:} $t=1$, $loss=0.0$, $y_{0}=d$\\
\textbf{1.Training process}
\begin{enumerate}
 \item Generate different velocity models that have similar geological structures.
 \item Synthesize seismic data using the finite-difference scheme.
 \item Input all data pairs into the  network and use the Adam algorithm to update the parameters.

    \begin{algorithmic}
    \FOR {t=1:1:T and $(data,models)$ in $traing$ $set$}
    \FOR {j=1:1:num/h}
    \FOR {i=1:1:l-1}
    \STATE $y_{i}\leftarrow B(R(K_{(2i-1)}\ast y_{i-1}+b_{(2i-1)}))$\\
    \STATE $m_{i}\leftarrow B(R(K_{(2i)}\ast y_{i}+b_{(2i)}))$\\
    \STATE $y_{i}\leftarrow M(m_{i})$
    \ENDFOR
    \STATE $y_{l}\leftarrow B(R(K_{(2l-1)}\ast y_{l-1}+b_{(2l-1)}))$\\
    \STATE $y_{l}\leftarrow B(R(K_{(2l)}\ast y_{l}+b_{(2l)}))$\\
    \FOR {i=l-1:-1:1}
    \STATE $y_{i}\leftarrow B(R(K_{(2l+3(l-i)-2)}\ast_{\uparrow} y_{i+1}+b_{(2l+3(l-i)-2)}))$\\
    \STATE $m_{i}\leftarrow B(R(K_{((2l+3(l-i)-1))}\ast C(y_{i},m_{i})+b_{(2l+3(l-i)-1)}))$\\
    \STATE $y_{i}\leftarrow B(R(K_{(2l+3(l-i))}\ast m_{i}+b_{(2L+3(l-i))}))$
    \ENDFOR
    \STATE $\widetilde{v}\leftarrow B(R(K_{(5l-2)}\ast y_{1}+b_{(5l-2)}))$\\
    \STATE $loss=L_{h}(\widetilde{v},v)$\\
    \ENDFOR
    \STATE $\Theta_{j+1}\leftarrow Adam(\Theta_{j};lr;loss)$
    \ENDFOR
   \end{algorithmic}

\end{enumerate}
\textbf{2.Prediction process}
\begin{enumerate}
  \item Synthesize seismic data for different velocity models in the same way as that used for generating the training seismic data.
  \item Input new seismic  data into the learned network for prediction.
\end{enumerate}
\textbf{Output:} Predicted velocity model $v^{\ast}$
\label{alg1}
\end{algorithm}

\subsection{Architecture of the  {{\sout{DLVMB}}} network }
\begin{figure*}
  \hspace{-0.8cm}
  \includegraphics[width=1.1\textwidth]{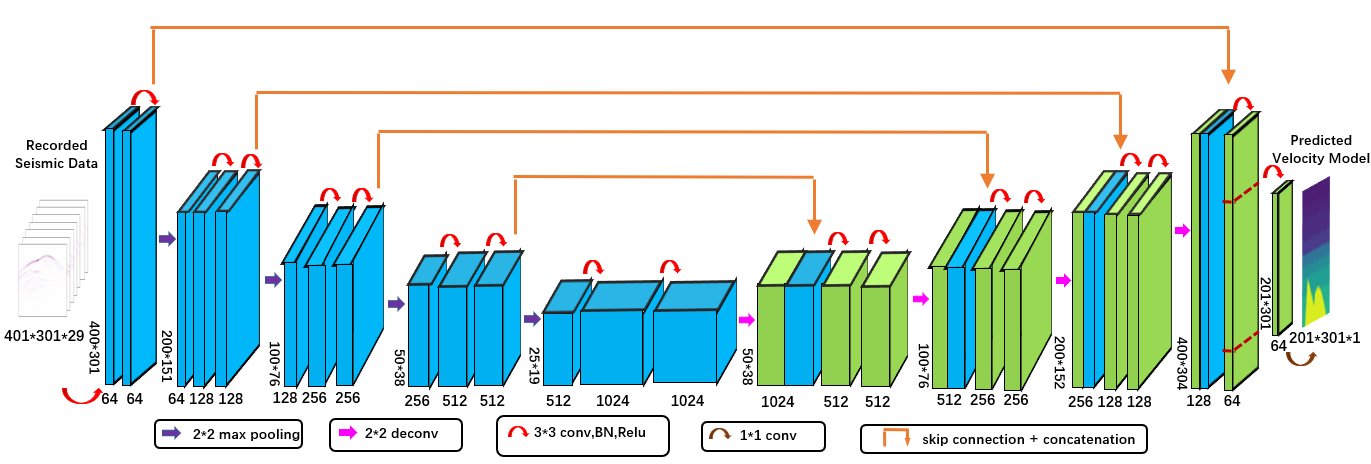}
  \caption{Architecture of the network used  for seismic velocity inversion. Each blue {{and green}} {{\sout{box}}} {{cube}} corresponds to a multi-channel feature map. The number of channels is shown on bottom of the {{\sout{box}}} {{cube}}. The x-z size is provided at the lower left edge of the {{\sout{box}}} {{cube}} (example shown for $25\times 19$ in lower resolution).  The arrows denote the different operations, and the size of the corresponding parameter set is defined in each box.  The abbreviations shown in the explanatory frame,  i.e., conv, max--pooling, BN, Relu, deconv and skip connection + concatenation, are defined in Table \ref{tab2}.}
  \label{fig8}
\end{figure*}

To achieve automatic seismic VMB from the raw seismic data, we adopted and modified the UNet \citep{Ronneberger2015} architecture, which is a specific network built upon the concept of the FCN. Figure \ref{fig8} shows the detailed architecture of the proposed  network. It consists of a contracting path (left) used to capture the geological features and a symmetric shape of an expanding path (right) that enables precise localization. {\yc{This symmetric form is an encoder--decoder structure and employing a contraction--expansion structure
based on the max-pooling and the transposed convolution. The effective receptive field of the network increases as the input goes deeper into the network, when a fixed size convolutional kernel ($3*3$ in our case) is given.}} The numbers of channels in the left path are 64, 128, 256, 512, and 1024, as {\yc{the}} network depth increases. Skip layers are adopted to combine the local, shallow feature maps in the right path with the global, deep feature maps in the left path. We summarize the definitions of the different operations in Table \ref{tab2}, where $K$ and $\overline{K}$ denote the convolutional kernels. The mean and standard deviation in the batch normalization  were calculated per dimension over the mini-batch. $\varepsilon$ is a value added to the denominator for numerical stability, $\gamma$ and $\beta$ are also learnable parameters; however, they were not used in our method.

We made two main modifications to the original UNet to fit the seismic VMB. First, the original UNet, proposed in the image-processing community, reads input images in RGB color channels that represent the information from the input images. {{\sout{while for processing the seismic data here}}} {\yc{To process the seismic data}}, we assigned different shot gathers, generated at different source locations but from the same model as channels for the input. Therefore, the number of input channels is the same as the number of sources for each model. The multi-shot seismic data were fed into the network together to improve data redundancy. Second, {{\sout{different from}}} {\yc{in}} a usual UNet, the outputs and inputs are in the same (image) domain.  However, for our goal, we expected the network to realize the domain projection, i.e., to transform the data from the (x, t) domain to the (x, z) domain and {{\sout{tag the geologic bodies simultaneously}}} {\yc{to build the velocity model simultaneously}}. To complete this, the size of feature maps obtained by the final $3*3$ convolution was  truncated to be the same size as the velocity model, and the channel of the output layer was modified to 1.  This was done so that the neural network could train itself during the contracting and expanding processes to map the seismic data to the exact velocity model directly. {\yc{The main body of the network is similar to that in the original UNet,}} and  23 convolutional layers in total are used in the network. {{\sout{For different tasks, i.e., velocity inversion and structure detection, the number of channels of the first and last layer was different, i.e.,  the first layer had 29 channels and last layer had 1 channel for the first task, but in terms of second task, the number of channels was 21 and 9 for first and last layer, respectively.}}}

\section{Numerical experiments and results}

{\yc{In this section, the data preparation, including the model (output) design and data (input) design for training  and testing datasets, is first presented. Subsequently, we use the simulated training dataset to train the network for velocity inversion, and we predict other unknown velocity models by the valuable learned network. Further, for SEG salt model training, the trained network for simulated models  is regarded as the initialization; this pre-trained network is a common approach used in transfer learning \citep{Pan2010}.  The  testing process on the SEG  dataset is also performed.}}  {{\sout{As stated earlier, inversion is our primary objective. However, evaluating the quality of velocity-model building, which included two-aspect applications, produced by our approach is important as it will condition the ultimate quality of our results. In terms of the velocity inversions studies, we first trained the network just with the simulated velocity models, then predicted the other input seismic traces. After that, we regarded the trained network as the pre-trained network to continue to train with the SEG salt velocity models and tested new model again.}}}  We  compare the numerical results between our method and FWI. {{\sout{For all of the numerical examples, the hyper-parameter selections are shown in Table 4, which depends on the empirical experiments and experimental equipments.}}}  The numerical experiments are performed on an HP Z840 workstation with a Tesla K40 GPU, 32 Core Xeon CPU, 128 GB RAM, and an Ubuntu operating system that implements PyTorch ({\tt http://pytorch.org}).

\subsection{Data preparation {{\sout{for training set}}}}
To train an efficient network, a suitable large-scale  training set, i.e.,  input--output pairs, is needed. In a typical FCN model, training outputs are provided by some of the labeled images. {\yc{In this paper, 2D synthetic  models are utilized for testing the data-driven  method. Two types of  velocity models are provided for numerical experiments:  2D simulated  models and  2D SEG salt models extracted  from a 3D salt model ({\url {https://wiki.seg.org/wiki/SEG/EAGE_3D_modeling_Salt_Model_Phase-C_1996}})}}. {{\sout{due to  no}}}  {{\sout{currently, to the research environment of to the authors}}}  Each velocity model is unique.

{{\sout{In this study, we used the DLVMB network to solve the problem of p-wave velocity inversion. Additionally, as a particular example, we also used our method to automatically detect geological structures. According to the definition of the two applications, we used different approaches to generate data set for these two tasks,  i.e., inversing velocity  and identifying boundaries and locations of  geological structures (e.g., salt domes and faults).}}}

{{\sout{This is the first step we try to  allow deep learning work for geological application. For reducing the difficulty of this problem in the beginning, only  p-wave velocity was used to describe characteristics of subsurface structures.}}}

{\yc{\textit{Training dataset}}}

\begin{figure*}
  \centering
  \includegraphics[width=1.0\textwidth]{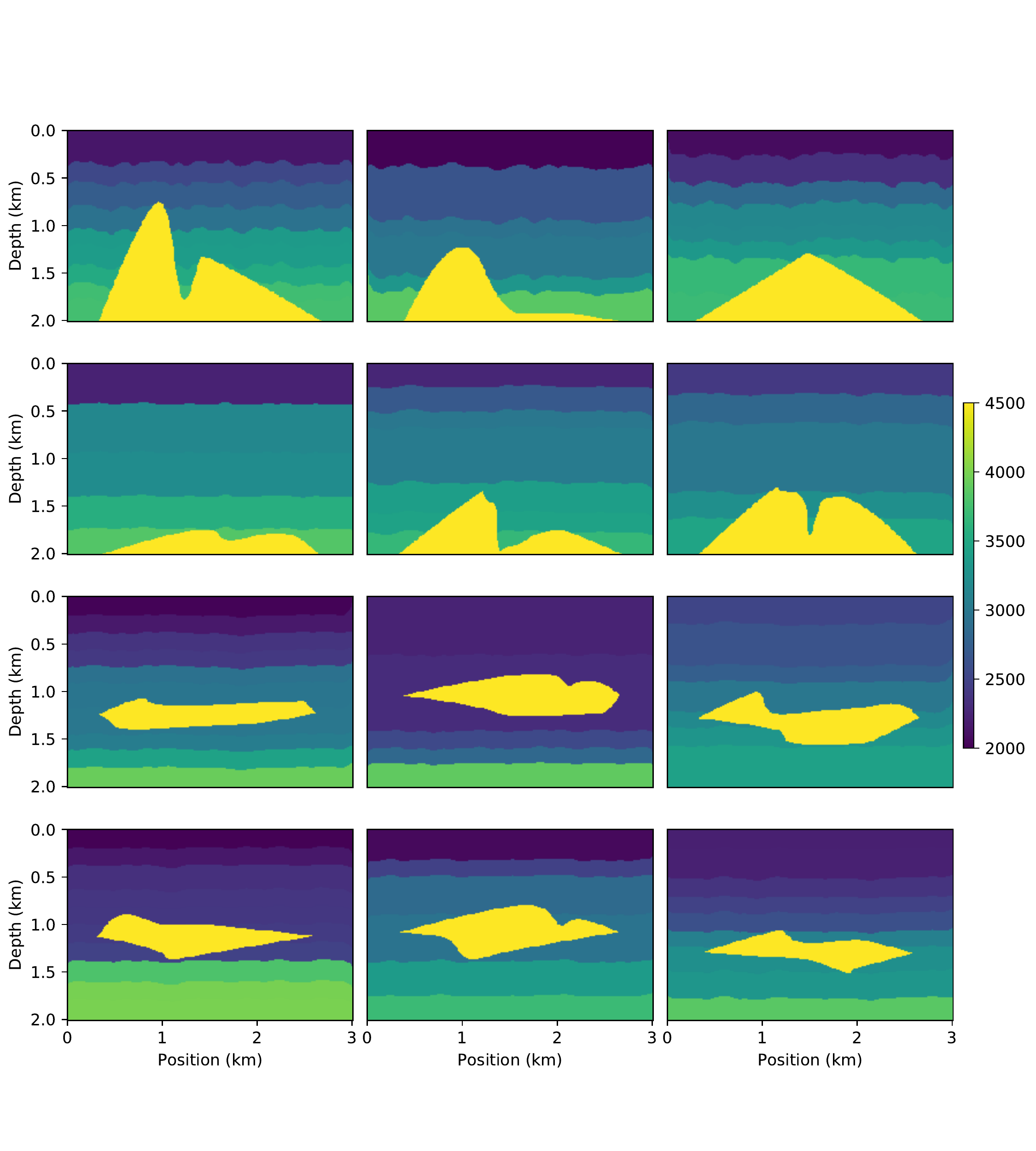}
  \caption{Twelve representative samples from 1600  simulated training velocity models.}
  \label{fig6-1}
\end{figure*}

\begin{figure*}
  \centering
  \includegraphics[width=1\textwidth]{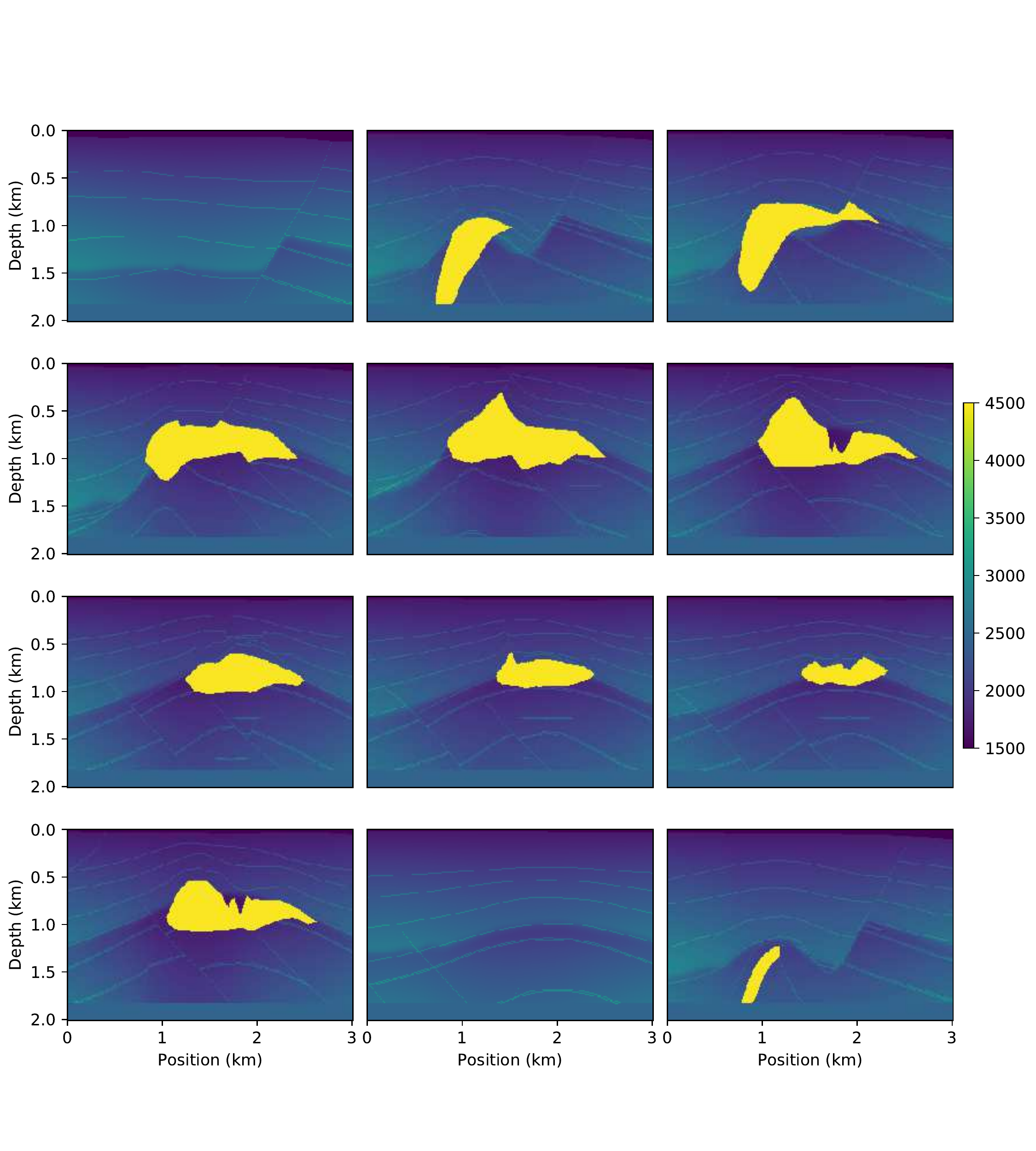}
  \caption{Twelve representative samples from 130  SEG-salt training velocity  models.}
  \label{fig6-2}
\end{figure*}

\begin{figure*}
  \centering
  \subfigure[]{\label{fig6-3}
  \includegraphics[width=1\textwidth]{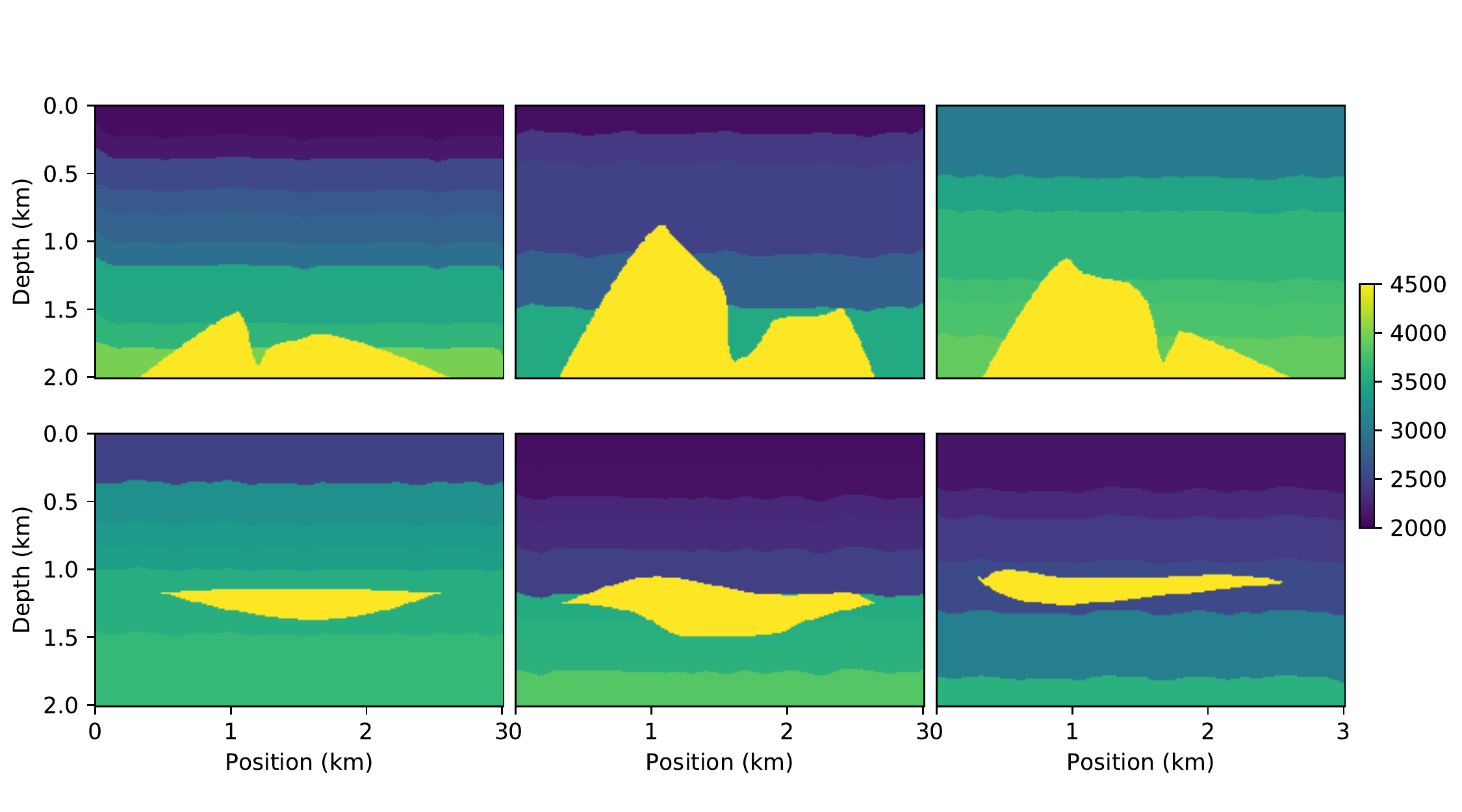}}
  \hspace{-0.2cm}
  \subfigure[]{\label{fig6-4}
  \includegraphics[width=1\textwidth]{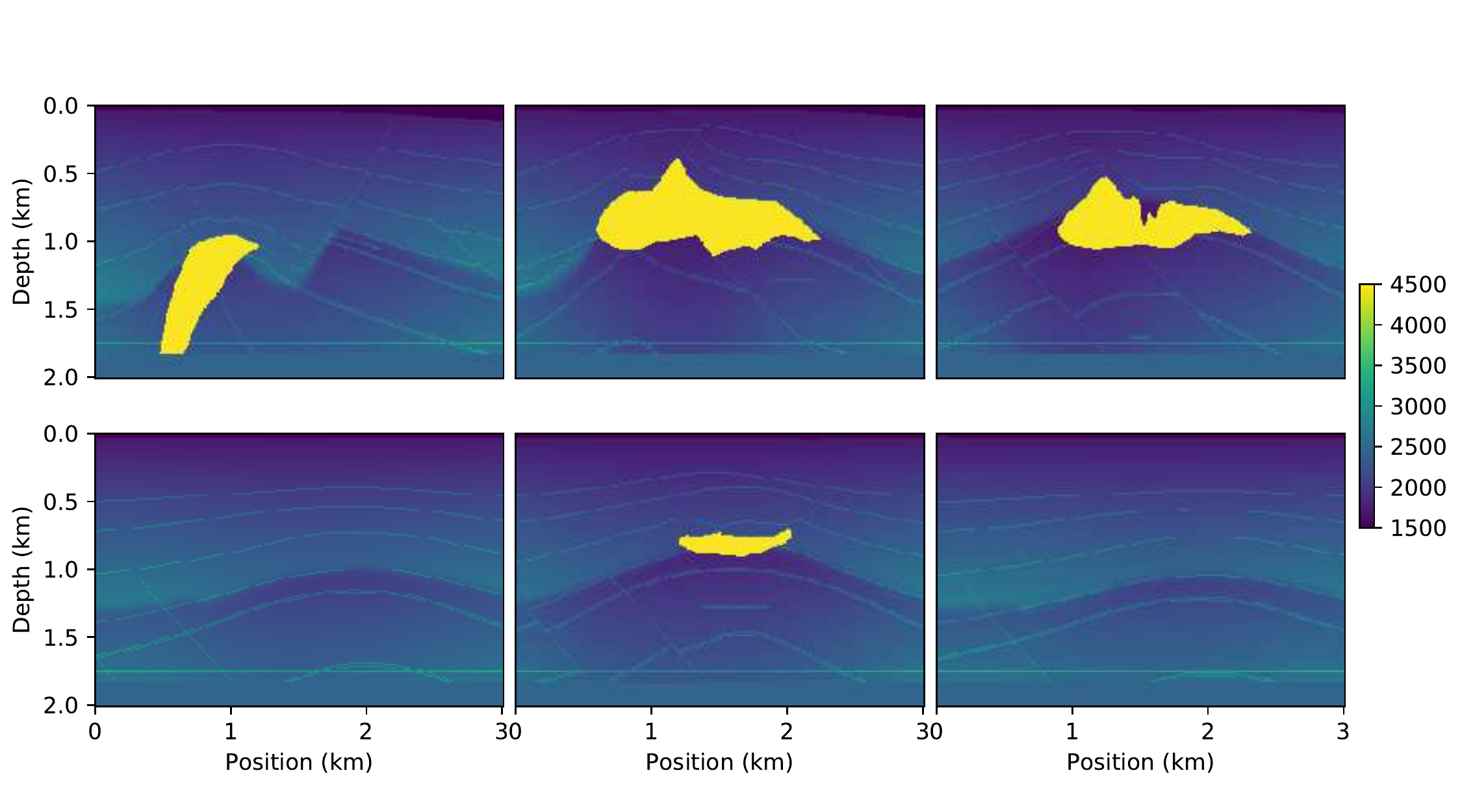}}
  \caption{Typical samples from testing dataset for velocity inversion: (a) six velocity models of the simulated dataset; (b) six  velocity models of the Society of Exploration Geophysics (SEG) dataset. }
  \label{fig6-0}
\end{figure*}

{\yc{\textit{Model (output) design:}}} To explore and prove the available  capabilities of DL for seismic waveform inversion,  we first {{\sout{attempted to}}} generated random velocity models  with smooth interface curvatures and increased the velocity values with depth. For the sake of simplicity, we assumed that each  model had 5 to 12 layers as the background velocity and that the velocity values of each layer ranged arbitrarily from 2000 m/s to 4000 m/s. A salt body with an arbitrary shape and position was embedded into each model, each having a constant velocity value of 4500 m/s. The size of each velocity model used $x\times z$ = $201\times 301$ grid points with a spatial interval $\triangle{x}$ = $\triangle{z}$ = 10 m. Figure \ref{fig6-1} shows 12  models from the simulated training dataset, and Figure \ref{fig6-3} shows 6 examples of the testing dataset. {\yc{In our work, the simulated training dataset contained 1600 velocity samples.}} To better apply our new method for inversion, a 3D salt velocity model from the SEG reference website was utilized for obtaining the 2D salt models. This type of model had {\yc{the}} same size {{\sout{with}}} {\yc{as the}} {{
\sout{synthetic}}} {\yc{simulated}} models, and the values ranged from 1500 m/s to 4482 m/s. Figure \ref{fig6-2}  {{\sout{are}}} {\yc{shows}} the 12 representative examples of the SEG salt models from the training dataset, and  Figure \ref{fig6-4} {{\sout{are}}} {\yc{shows}} the 6 models from the testing dataset. Owing to the limited extraction,  130 velocity models were included in SEG training dataset.

\textit{{\yc{Data (input) design:}}} To solve the acoustic wave equation, we used the time-domain stagger-grid finite-difference scheme that adopts a second-order time direction and eighth-order space direction \citep{Ozdenvar1997,Hardi2016}. For each velocity model, 29 sources were evenly placed, and shot gathers were simulated sequentially. The recording geometry consisted of 301 receivers evenly placed  at a uniform spatial interval. {\yc{The detailed parameters for forward modeling are shown in Table \ref{tab4}}.} {{\sout{We used {{\sout{as a source wavelet}}} a Ricker {\yc{source}} wavelet with a dominant frequency of 25 Hz, the sampling time interval {{\sout{was}}} {\yc{of}} 0.001 s, and the maximum {{\sout{travel}}} {\yc{recording}} time {{\sout{was}}} {\yc{of}} 2 s.}}} The perfectly matched layer (PML) \citep{D.Komatitsch2003}  absorbing boundary condition was adopted to reduce {{\sout{unwanted}}} {\yc{unphysical}} reflection on the left, right{\yc{,}} and bottom edges. Additionally,  to verify the stability of our method,  we added Gaussian noise,  with zero mean and standard derivation of $5\%$, to each testing seismic data. Moreover, we {{\sout{magnified the amplitude of seismic data to two times larger}}} {\yc{made the amplitude of the seismic data two times higher}}. The noisy or magnified data were also used as inputs and were fed into the network  to invert the velocity values.

{\yc{\textit{Testing dataset}}}

The ground-truth velocity models of the testing dataset had  geological structures similar to those of the training dataset owing to the usage of supervised learning method. All of the velocity models  for prediction were not included in the training dataset and were unknown in the prediction process.  The input seismic data for prediction were also obtained by using the same method as that used for generating the inputs for the training dataset. For simulated models and SEG salt models, the testing dataset was composed of 100 and 10 velocity samples, respectively.

\subsection{{{\sout{Results of the}}} {\yc{Inversion for simulated dataset}}}

The first inversion case  was performed for  2D {{\sout{synthetic}}} {\yc{simulated}} velocity models. During the training stage, the training batch for each epoch was constructed by randomly choosing 10 samples of velocity-model dimension $201\times 301$ from the training dataset. In each pair data, the dimension of one-shot seismic data was downsampled to $400 \times 301$. {{\sout{The learning rate was set to 0.001, and the learned DLVMB}}} The network deemed to work better was selected when {\yc{the hyper-parameters were set as shown in Table \ref{tab3}}} based on the training dataset and experimental guidance \citep{Bengio2012}. The mean squared error between the prediction  velocity  values and ground-truth velocity values is shown in Figure \ref{fig9-2}.  Figure  \ref{fig12-10}--Figure \ref{fig12-12} show  three exemplified results of the {{\sout{DLVMB}}} proposed method. Visually, a generally good match was achieved between the predictions and the corresponding ground-truth.

In this case study, a comparison between our method and FWI was performed.  We used the same parameter setting as that used to generate the training seismic data for the time-domain forward modeling. Multi-scale frequency-domain inversion strategy \citep{Sirgue2008} was adopted. The selected inversion frequencies were 2.5 Hz, 5 Hz, 10 Hz, 15 Hz, and 21 Hz, based on the research of  \cite{Sirgue2004}. An adjoint-state based gradient descending method was adopted in  this experiment \citep{Plessix2006}. The observed data of FWI was same as the seismic data we used for prediction.  In addition, the true velocity model smoothed by the Gaussian smooth function was taken as the initial velocity model, as  shown in Figure \ref{fig12-4}--Figure \ref{fig12-6}. The numerical experiments of FWI were performed on a computer cluster with four Tesla K80 GPU units, and a central  operating system.  Figure \ref{fig12-7}--Figure \ref{fig12-9} shows the results of FWI. All subfigures have the same colorbar, and the velocity value ranges from 2000 m/s to 4500 m/s. In this scenario, the FCN-based inversion method showed comparable results and preserved most of the geological structures.

To quantitatively analyze the accuracy of the predictions, we chose two horizontal positions, x = 900 m and x = 2000 m,  and we plotted the prediction (blue), FWI (red), and ground-truth (green) velocity values in the velocity versus depth profiles shown in Figure \ref{fig13}. Most prediction values matched well with the ground-truth values. Moreover, in Figure \ref{fig23}, a comparison of the shot records of the 15th receiver is displayed, including the observed data using the ground-truth velocity model, reconstructed data obtained by simulating the inversion result of FWI, and reconstructed data obtained by forward modeling the  prediction velocity model of our method. The reconstructed data with {{\sout{DLVMB}}} predictions obtained by the proposed method also matched well with the observed data.

The process of FWI for one {{\sout{synthetic}}} {\yc{simulated}} model inversion incurred a GPU time of 37 min. In contrast, after training the {{\sout{DLVMB}}} {\yc{FCN-based inversion}} network {\yc{with 1078 min}}, the GPU time per prediction in our simulated models was only 2 s on the lower-equipment machine, which is more than 1000 times less than that for FWI.

{{\sout{For further exploiting, we showed more promising experiments}}} {\yc{To further validate the capability of the proposed method, more test results under more realistic conditions should be performed}}. Here, we {{\sout{just}}} present experiments {{\sout{related to}}} {\yc{in which}} seismic data {\yc{are}} contaminated  with random noise or the seismic amplitude {\yc{is doubled}}. {{\sout{for an example}}}. {{\sout{For the noisy seismic data, which was generated by adding Gaussian noise ($\mu=0, \sigma=5\% *$ minimum amplitude of each shot), was put into the trained network for prediction. Because the seismic data can be denoised before inversion, so we just added small-scale random noise to test the sensitivity to noise of our method.}}} {\yc{The noisy data were generated by adding  zero-mean  Gaussian noise  with a standard deviation of $5\%$.}}  Figure \ref{fig12-31}--Figure \ref{fig12-33} show the prediction results using the proposed method with  noisy inputs;  examples are shown in second column of Figure \ref{fig20}. A comparison of the predictions (i.e., the results shown in Figure \ref{fig12-10}--Figure \ref{fig12-12}) and the clean inputs (e.g., examples shown in first column of Figure \ref{fig20}) revealed that our method still provides acceptable results. However, compared with the ground truth, some parts of the predictions are not close to the true values, particularly the superficial background layers. This may have been caused by perturbations. In future research, the sensitivity to other type noise will be considered such as coherent noise and multiples.

Similarly, to test the sensitivity to amplitude, another test was performed in which the amplitude of the testing seismic data was doubled; examples are shown in third column of Figure \ref{fig20}. In this test, the processed data were applied as the input for prediction; the performance comparison is displayed in Figure \ref{fig12-34}--Figure \ref{fig12-36}. The prediction velocities using the processed inputs with higher amplitudes were consistent with the predictions using the original inputs. This is in compliance with the theoretical analysis and indicates that our proposed method achieves velocity inversion adaptively and stably.

\subsection{{\yc{Inversion for SEG salt dataset}}}

To further show the  outstanding ability of the {{\sout{DL}}} proposed method,  the  trained network for the simulated dataset was utilized as the initialized network, which is one approach of transfer learning to train the SEG salt dataset. In the training process, the number of epochs  was set to 50, and each epoch had 10 training samples (i.e., training mini-batch size). The other hyper-parameters for learning were same as {{\sout{DLVMB}}} those used for {{\sout{synthetic}}} {\yc{simulated}} model inversion. Figure \ref{fig9-3} shows the mean-squared error  of the SEG training  dataset. The loss converged to zero when  only 130 models were used for training. Similar to the test above, a  comparison was performed between the  proposed  method and FWI with the same algorithm as that used in the experiments for simulated models. In this FWI experiment, the selected inversion frequency was 2.5 Hz, and the other three values ranged from 5 Hz to 15 Hz with a uniform frequency interval of 5 Hz. The initial velocity models were also obtained by using the Gaussian smooth function shown in Figure \ref{fig13-4}--Figure \ref{fig13-6}. Figure \ref{fig14} describes  all performances of the numerical experiments, in which all subfigures have the same colorbar, and the value is from 1500 m/s to 4500 m/s.   The comparison results of velocity values in the velocity versus depth profiles are displayed in Figure \ref{fig15}. In this test, compared with FWI method, the proposed {{\sout{DLVMB}}}  method yielded a slightly lower performance, which could be attributed to  the small number of training datasets. However, the predictions using the pre-trained initialized network (i.e., transfer learning in our study) were better than those obtained using the random initialized network.  {\yc{Moreover, the results of our method could  serve as the initial  models for FWI or travel time tomography. They could also be used for on-site quality control during  seismic survey development.}}

The additional experiments using SEG salt datasets  under more realistic conditions are also performed. When the seismic data were contaminated with noise, as shown  in the second column of Figure \ref{fig21}, most prediction values obtained by the {{\sout{DLVMB}}} inversion network  shown in Figure \ref{fig13-31}--Figure \ref{fig13-33} were close to ground-truth velocities, but were slightly lower  than the predictions obtained using the clean data. However, the prediction results shown in Figure \ref{fig13-34}--Figure \ref{fig13-36}  using  the inputs with higher amplitudes, as  shown in third column of Figure \ref{fig21},  made no differences. In this test, because the training dataset  of the SEG salt models was less than that for the simulated models, the performance of the proposed method for the SEG dataset was not outstanding. Therefore, in  future work, we will augment the diversity of the training set gradually and take advantage of the transfer learning to apply this novel method on other complex samples.

{\yc{For one SEG salt velocity-model inversion, the process of FWI  incurred a GPU time of 25 min.}} In comparison, the training time of our method for all model inversion was 43 min; the GPU time per prediction in the SEG salt dataset was 2 s on the lower-equipment machine. A comparison of the time consumed for training and prediction process is shown in Table \ref{tab7}.

In summary, the numerical experiments provide promising evidence for the feasibility of our {{\sout{DLVMB}}} proposed method for velocity inversion {{\sout{and structure detection}}} from the raw input of seismic shot gathers directly without the need for initial velocity models. This indicates that the neural network can effectively approximate  non-linear mapping even when the inputs have perturbation. Compared with conventional FWI, the computational time of the proposed method is fast because it does not involve the {\yc{iterations to search for optimal solutions}}. The main computational costs  are incurred mostly during the training stage, which is only once during the model setup; this can be handled off-line in advance. After training, the prediction costs are negligible.  Thus, the FCN-based inversion method makes the overall computational time a fraction of that needed for traditional physical-based  inversion techniques.

\section{Discussion}

From an experimental perspective, the numerical results demonstrate that our proposed method presents promising capabilities of DL for velocity inversion. The objective of the research is to apply the latest breakthroughs in data science, particularly in DL techniques.  Although the indications of our method are inspiring, many factors can affect its performance, including the choice of training dataset; the selection of hyper-parameters such as learning rate, batch size, and training epoch; and the architecture of the neural network. For our purposes, we focused on the profound understanding of DL
applied for seismic inversion.  Therefore a discussion on  more impressive results and the advantages and  disadvantages of the new method is  provided in this section.

\textit{(a) How does the training dataset affect the network?}

The limitation of our approach is that the capability of the network relies  on the dataset. In general, the models to be trained should involve structures or characteristics similar to those contained in the predictions. That is, the supervised learned network for prediction is limited to the choice of the training dataset, and the amount of data required for training depends on many factors. In most cases, a large amount of large-scale and diverse training samples results in a more powerful network. Moreover, the time consumed for training process is longer. One representative test of our proposed method for SEG salt models was conducted. In our experiments, predictions without salt are also presented. A comparison of the results between our method and   FWI are shown in Figure \ref{fig16}. Our method yielded a lower performance than FWI. In particular, the  sediment was vague because only 10 training samples without salt were utilized for training. Thus, the capability of the network to learn these models is  lower than that for other salt models.

In addition, according to the simple geological structures contained in the simulated dataset, such as several smooth interfaces, increasing background velocity, and constant-velocity salt, relatively high similarity was noted between the training and testing datasets. According to experimental guidance \citep{Bengio2012} and the other similar research \citep{Wangw2018,Wu2018},  the number of simulated training datasets was set to 1600.  The effect of the of training dataset number on the network  will be investigated in the future.

In training stage of the DL method, one needs a training dataset that including lots of training pairs. The observed data could be a 1-shot gather, or other number shot gathers. In our method, the shot numbers were fixed and were specific to the network such that 29 available shot seismic datasets were used for training the 29-shot network; in the prediction step, the same was considered for the 29 shot gathers. For further exploration, a comparison of performances with only 1-, 13-, 21-, 27-, and 29-shot training data is shown in Figure  \ref{fig11} owing to the constraint of the GPU memory. In Figure \ref{fig11-1}, the comparison mean loss revealed that all of the training losses in different cases  converged to zero along the epoch number. This indicates that our proposed method can be applied with arbitrary training shots  and may be an advantage over the traditional FWI.  Figure \ref{fig11-2}--Figure \ref{fig11-4} show  the testing performance for the mean loss, mean peak signal-to-noise ratio (PSNR), and mean structural similarity (SSIM).  The 1-shot case displayed a little bit unstable. The quantitative results may  be misleading, however, because  all testing evaluations are obtained for 10 selected network during training stage and take their average.  In our next work, we will investigate the effects of training shots to apply the novel method in a more realistic scenario.

\textit{(b) How can we apply the network when a lack of low frequencies exists in the testing data?}

The lack  of low frequencies in field data is a  main problem for practical application of FWI. However in the ML or DL methods, it is possible to learn the ``low frequency" from simulation data or prior-information data. Two other numerical experiments are provided to show the  performance of our method. As shown in Figure \ref{fig17},  all of the training datasets have low-frequency information, which is same as the original information used for prediction. However, the low frequency (i.e., 0--1/10 normalized Fourier spectrum) of the testing seismic data is removed by Fourier transform and the Butterworth high-pass filter. Then,  the reconstructed seismic data, shown in Figure \ref{fig17-6}, were used for prediction, the results of which are shown in Figure \ref{fig17-2}. In this case, the proposed method  predicted most parts of the velocity model. A comparison of the prediction shown in Figure \ref{fig17-1} with complete data shown in Figure \ref{fig17-5} revealed that the structure boundaries are less clear, and the background velocity layers are somewhat vague, which could be attributed to the low-frequency information.

In addition, the performance of the supervised learning method relies on a training set. Therefore, the new training seismic dataset missing the low frequencies, which was processed by using the same approach as that used for the testing data (Figure \ref{fig17-6}), and the corresponding ground-truth velocity models could be utilized to train the network. The prediction results are displayed in Figure \ref{fig18}. Visually, the predictions were slightly better than that shown in Figure \ref{fig17} but were still lower than those with complete data.

\textit{(c) Is the learned network robust and stable for any prediction?}

A general question often asked when learning is applied to some problems is whether the method can be generalized to other problems, e.g., whether a method trained on a specific dataset can be applied to another dataset without re-training.  Thus far, it has been difficult to test complex (e.g., SEG salt models) or real models by using the trained network directly because the performance of our proposed method relies on the datasets, and similar distribution is relatively weak between the two types different velocity models. In our work, transfer learning \citep{Pan2010}, i.e., a research problem in ML that focuses on storing knowledge gained while solving one problem and applying it to a different but related problem, was applied when the new training models are similar to the simulated models. The goal of using the pre-trained network as an initialization is to more effectively show the nonlinear mapping between the inputs and outputs rather than just allowing the machine to remember the characteristics of the dataset. A comparison of the training loss versus the number of epochs between random initial networks (i.e., the same as parameter initialization in UNet) and a pre-trained initial network (i.e., trained network for a simulated dataset) is shown in Figure \ref{fig24}. The network learned better with the pre-trained initialization in the same computational time.

\section{Conclusion}

In this study, we proposed a supervised end-to-end  DL method in a new fashion for velocity inversion that presents an alternative  to ``conventional" FWI formulation. In the proposed formulation, rather than performing local-based inversion with respect to subsurface parameters, we used a FCN to reconstruct these parameters. After a training process, the network is able to propose a subsurface model from only seismic data. The numerical experiments  showed impressive results in the potential of the DL in seismic model building and  clearly demonstrated that a neural network can effectively approximate the inverse of a non-linear operator that is very difficult to resolve.  The learned network still computes satisfactory velocity profiles when the seismic data are under more realistic conditions. Compared with  FWI, once the  network training is completed, the reconstruction costs are negligible. Moreover, little human  intervention is needed, and no initial velocity setup is involved. The loss function is measured in the model domain, and no seismograms are generated when using the network for prediction. In addition, no cycle-skipping problem exists.

The large-scale diverse training set plays an important role in the supervised learning method.  Inspired by the success of transfer learning and generative adversarial learning in computer vision, and the combination of  traditional methods and neural networks. We propose two possible directions for future work. The first is to generate more complex and realistic velocity models using a generative adversarial network, which is a type of semi-supervised learning network, based on the limited open dataset. Then, we can train the network with these complex datasets and apply the trained network to field data by transfer learning. The second  is to uncover the potential relationship between conventional approaches for inversion and specific networks. This approach enables to develop novel network designs that can reveal the hidden wave-equation model and invert  more complex geological structures  based on the physical systems. Further studies are required to adopt these methods to large problems, field data, and other applications.

\section{Acknowledgements}
The authors would like to thank the editors and reviewers for offering useful comments to improve this manuscript.  Thanks are extended to Dr. Wenlong  Wang for providing primary CNN code compiled with PyTorch and the suggestion of feeding multi-shot gathers into the network together to improve data redundancy.  This work is supported in part by the National Key Research and Development Program of China under Grant 2017YFB0202902, NSFC under Grant 41625017 and Grant 91730306, and the China Scholarship Council.

\clearpage

\bibliographystyle{seg}

\bibliography{DLIgeo}{}

\newpage

\listoffigures
\newpage

\listoftables

\clearpage

\clearpage

\clearpage

\clearpage

\clearpage

\clearpage

\clearpage

\clearpage
\begin{figure*}
  \centering
  \hspace{-0.3cm}
  \includegraphics[width=0.48\textwidth]{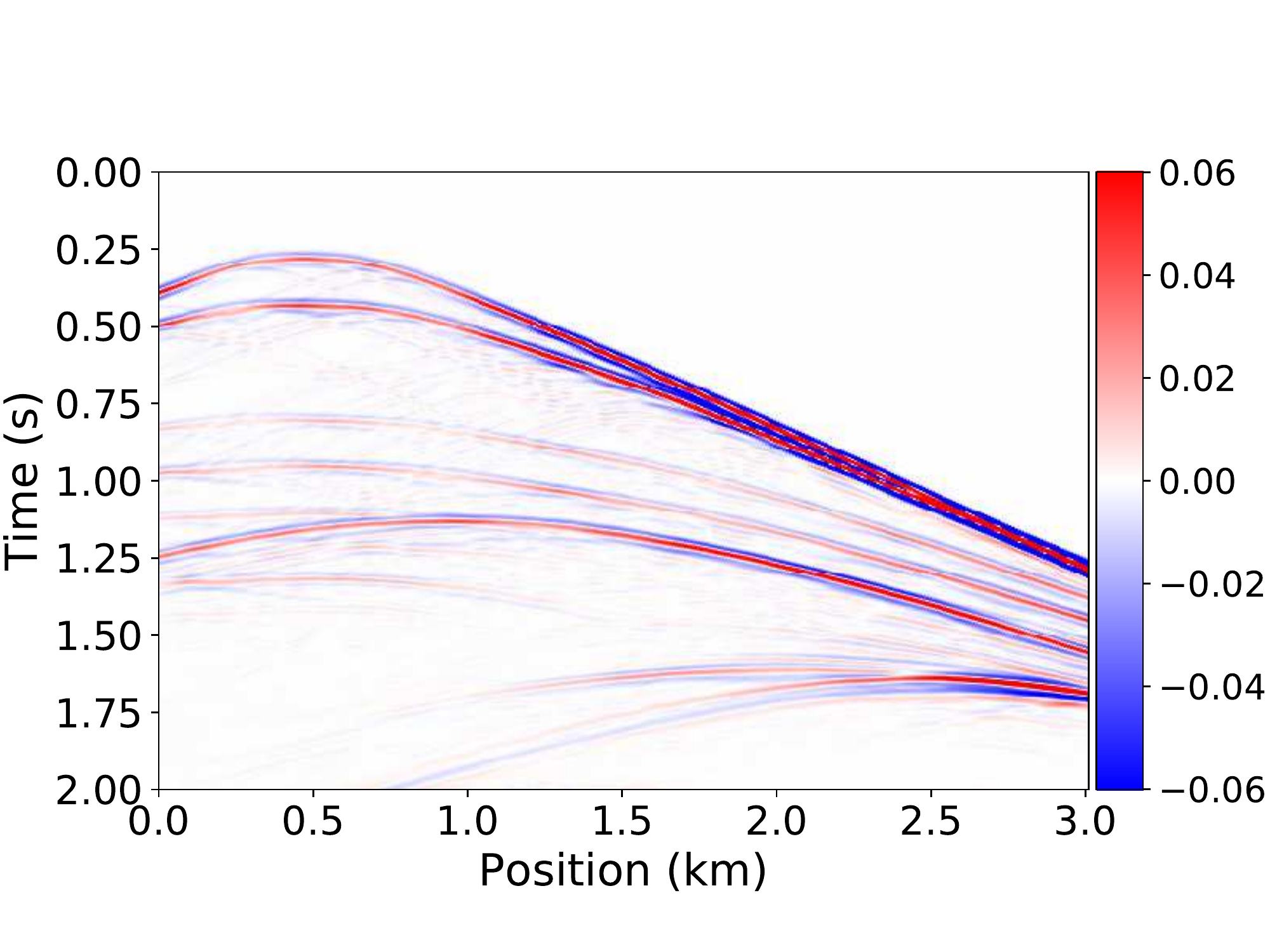}
  \hspace{-0.1cm}
  \includegraphics[width=0.48\textwidth]{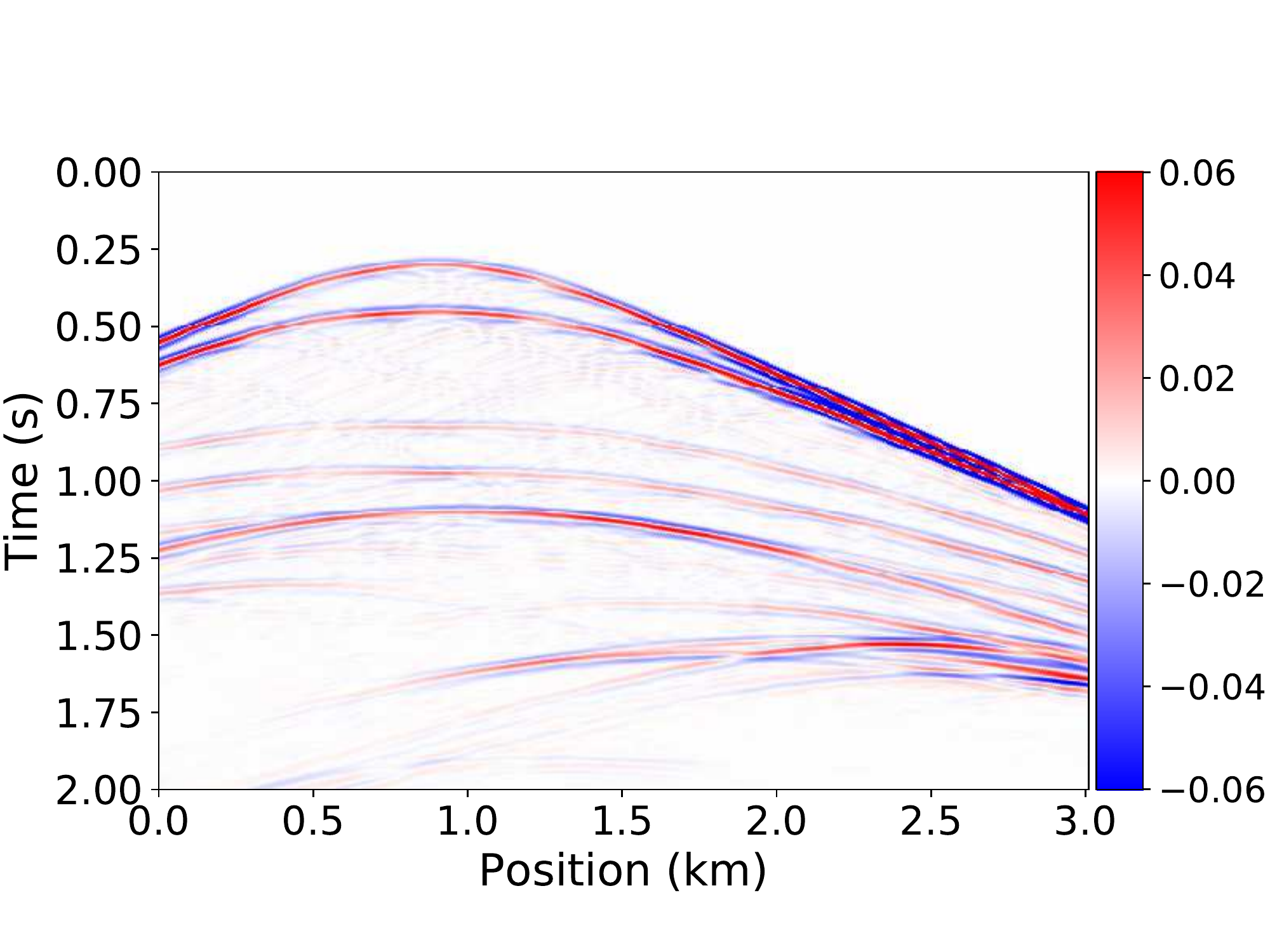}\\
  \hspace{-0.3cm}
  \includegraphics[width=0.48\textwidth]{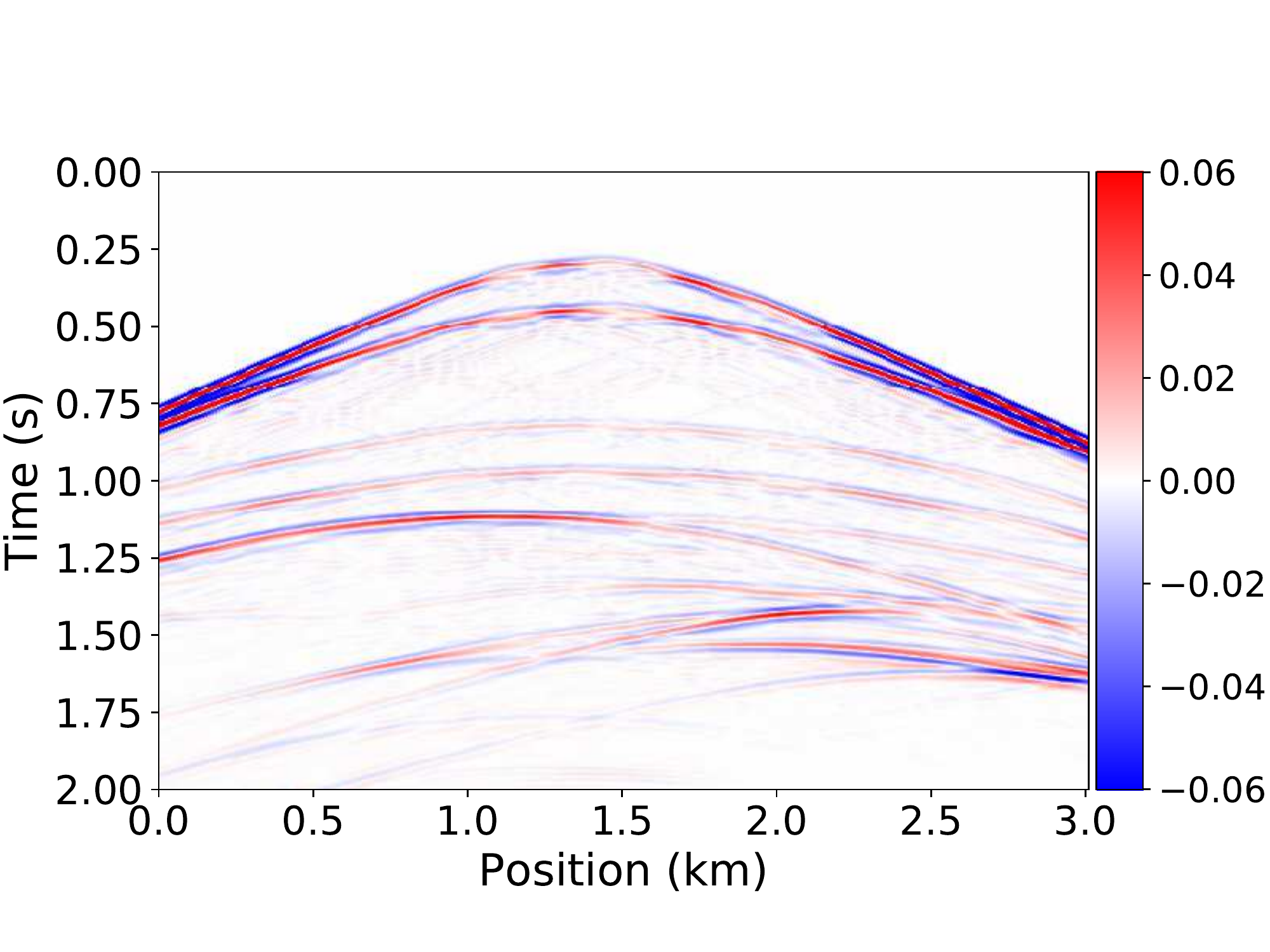}
  \hspace{-0.1cm}
  \includegraphics[width=0.48\textwidth]{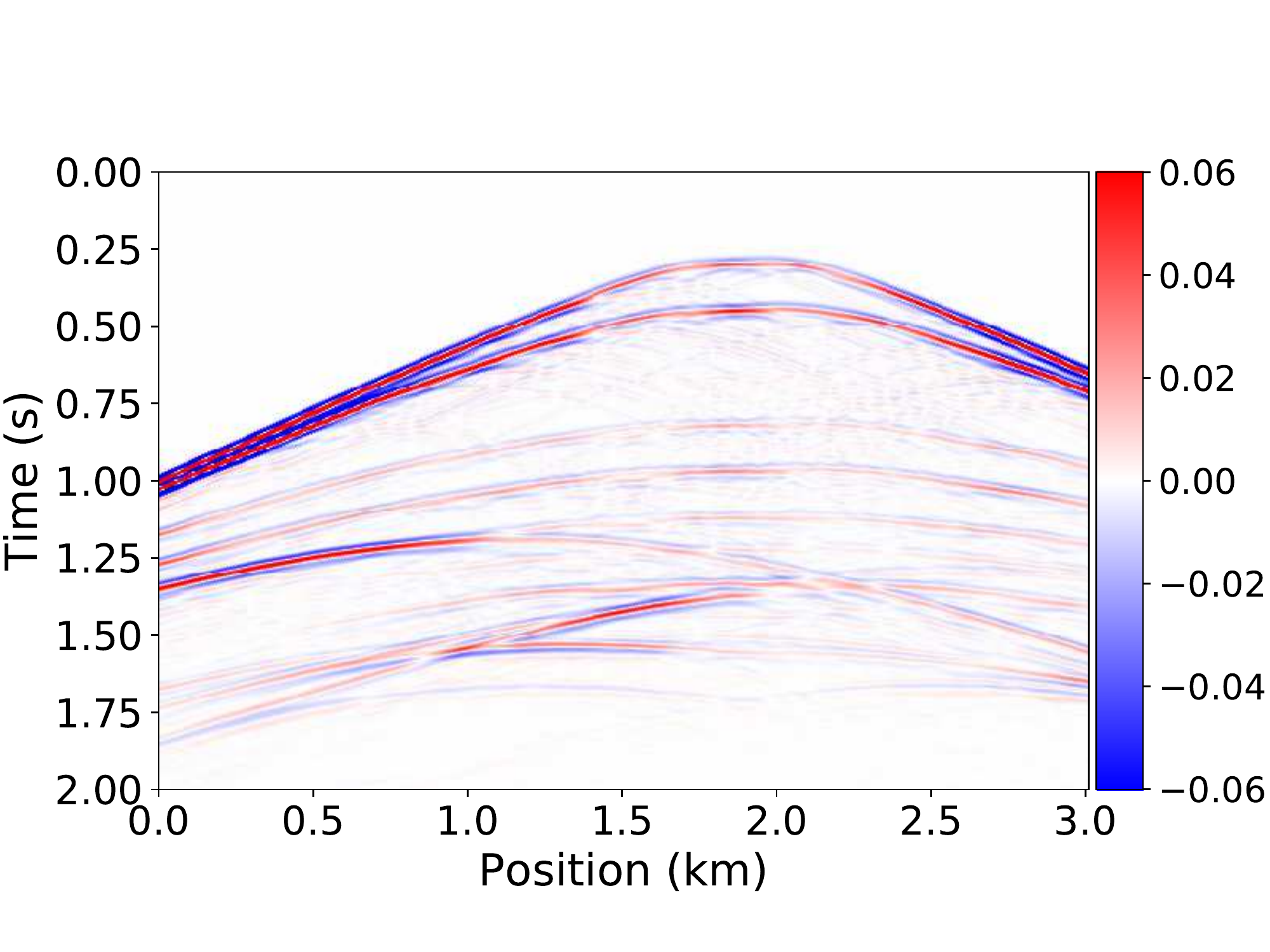}\\
  \hspace{-0.3cm}
  \includegraphics[width=0.48\textwidth]{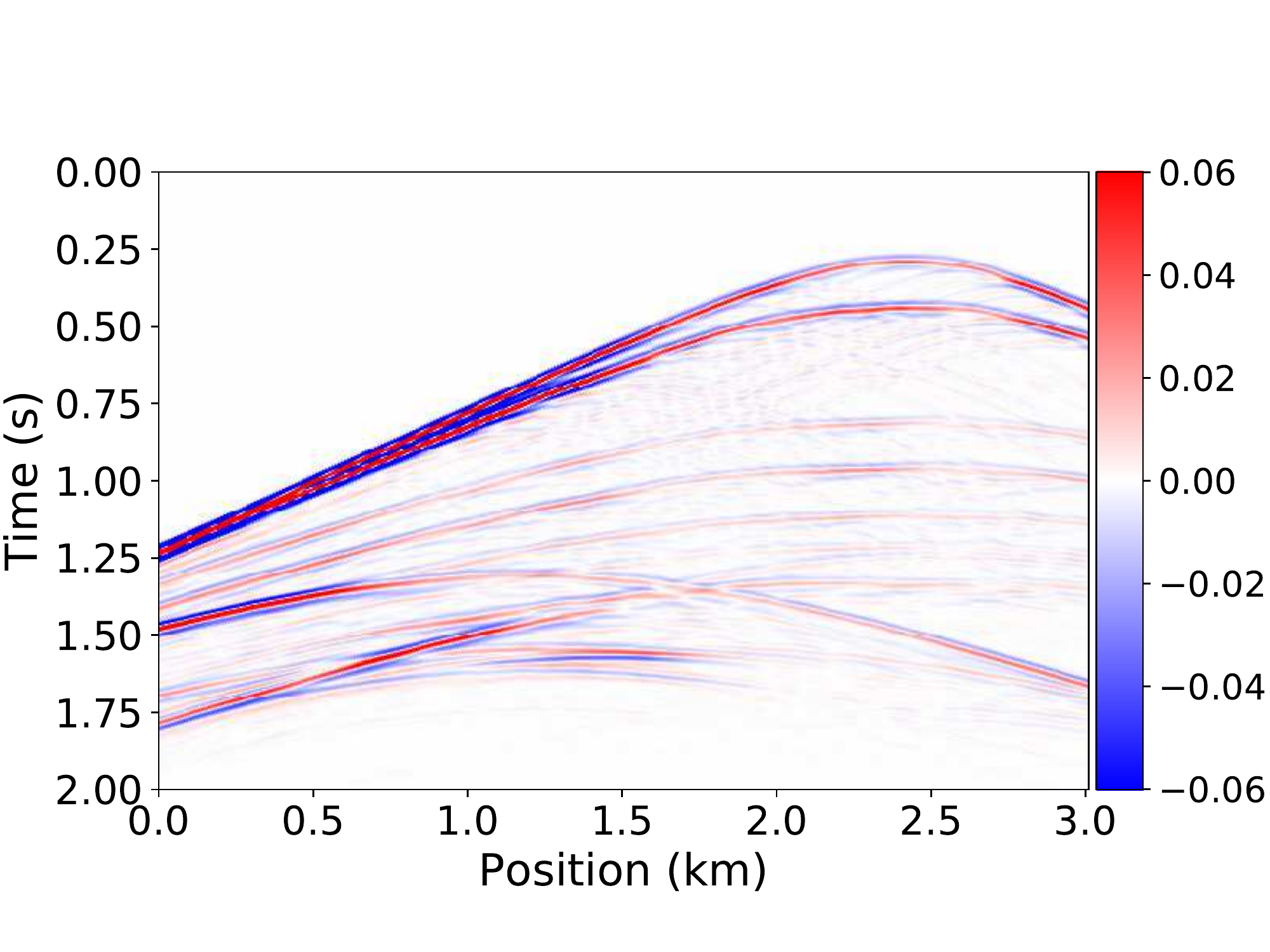}
  \hspace{-0.1cm}
  \includegraphics[width=0.48\textwidth]{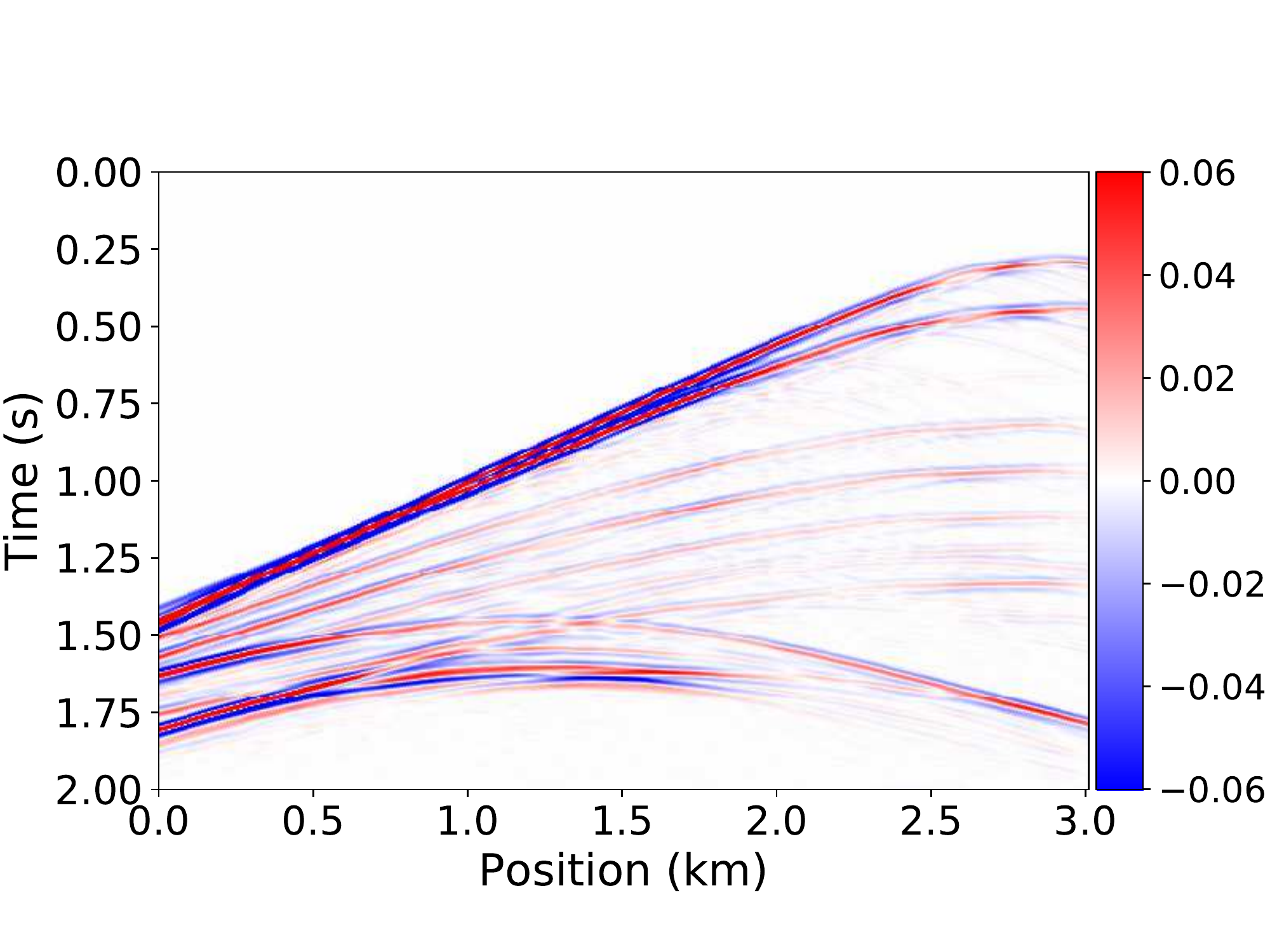}\\

  \caption{Six shots of the seismic data  generated by the finite-difference scheme.  The corresponding velocity model is the first model shown in Figure \ref{fig6-1}. }
  \label{fig7}
\end{figure*}

\clearpage
\begin{figure*}
  \centering
  \hspace{-0.8cm}
  \subfigure[]{\label{fig9-2}
  \includegraphics[width=0.53\textwidth]{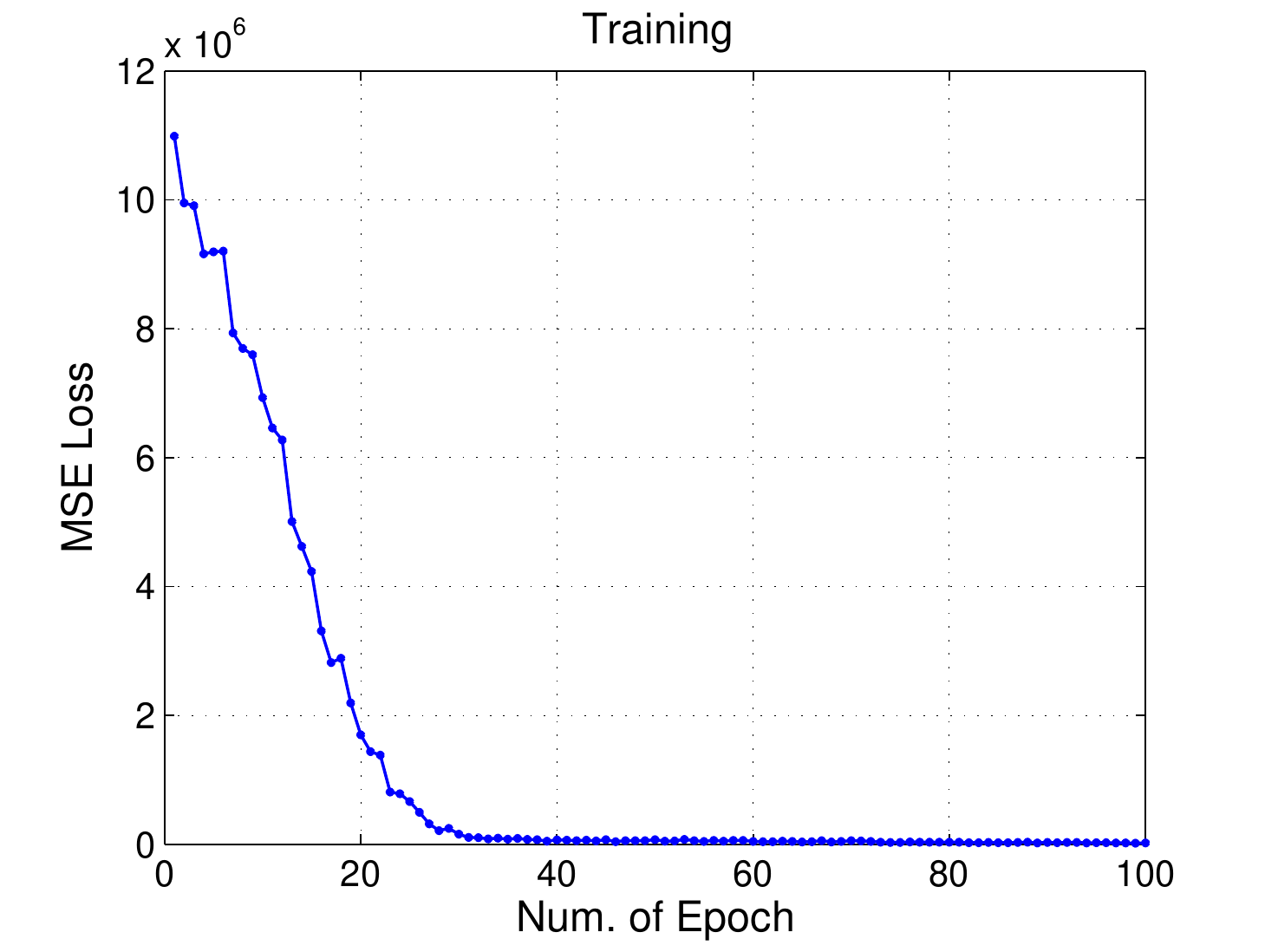}}
  \hspace{-0.8cm}
  \subfigure[]{\label{fig9-3}
  \includegraphics[width=0.53\textwidth]{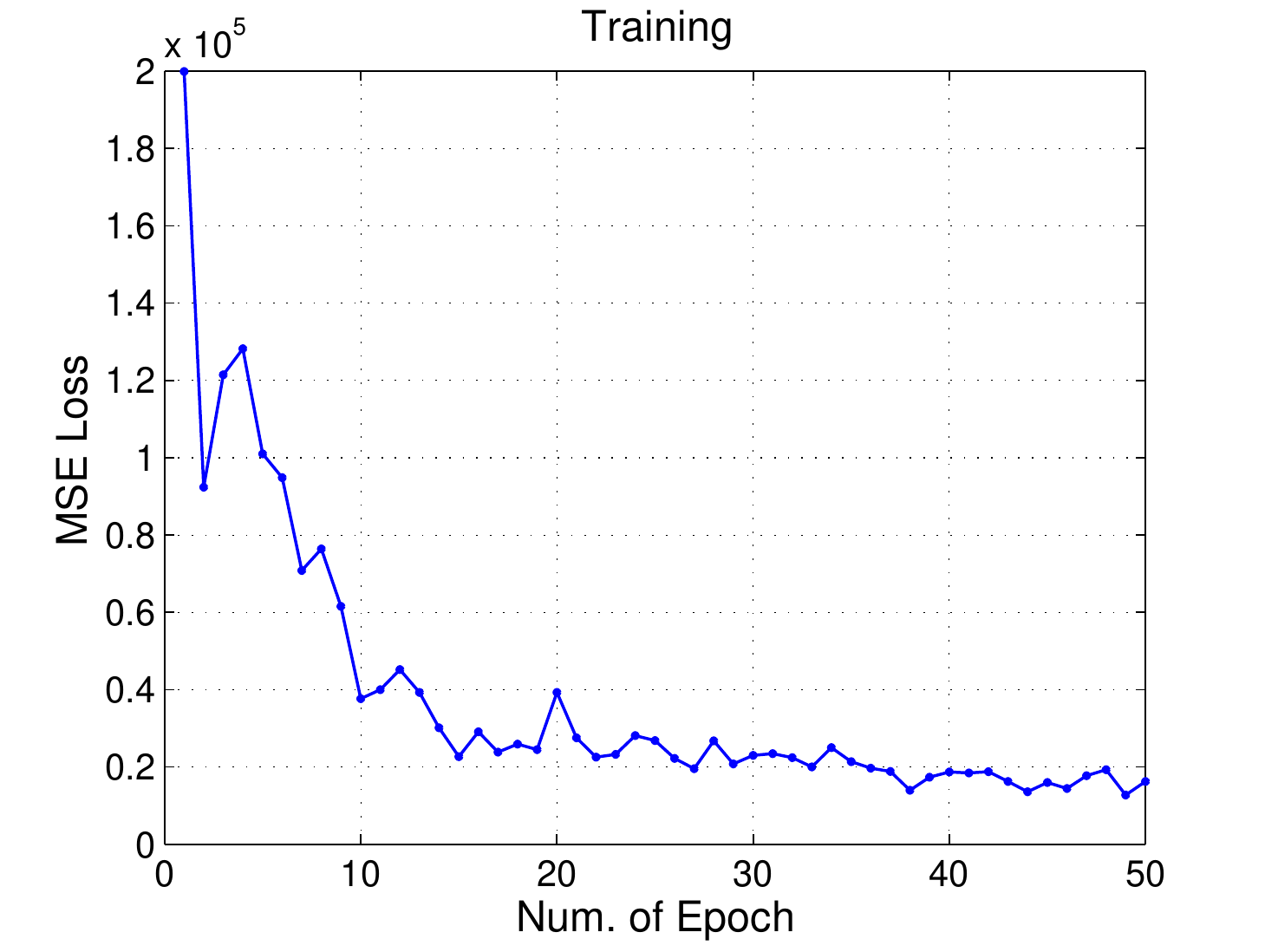}}

  \caption{Loss decreases during the training process:  (a) mean-squared error for the simulated velocity inversion; (b) mean-squared error for the SEG velocity inversion. }
  \label{fig9}
\end{figure*}

\clearpage
\begin{figure*}
\centering
  \hspace{-0.4cm}
  \subfigure[]{\label{fig12-1}
  \includegraphics[width=0.35\textwidth]{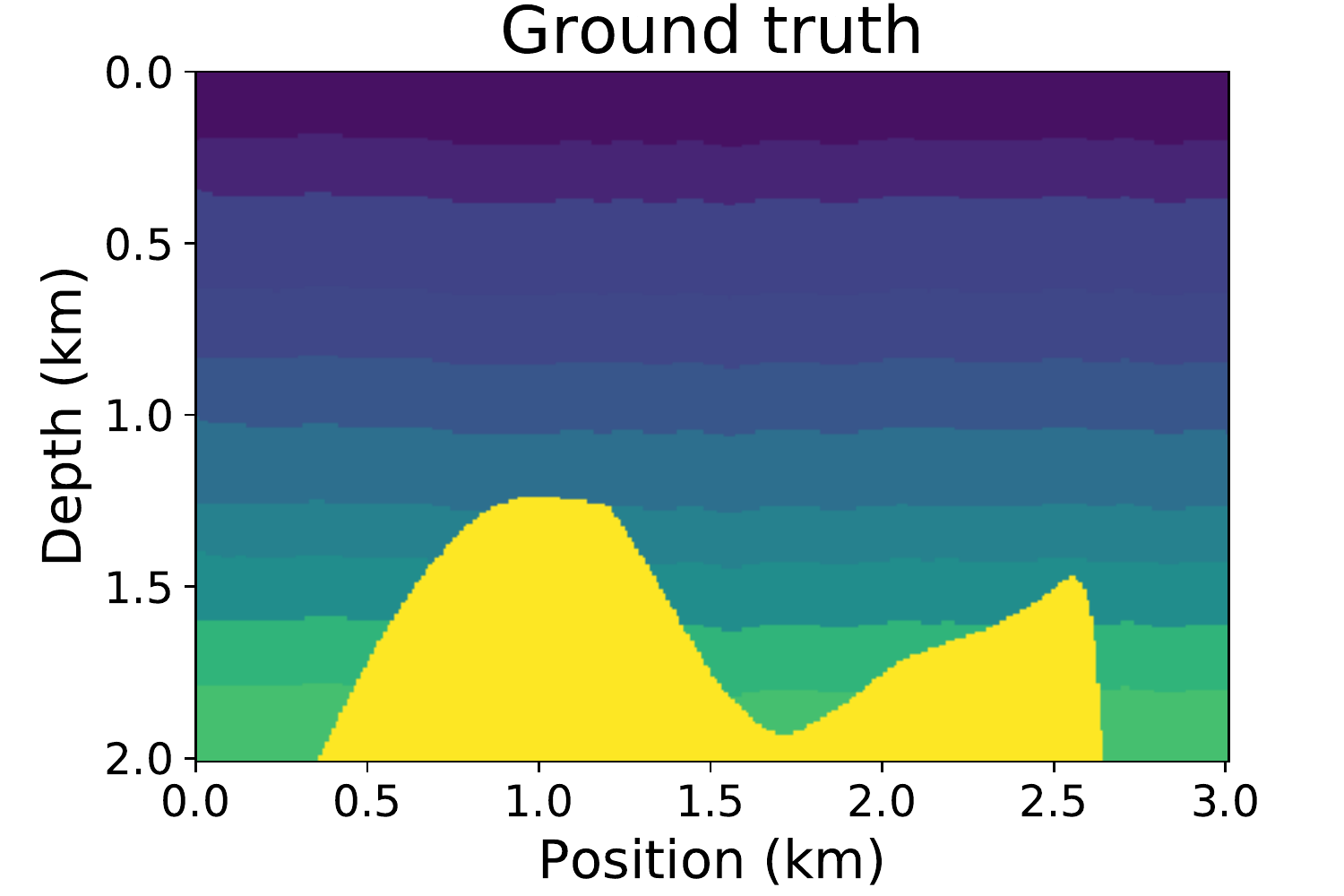}}
  \hspace{-0.7cm}
  \subfigure[]{\label{fig12-2}
  \includegraphics[width=0.35\textwidth]{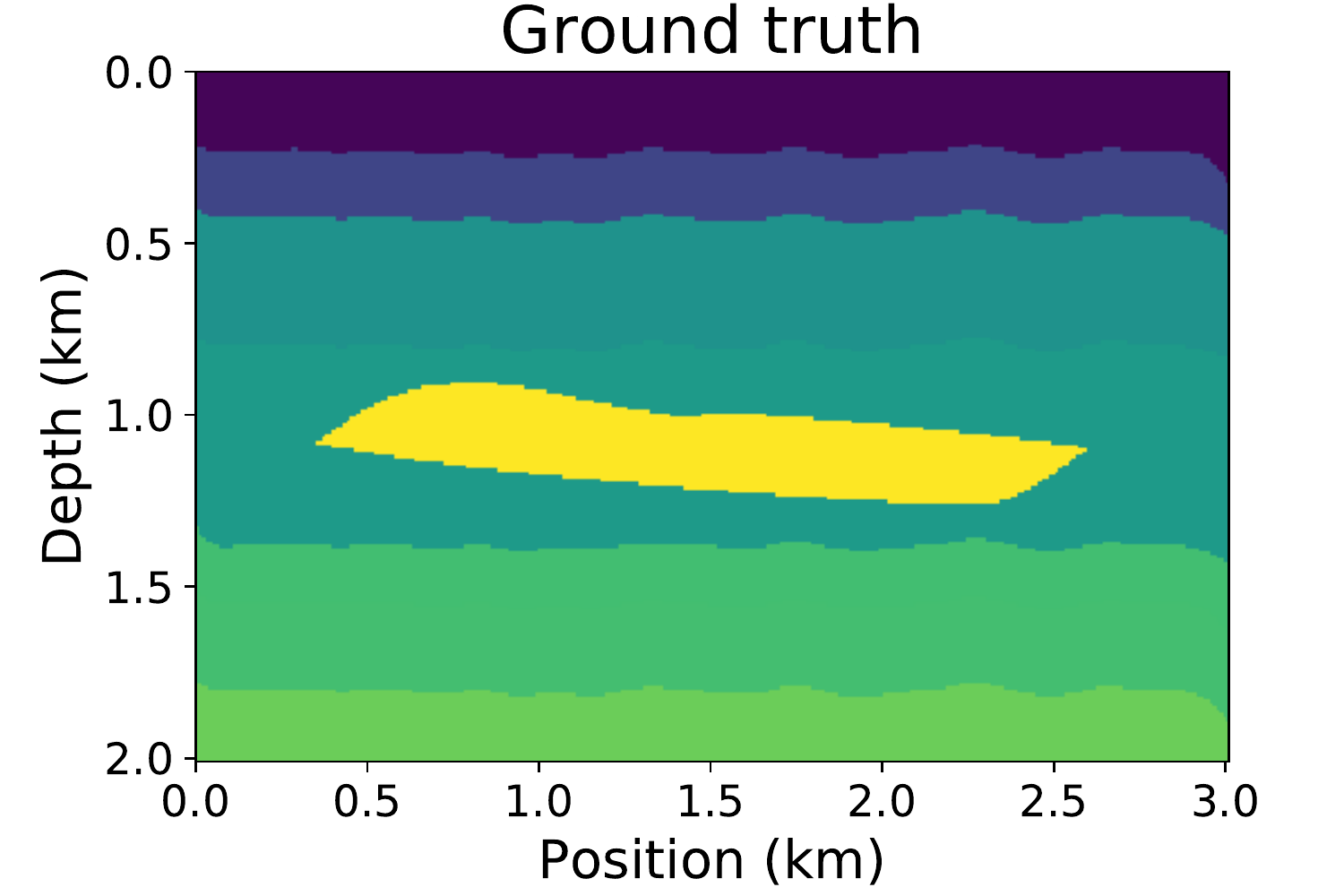}}
  \hspace{-0.7cm}
  \subfigure[]{\label{fig12-3}
  \includegraphics[width=0.35\textwidth]{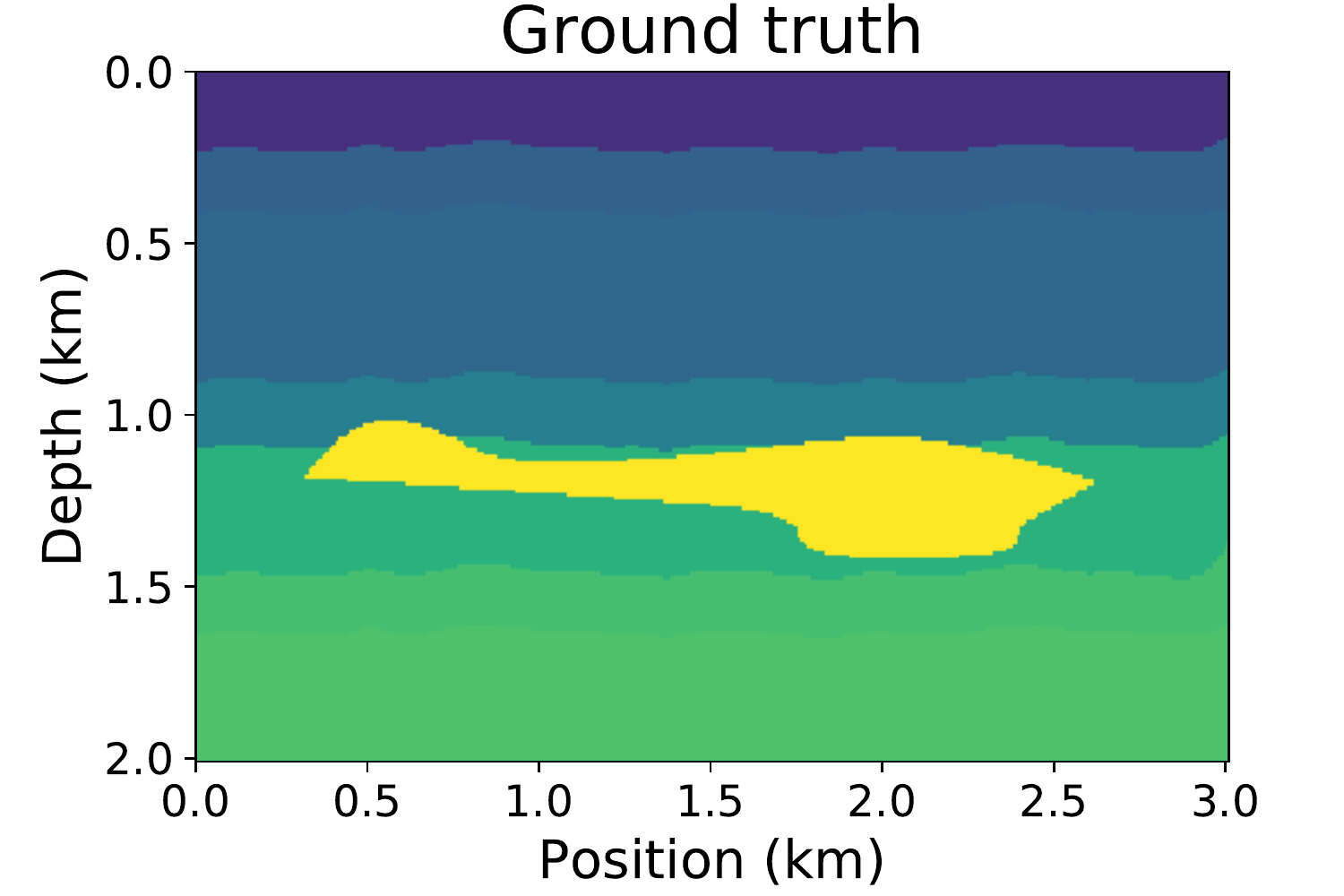}}

  \hspace{-0.4cm}
  \subfigure[]{\label{fig12-4}
  \includegraphics[width=0.35\textwidth]{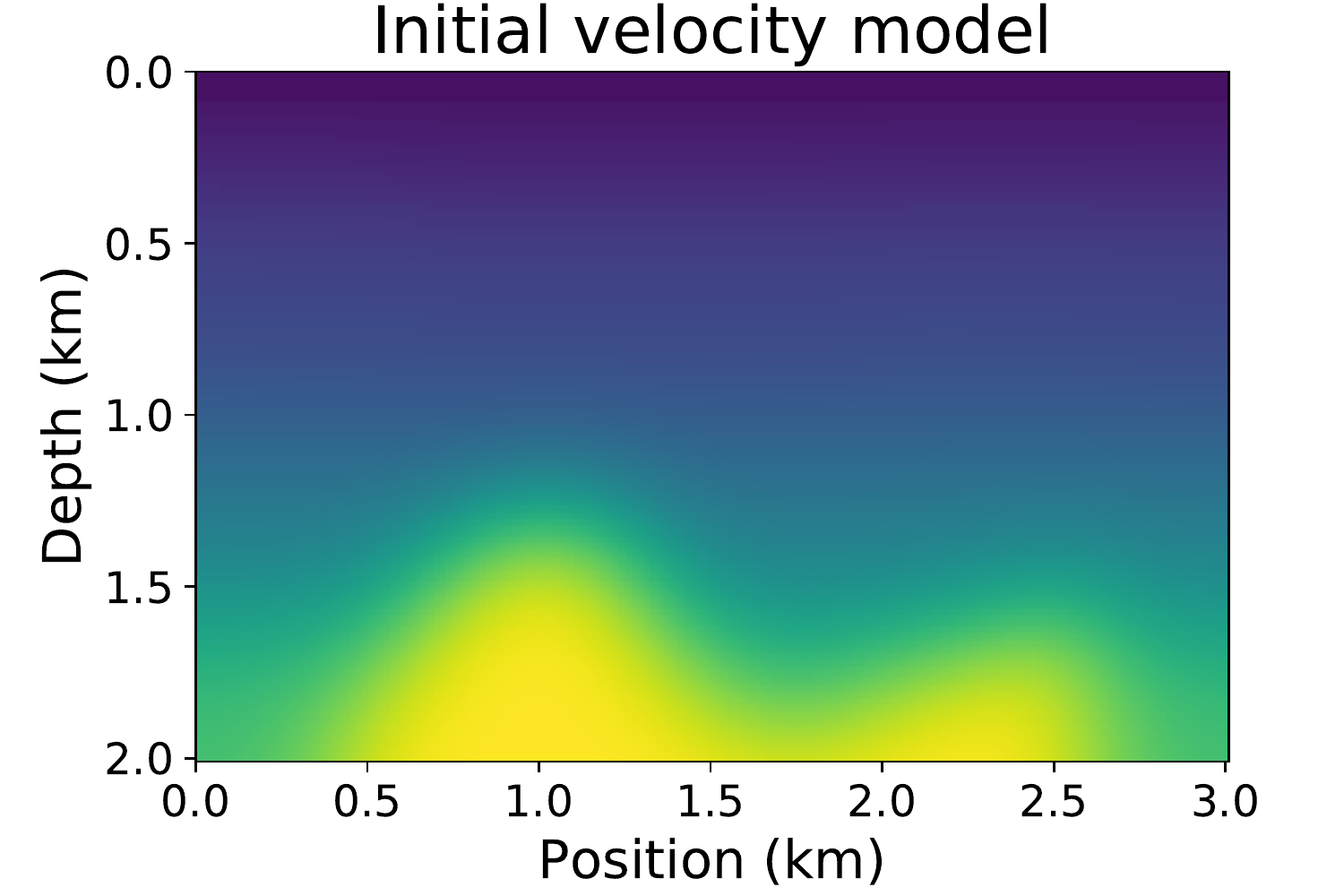}}
  \hspace{-0.7cm}
  \subfigure[]{\label{fig12-5}
  \includegraphics[width=0.35\textwidth]{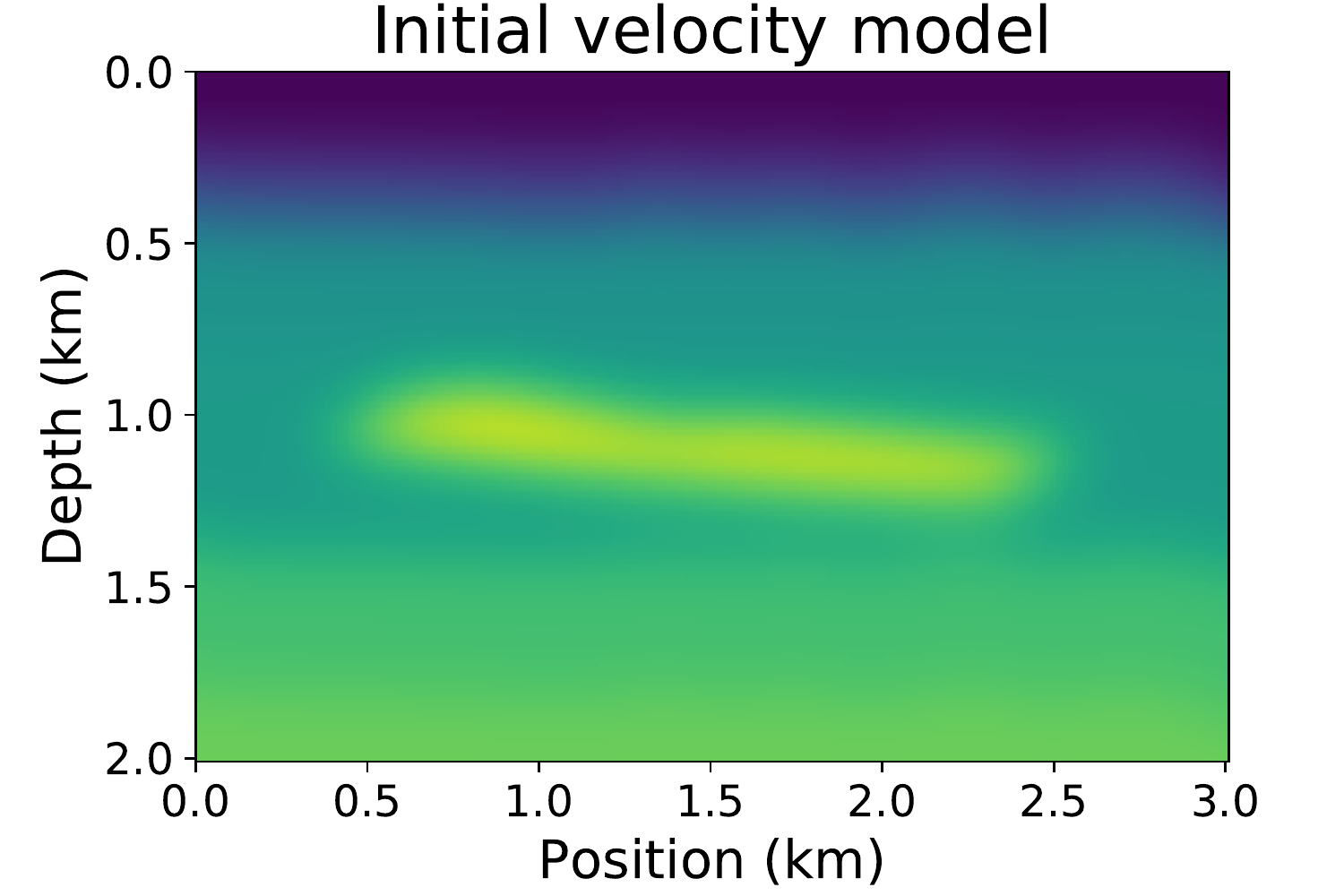}}
  \hspace{-0.7cm}
  \subfigure[]{\label{fig12-6}
  \includegraphics[width=0.35\textwidth]{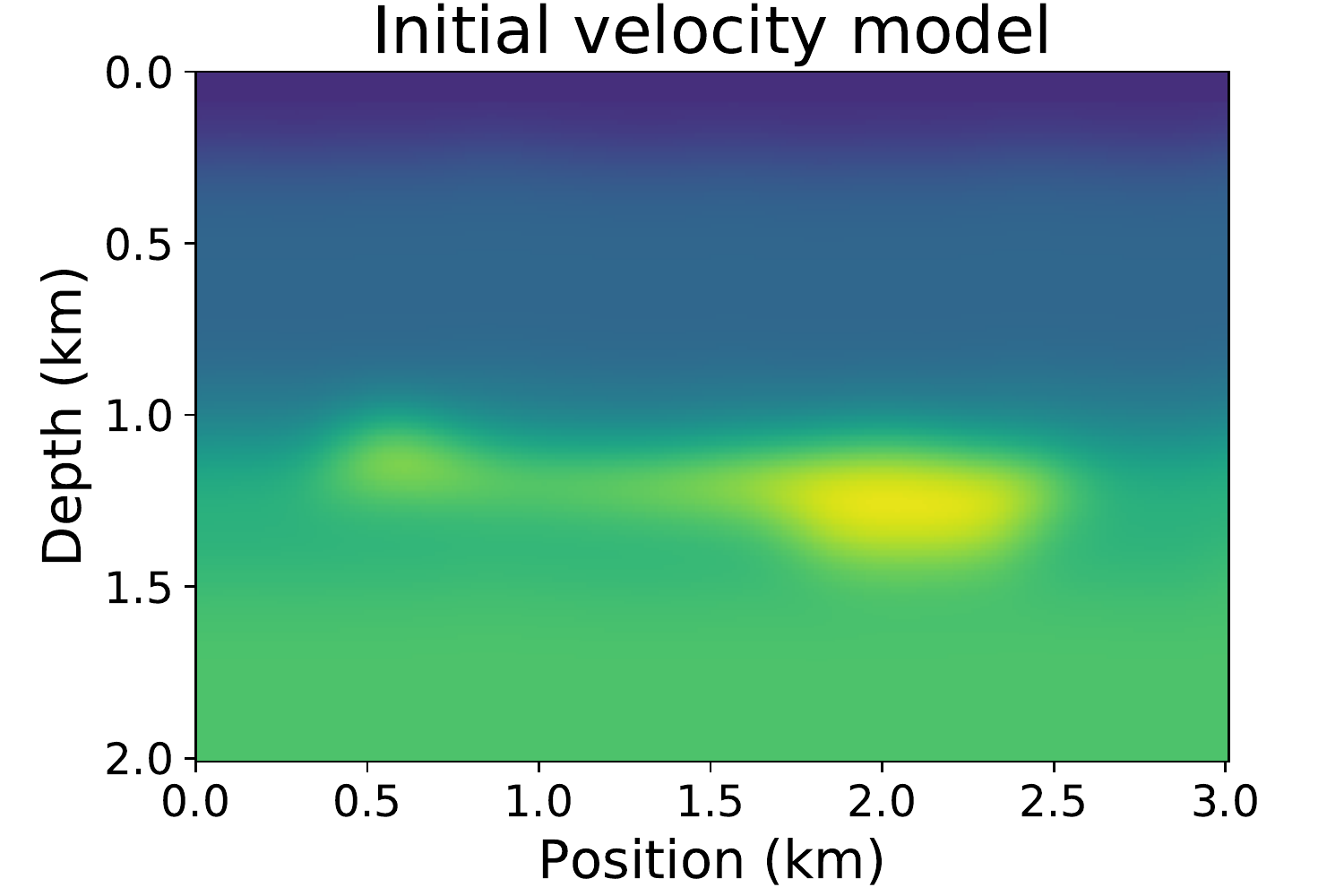}}

  \hspace{-0.4cm}
  \subfigure[]{\label{fig12-7}
  \includegraphics[width=0.35\textwidth]{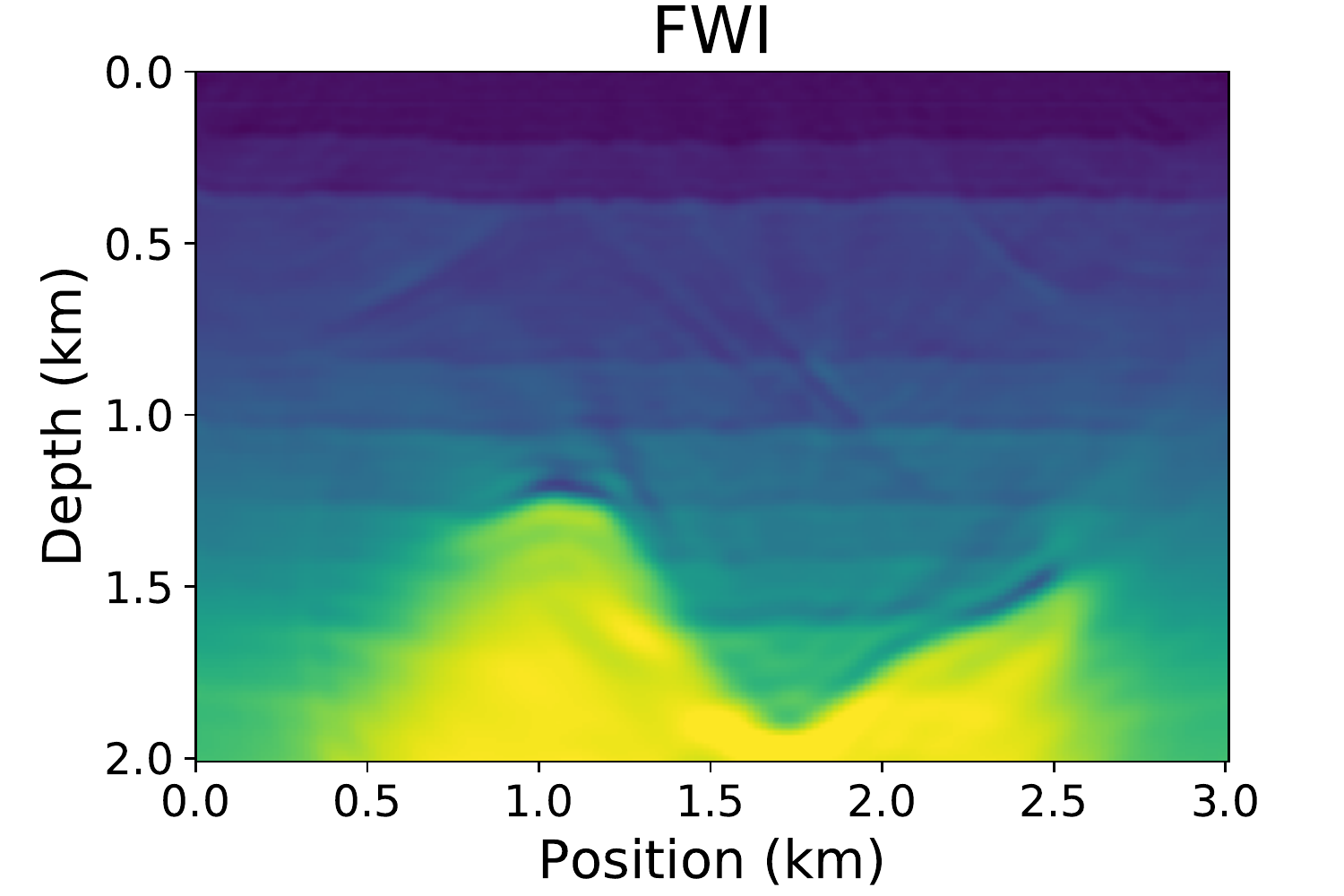}}
  \hspace{-0.7cm}
  \subfigure[]{\label{fig12-8}
  \includegraphics[width=0.35\textwidth]{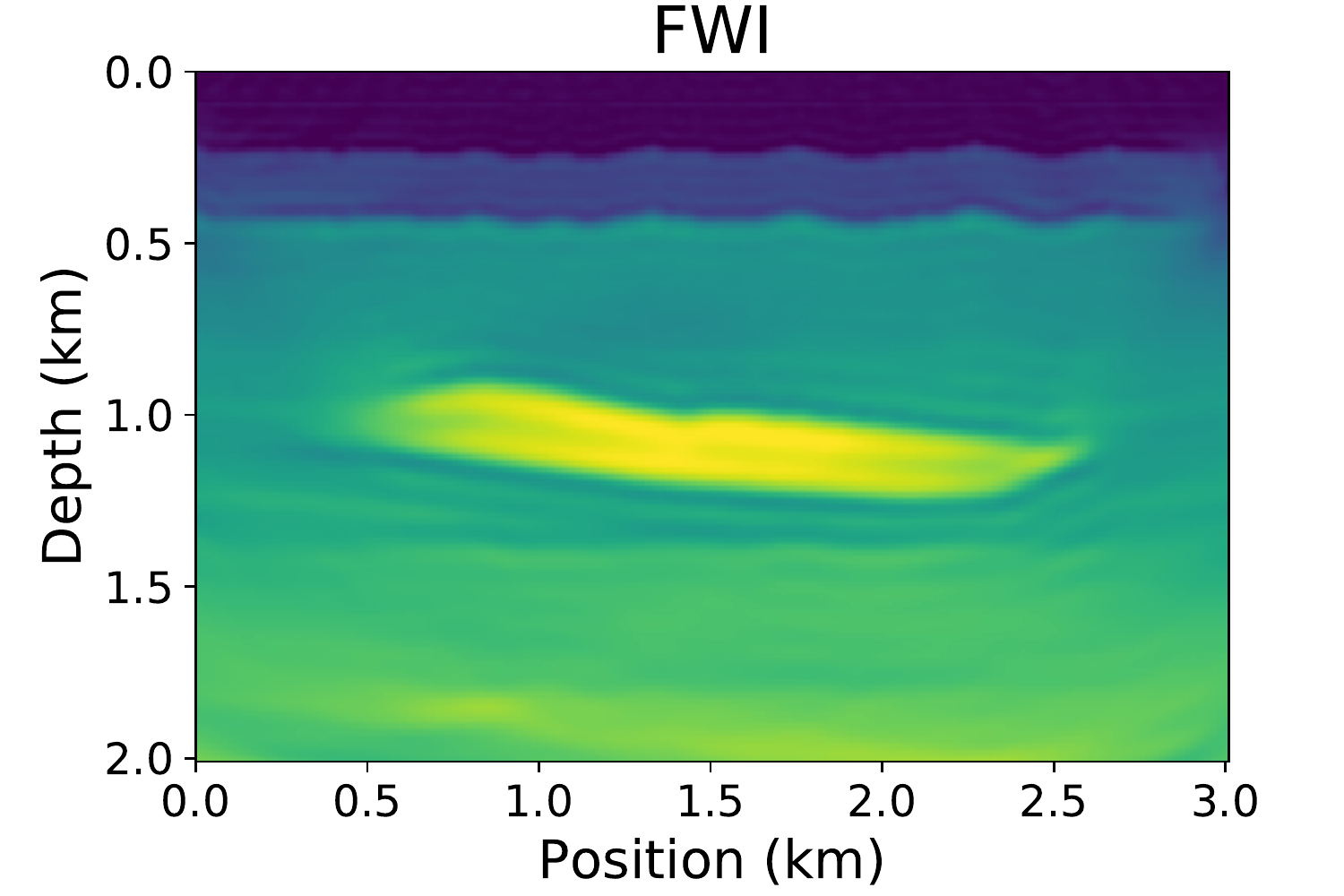}}
  \hspace{-0.7cm}
  \subfigure[]{\label{fig12-9}
  \includegraphics[width=0.35\textwidth]{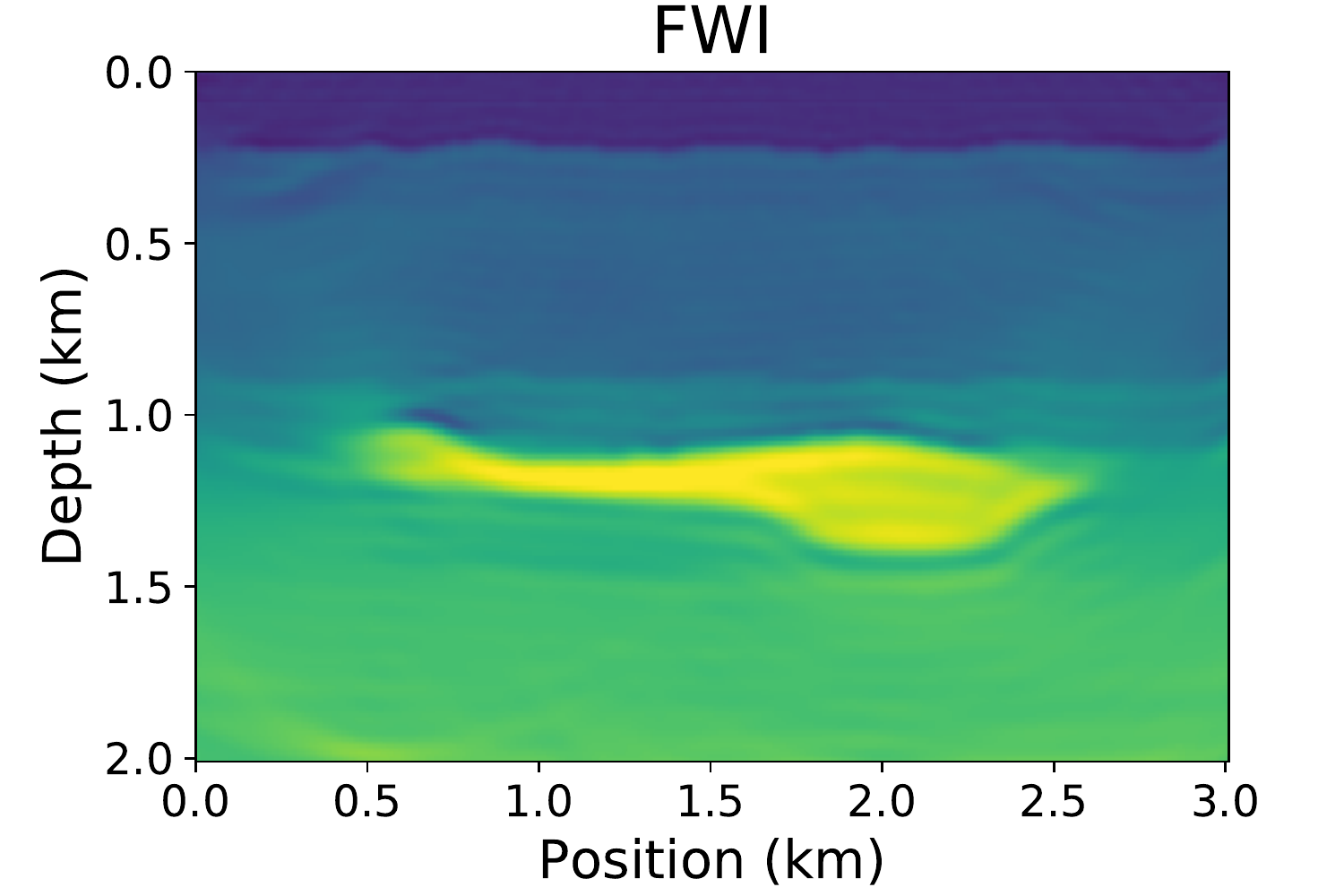}}

  \hspace{-0.4cm}
  \subfigure[]{\label{fig12-10}
  \includegraphics[width=0.35\textwidth]{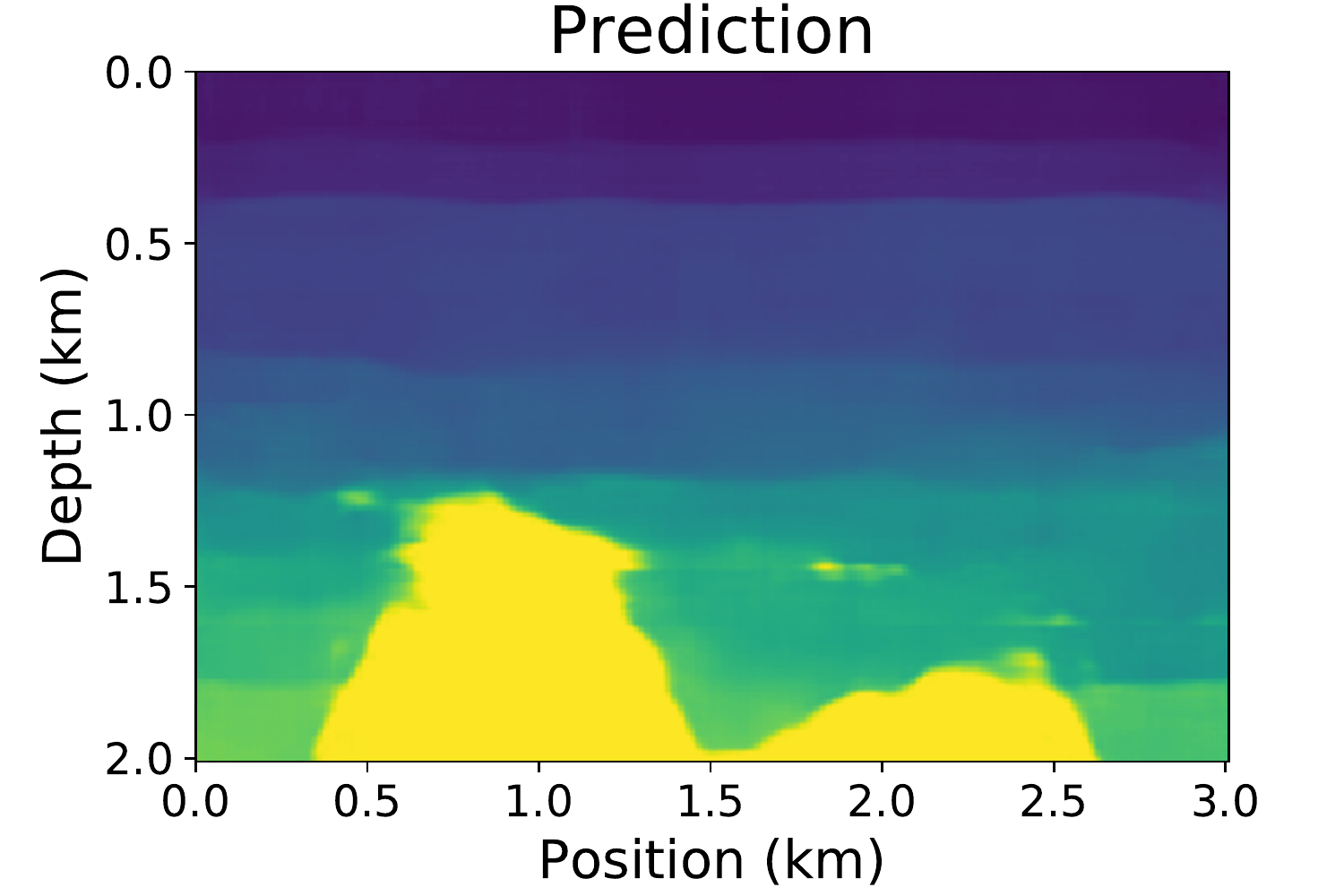}}
  \hspace{-0.7cm}
  \subfigure[]{\label{fig12-11}
  \includegraphics[width=0.35\textwidth]{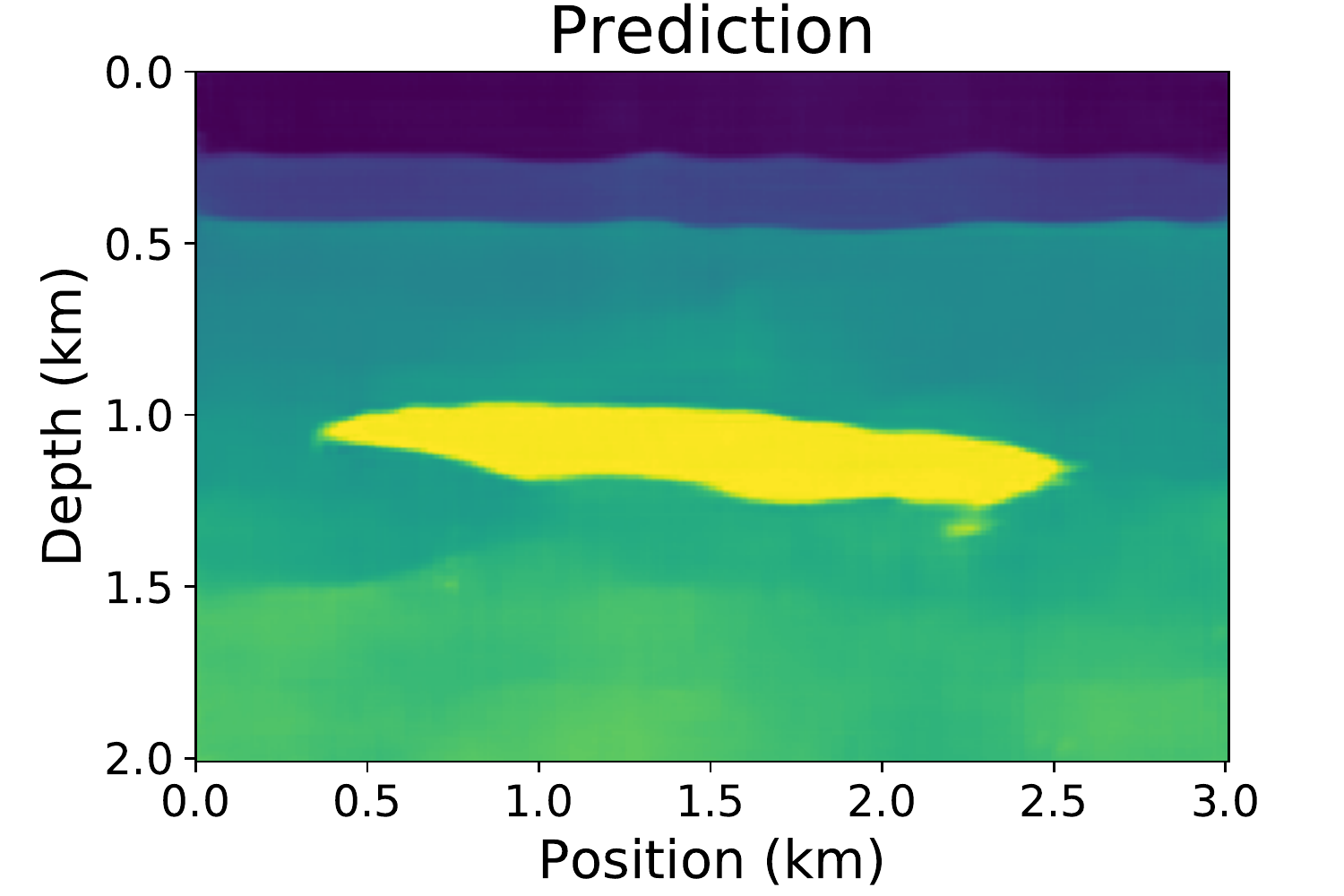}}
  \hspace{-0.7cm}
  \subfigure[]{\label{fig12-12}
  \includegraphics[width=0.35\textwidth]{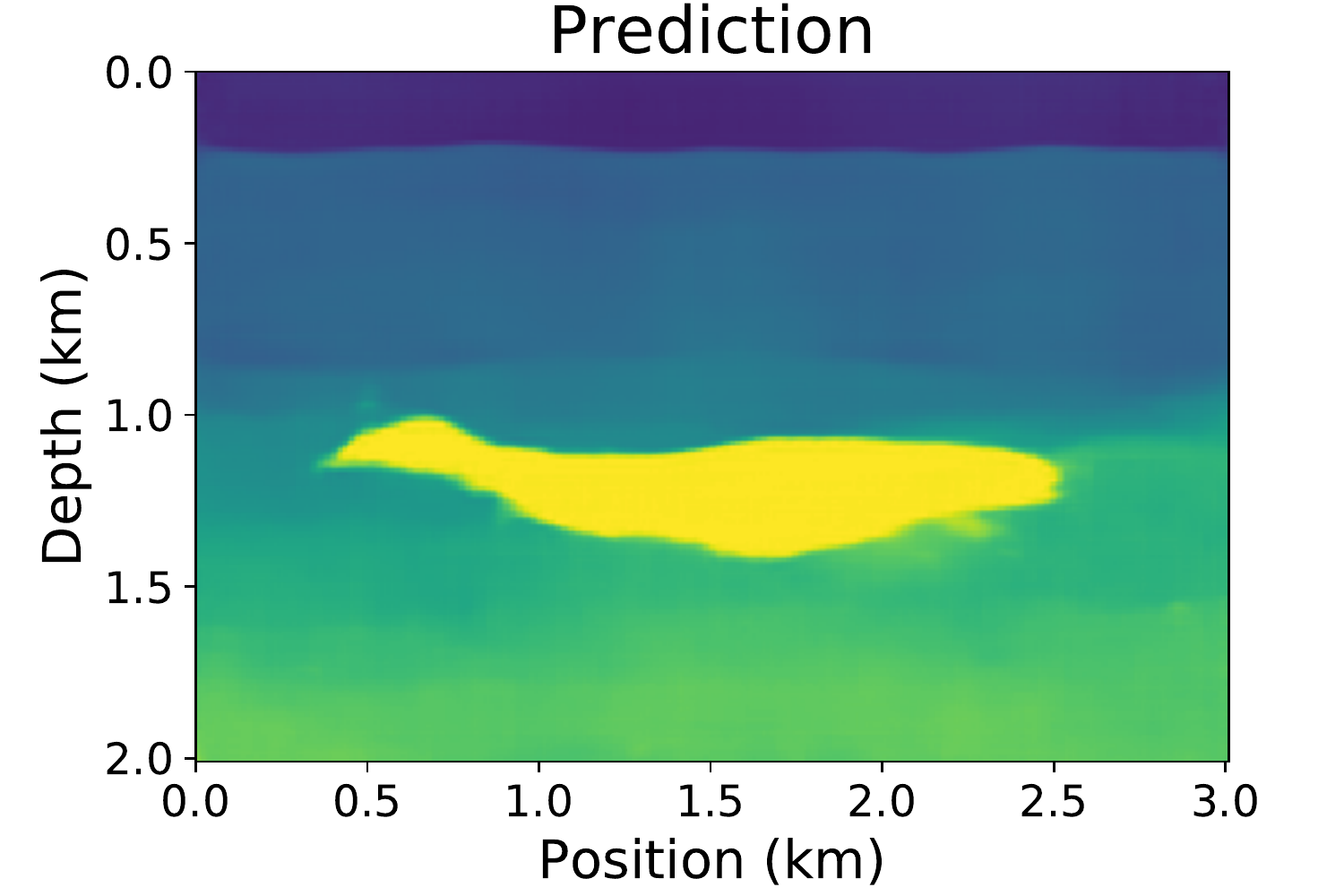}}
  \caption{Comparisons of the velocity inversion (simulated models): (a)--(c) ground truth; (d)--(f) initial velocity model of  FWI; (g)--(i) results of  FWI; (j)--(l): prediction of  our method.}
  \label{fig12}
\end{figure*}

\clearpage
\begin{figure*}
\centering
  \hspace{-0.4cm}
  \includegraphics[width=0.5\textwidth]{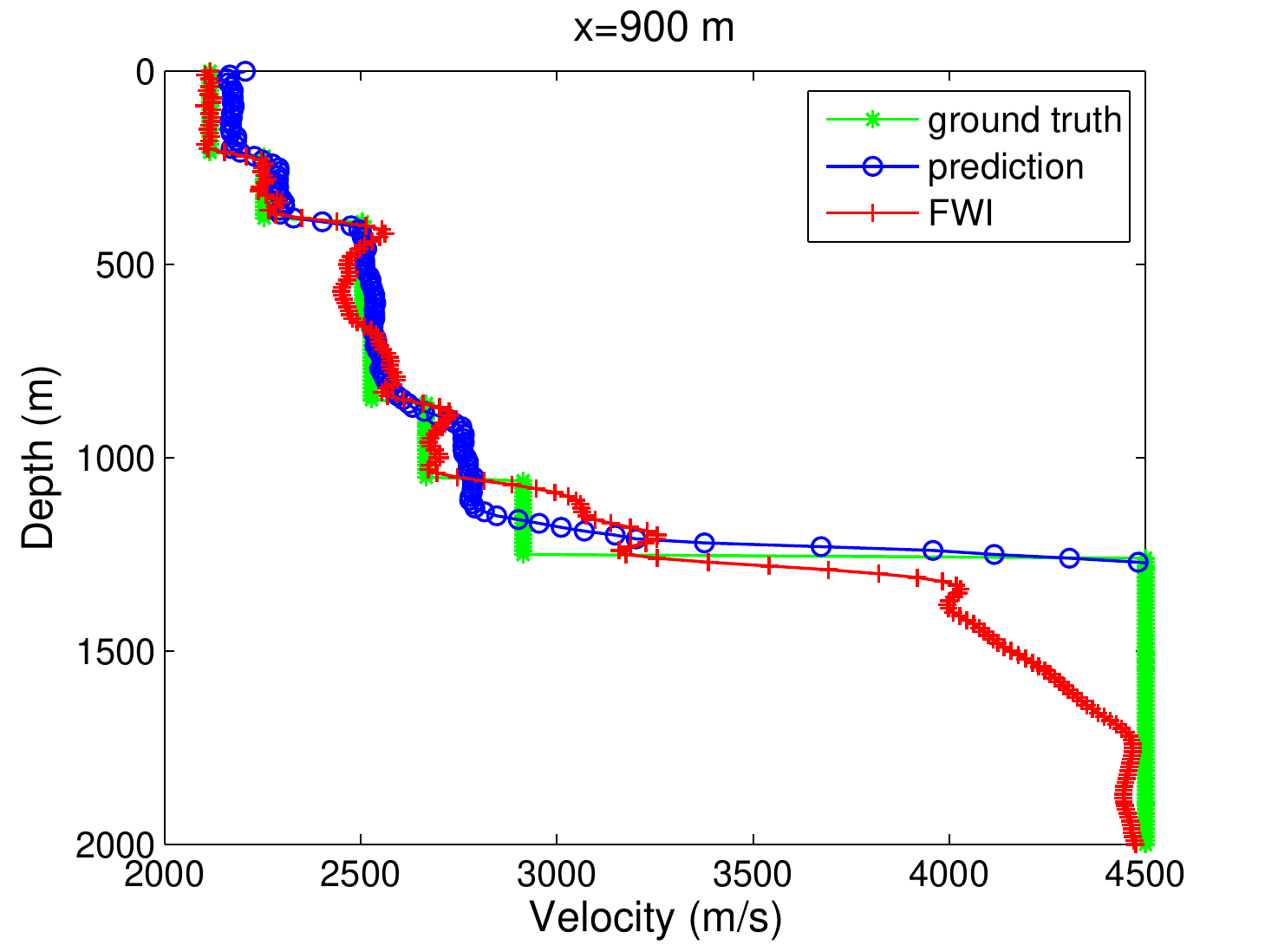}
  \hspace{-0.4cm}
  \includegraphics[width=0.5\textwidth]{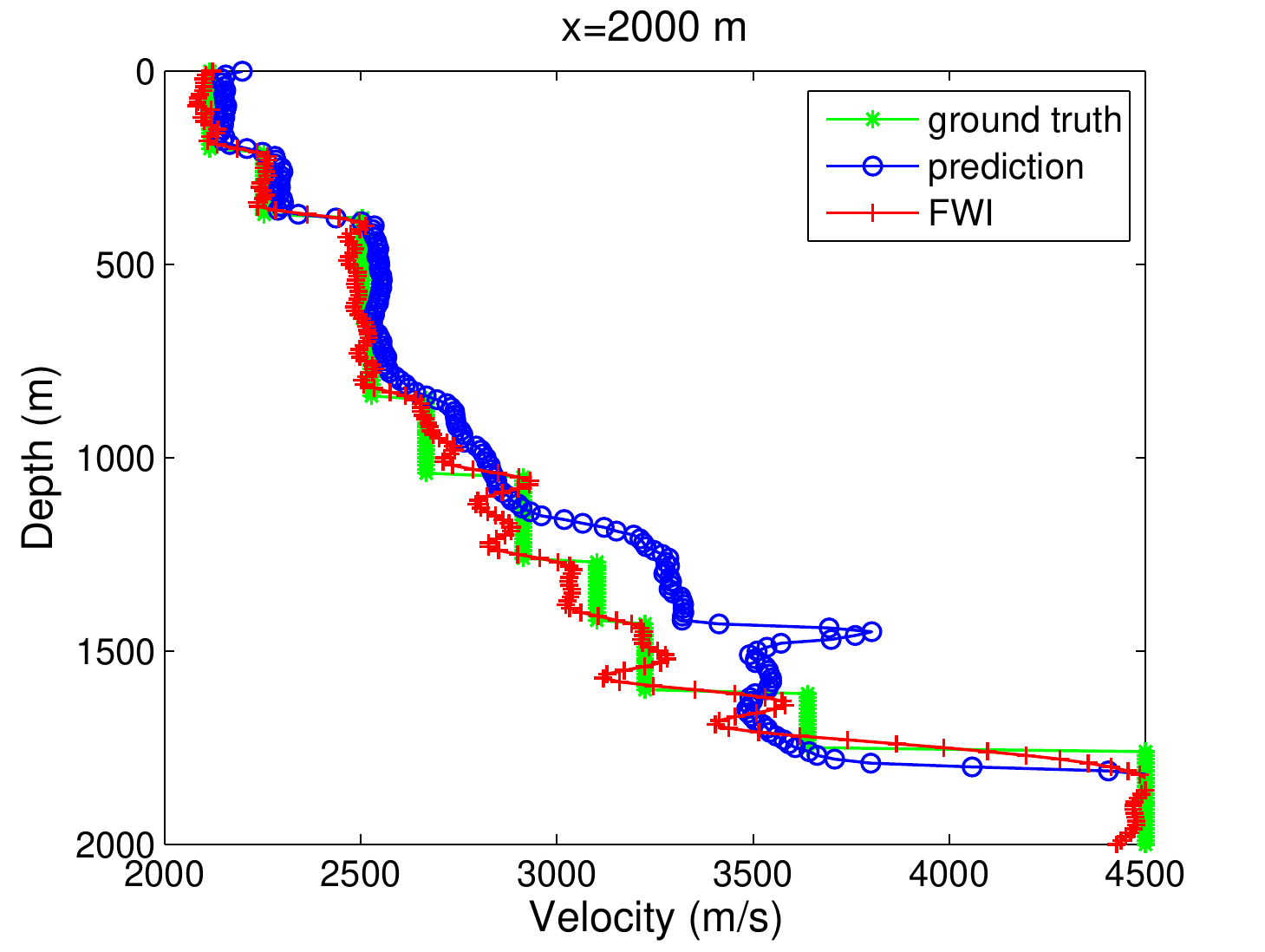}\\

  \hspace{-0.4cm}
  \includegraphics[width=0.5\textwidth]{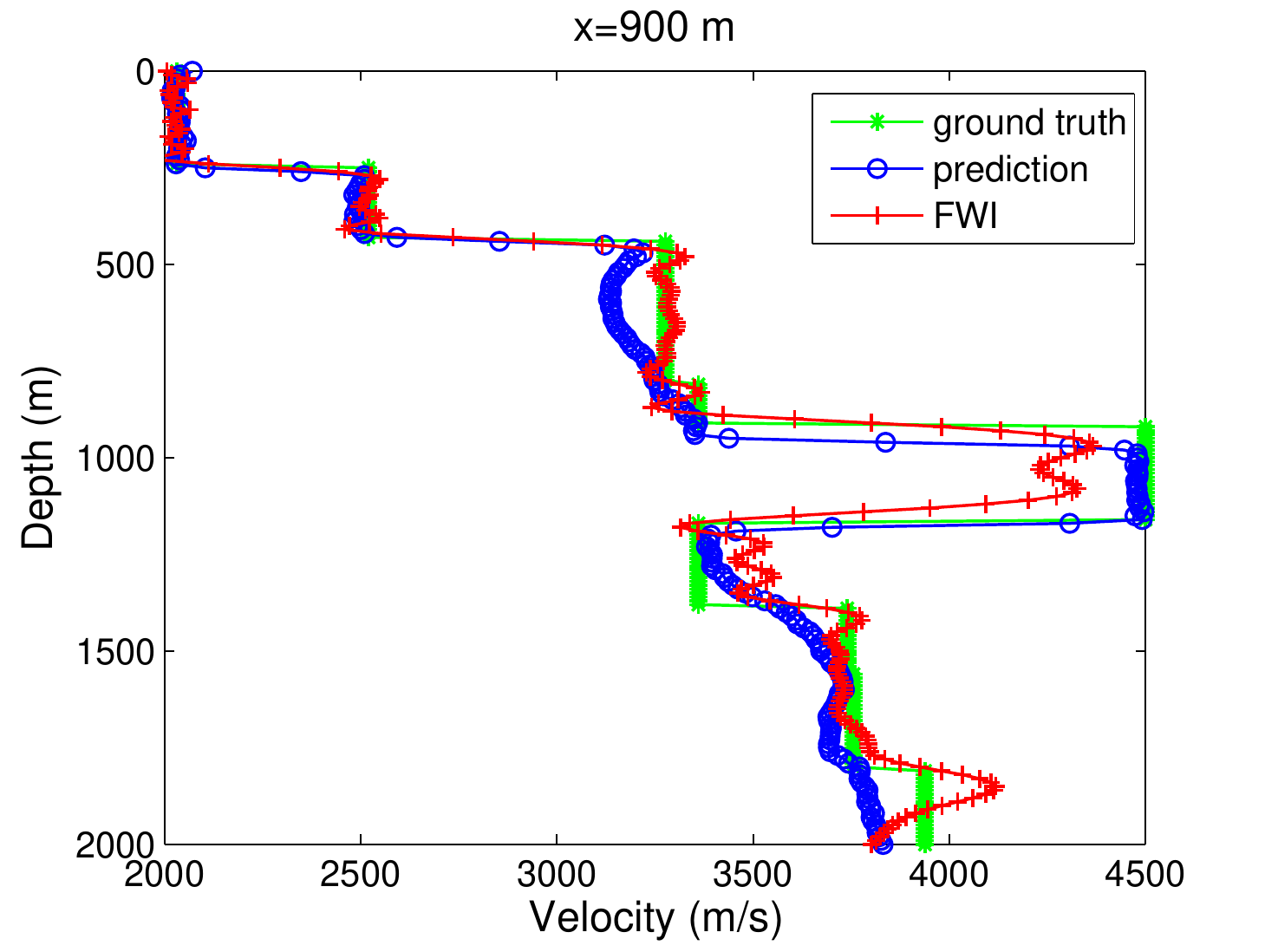}
  \hspace{-0.4cm}
  \includegraphics[width=0.5\textwidth]{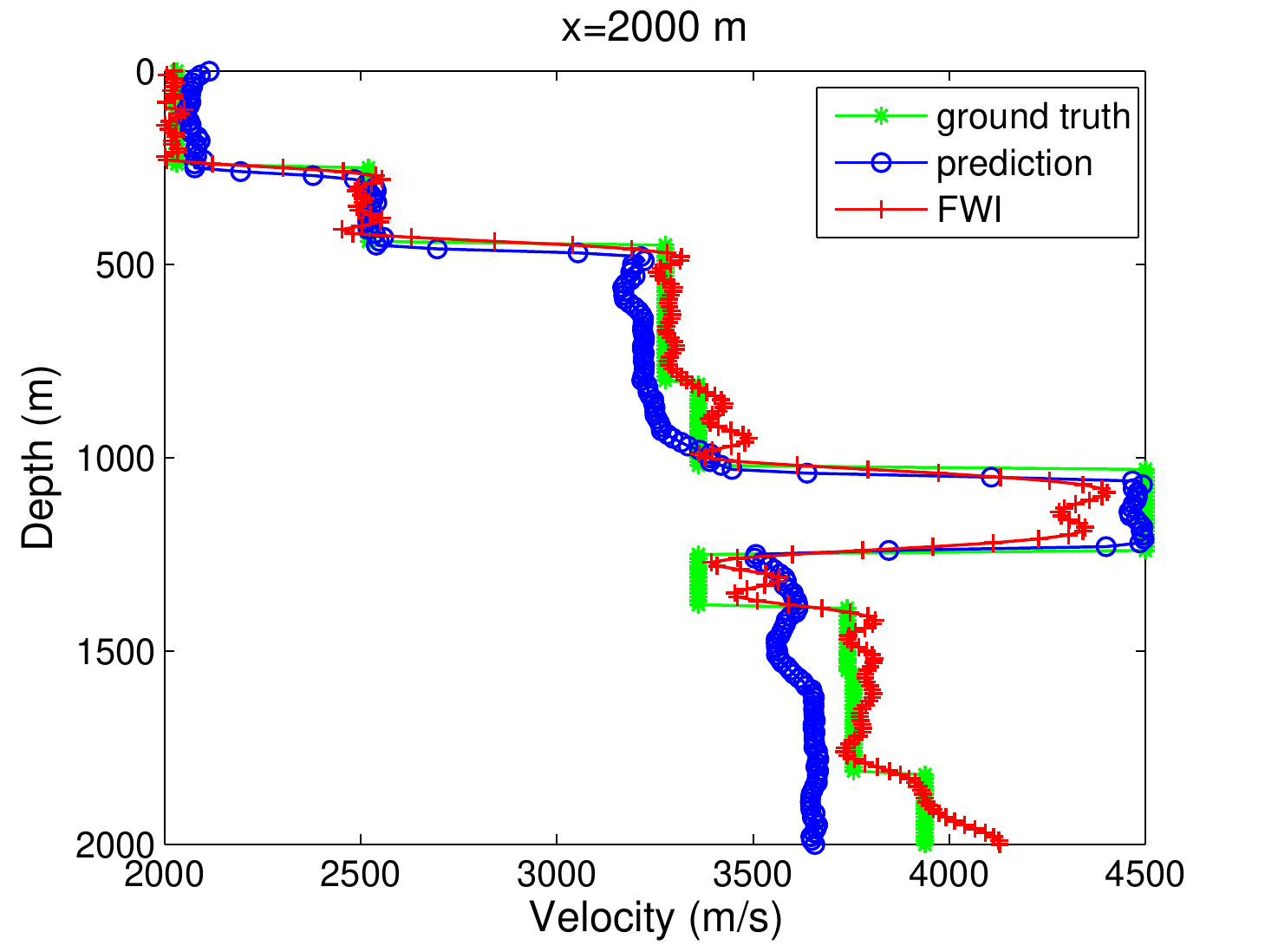}\\

  \hspace{-0.4cm}
  \includegraphics[width=0.5\textwidth]{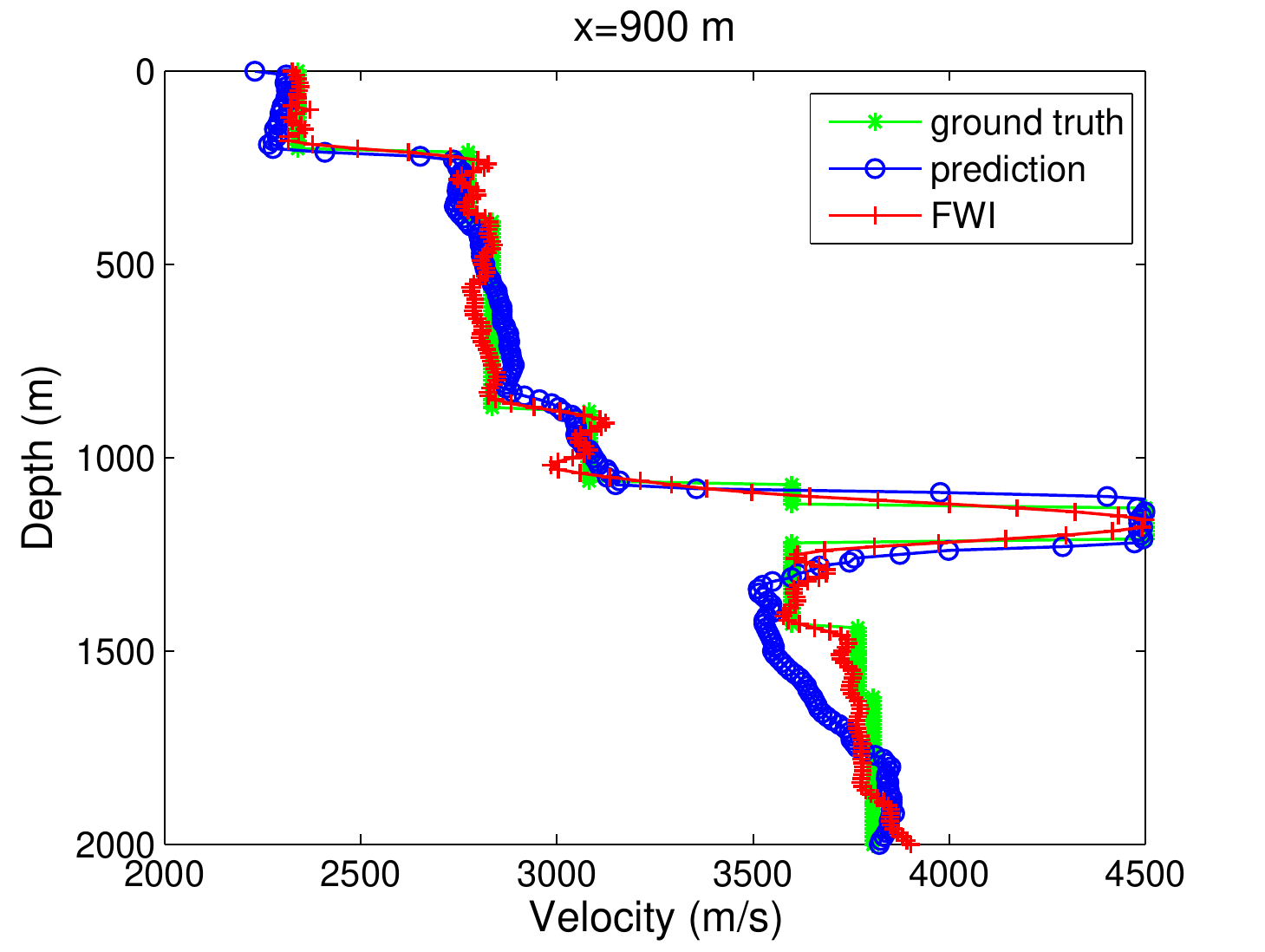}
  \hspace{-0.4cm}
  \includegraphics[width=0.5\textwidth]{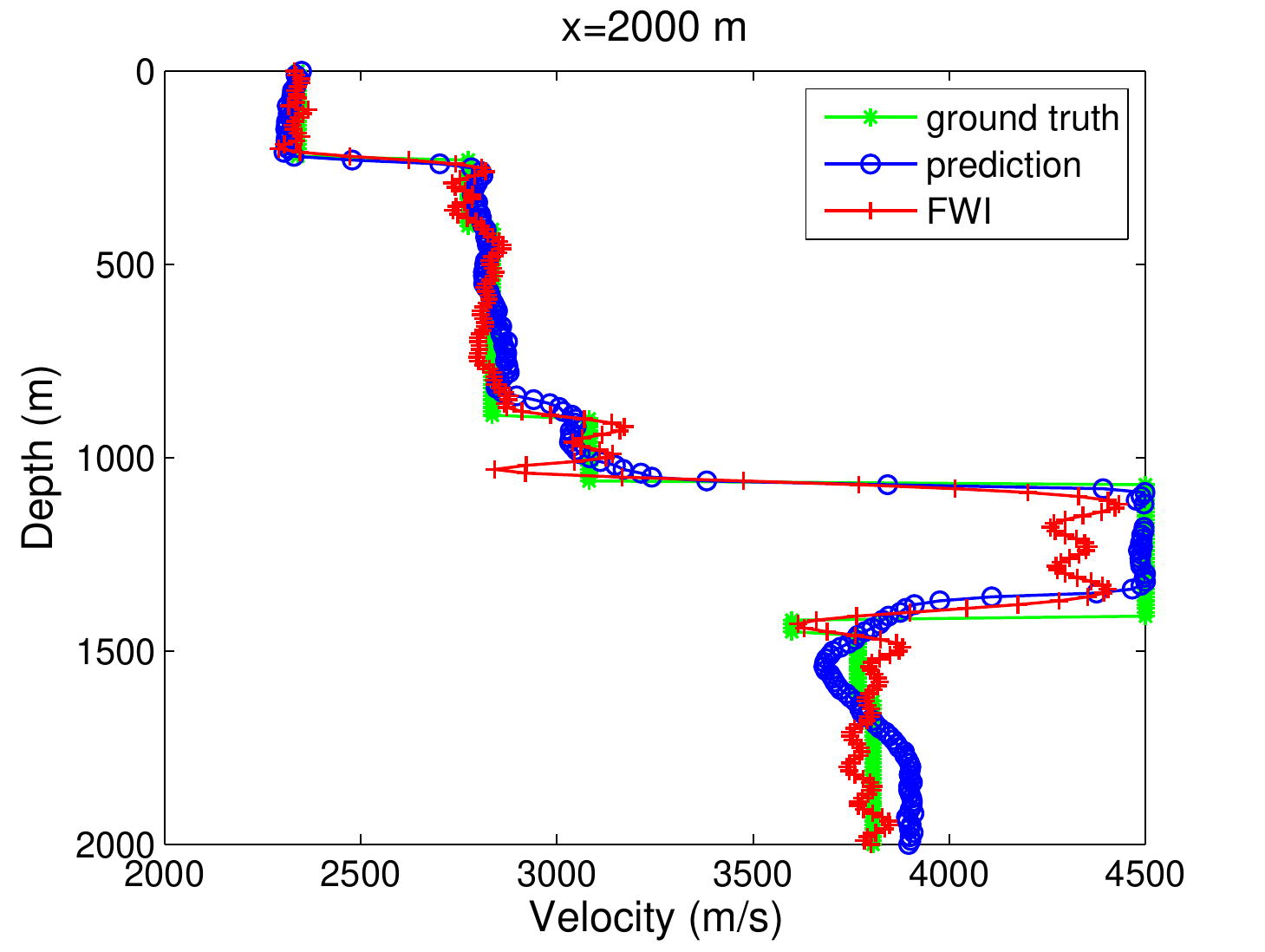}
  \caption{Vertical velocity profiles of our method and FWI. For the three test samples given in Figure \ref{fig12}, the prediction, FWI, and ground-truth velocities in the velocity versus depth profiles at two horizontal positions (x = 900 m, x = 2000 m) are presented in each row. }
  \label{fig13}
\end{figure*}

\clearpage
\begin{figure*}
\centering
  \includegraphics[scale=0.55]{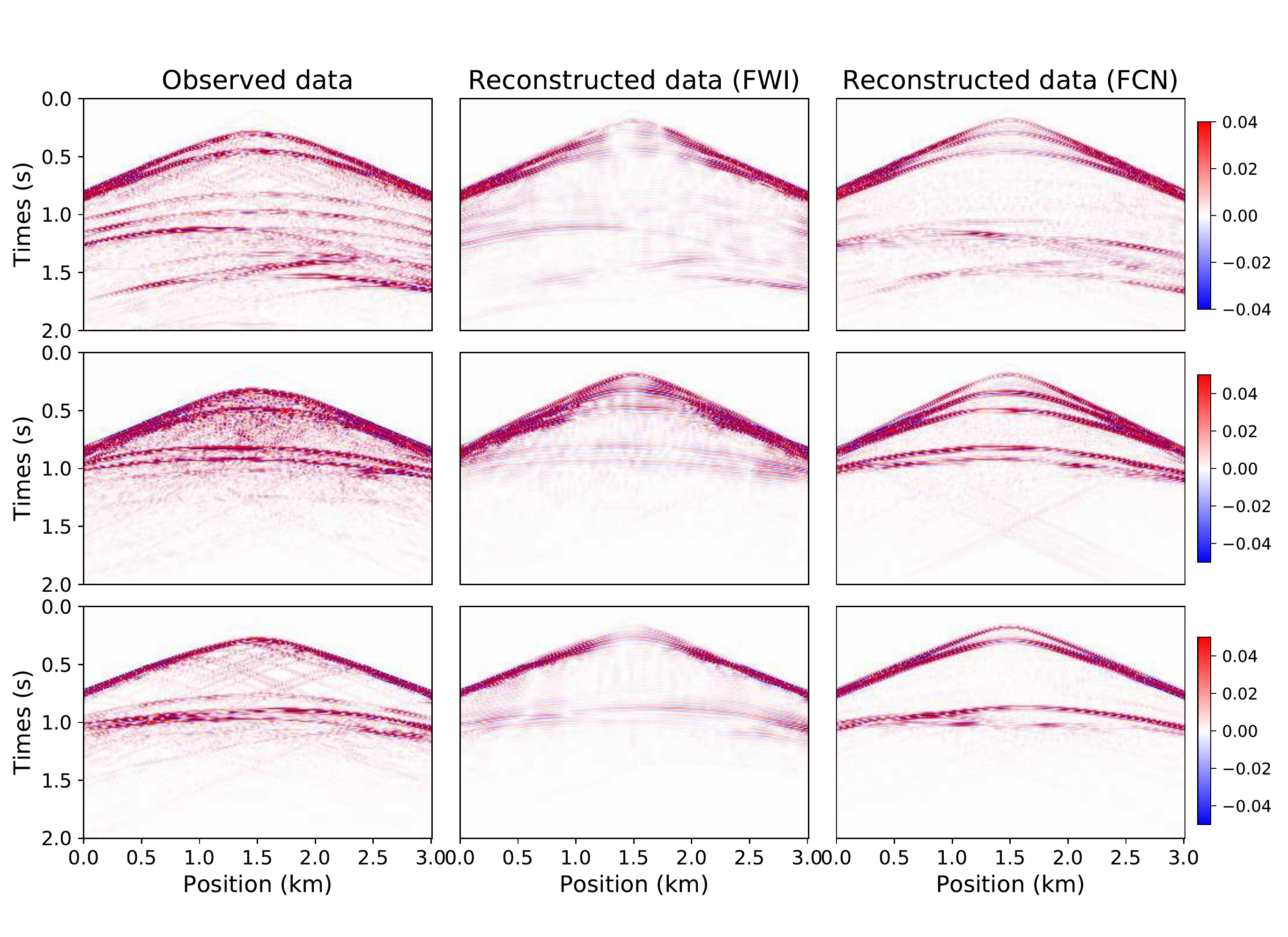}
  \caption{Shot records of the 15th receiver. Given from left to right in each row are the observed data according to the ground-truth velocity model (shown in Figure \ref{fig12-1}--Figure \ref{fig12-3}), reconstructed data by obtained by forward modeling of the FWI inverted velocity model (shown in Figure \ref{fig12-7}--Figure \ref{fig12-9}),and reconstructed data obtained by  forward modeling  the prediction of the FCN-based inversion method (shown in Figure \ref{fig12-10}--Figure \ref{fig12-12}).  }
  \label{fig23}
\end{figure*}

\clearpage
\begin{figure*}
\centering
  \hspace{-0.4cm}
  \subfigure[]{\label{fig12-31}
  \includegraphics[width=0.35\textwidth]{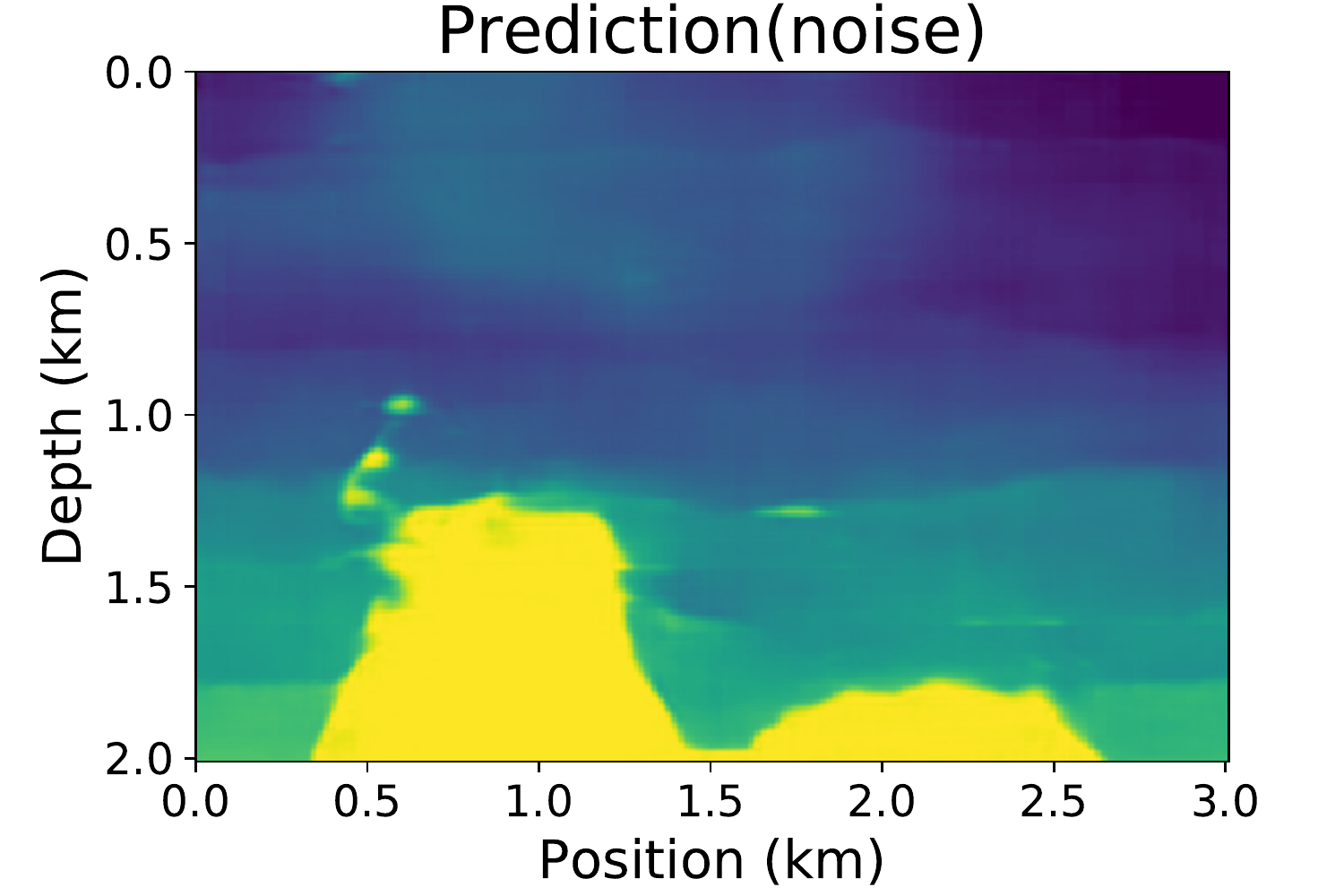}}
  \hspace{-0.7cm}
  \subfigure[]{\label{fig12-32}
  \includegraphics[width=0.35\textwidth]{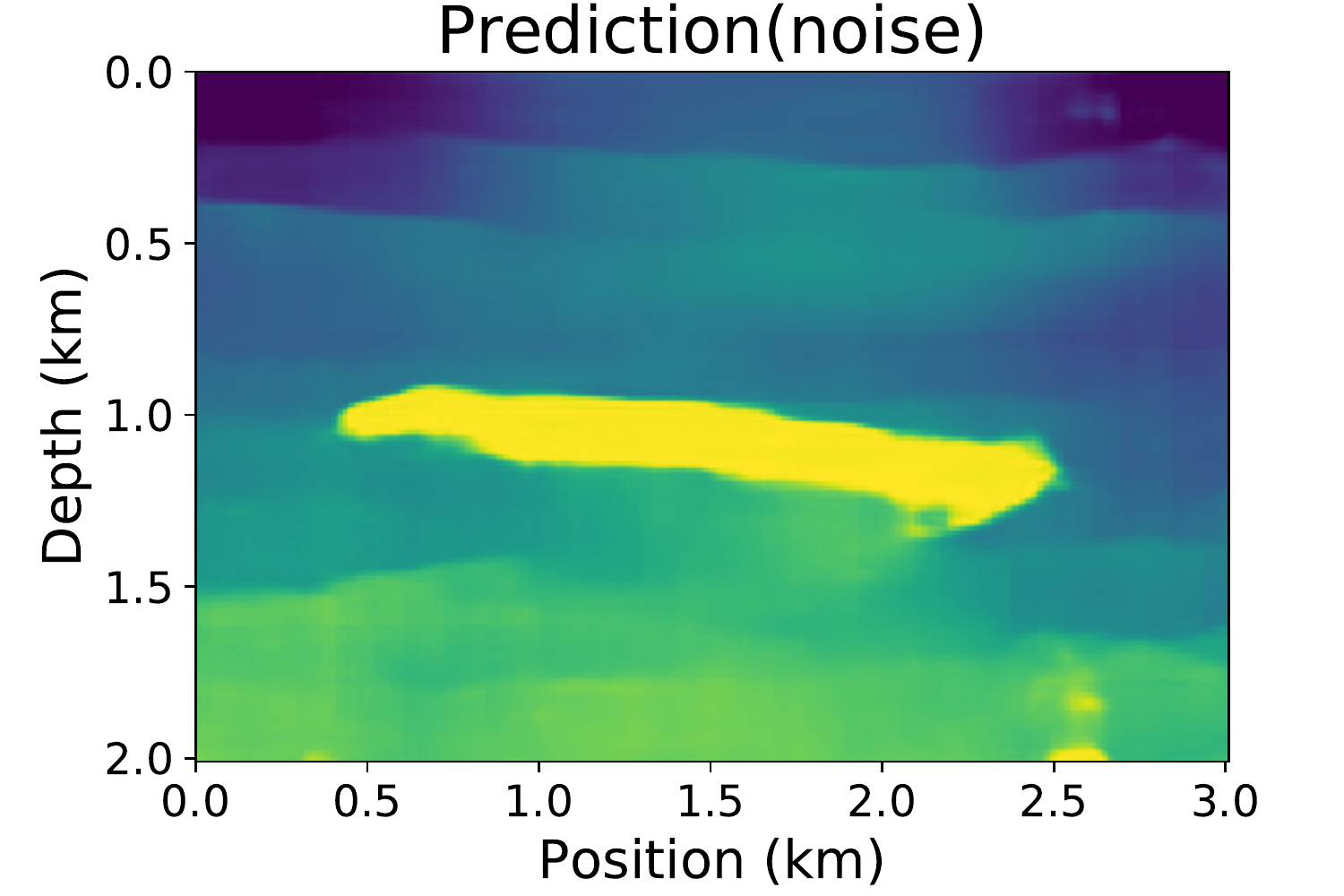}}
  \hspace{-0.7cm}
  \subfigure[]{\label{fig12-33}
  \includegraphics[width=0.35\textwidth]{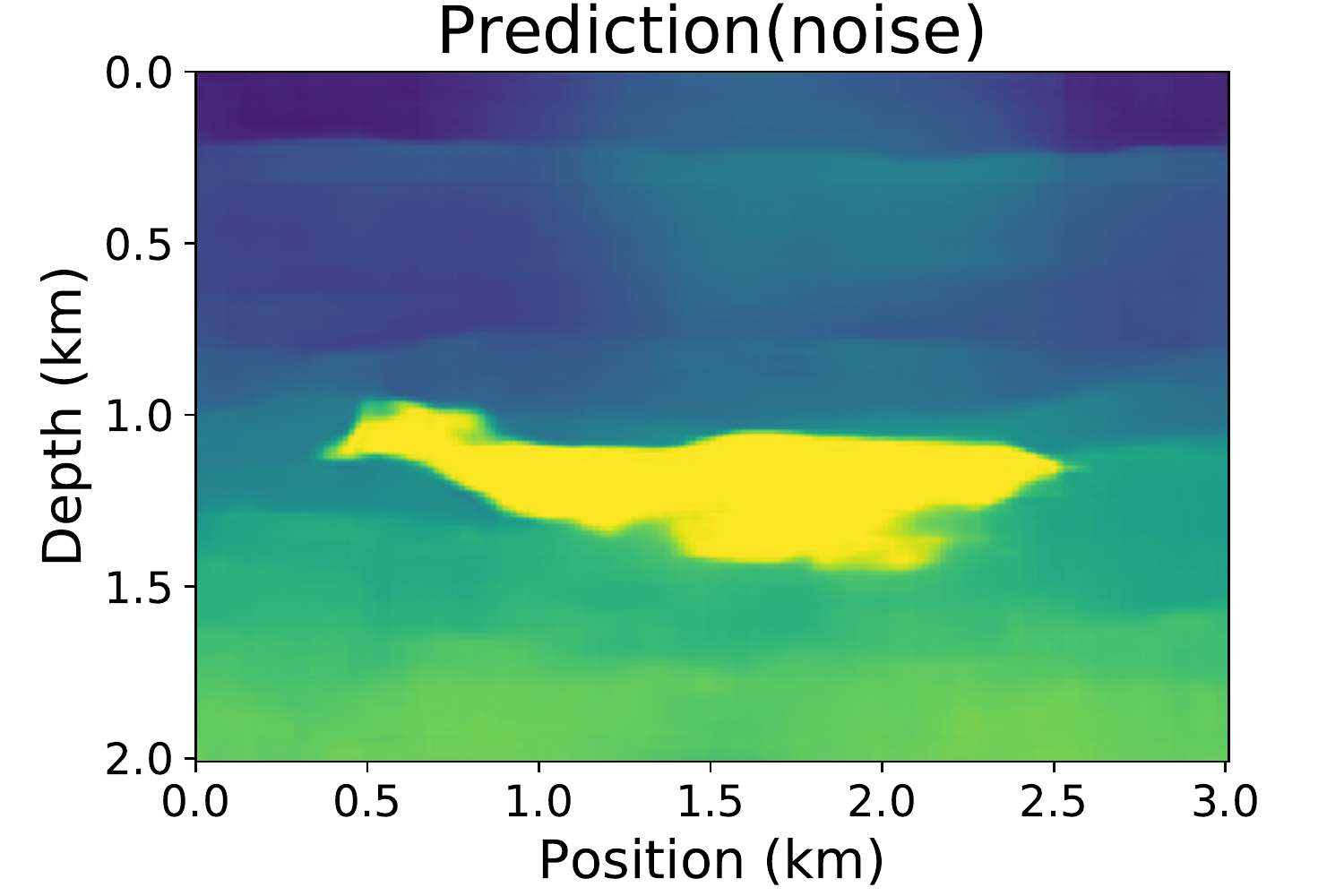}}

  \hspace{-0.4cm}
  \subfigure[]{\label{fig12-34}
  \includegraphics[width=0.35\textwidth]{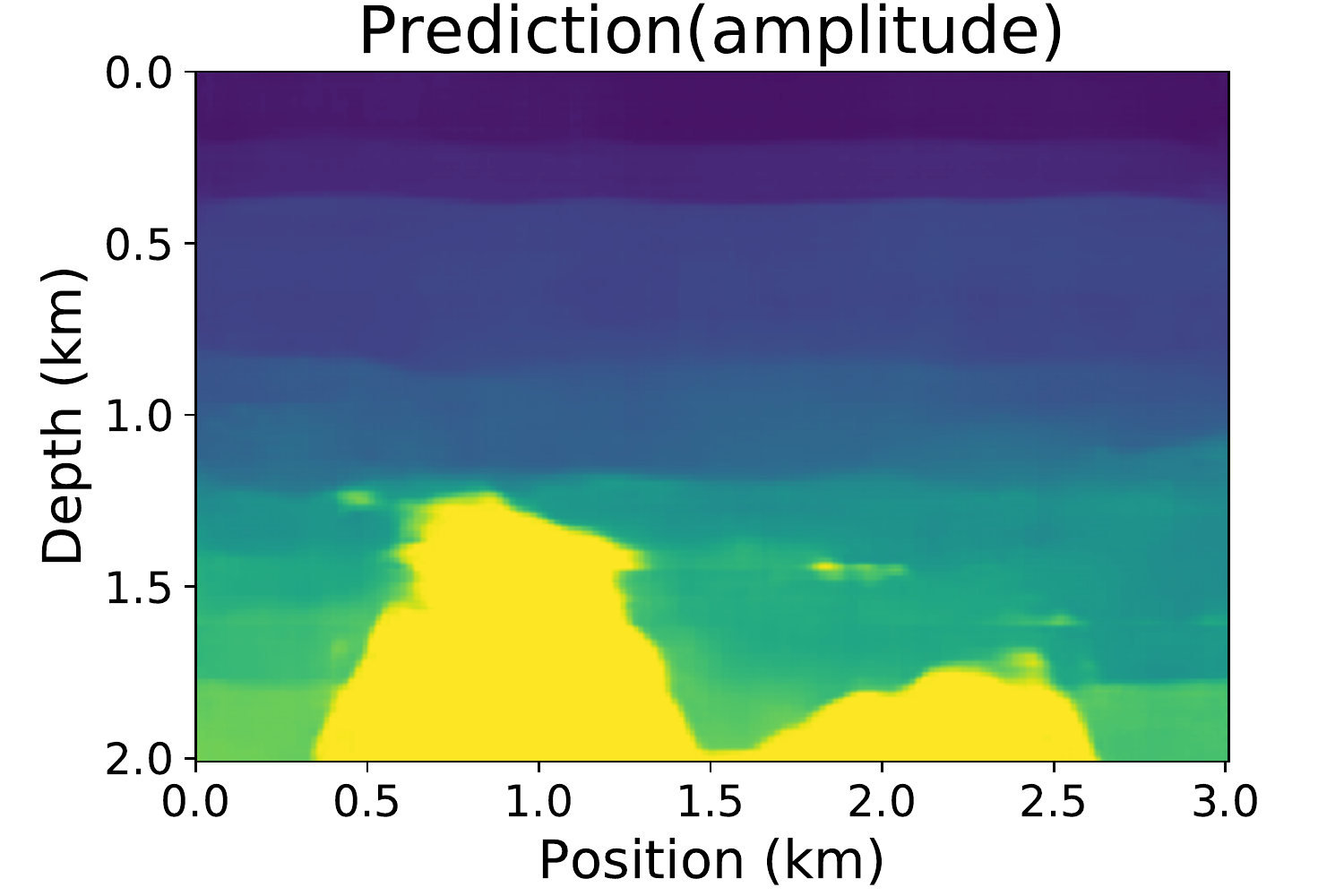}}
  \hspace{-0.7cm}
  \subfigure[]{\label{fig12-35}
  \includegraphics[width=0.35\textwidth]{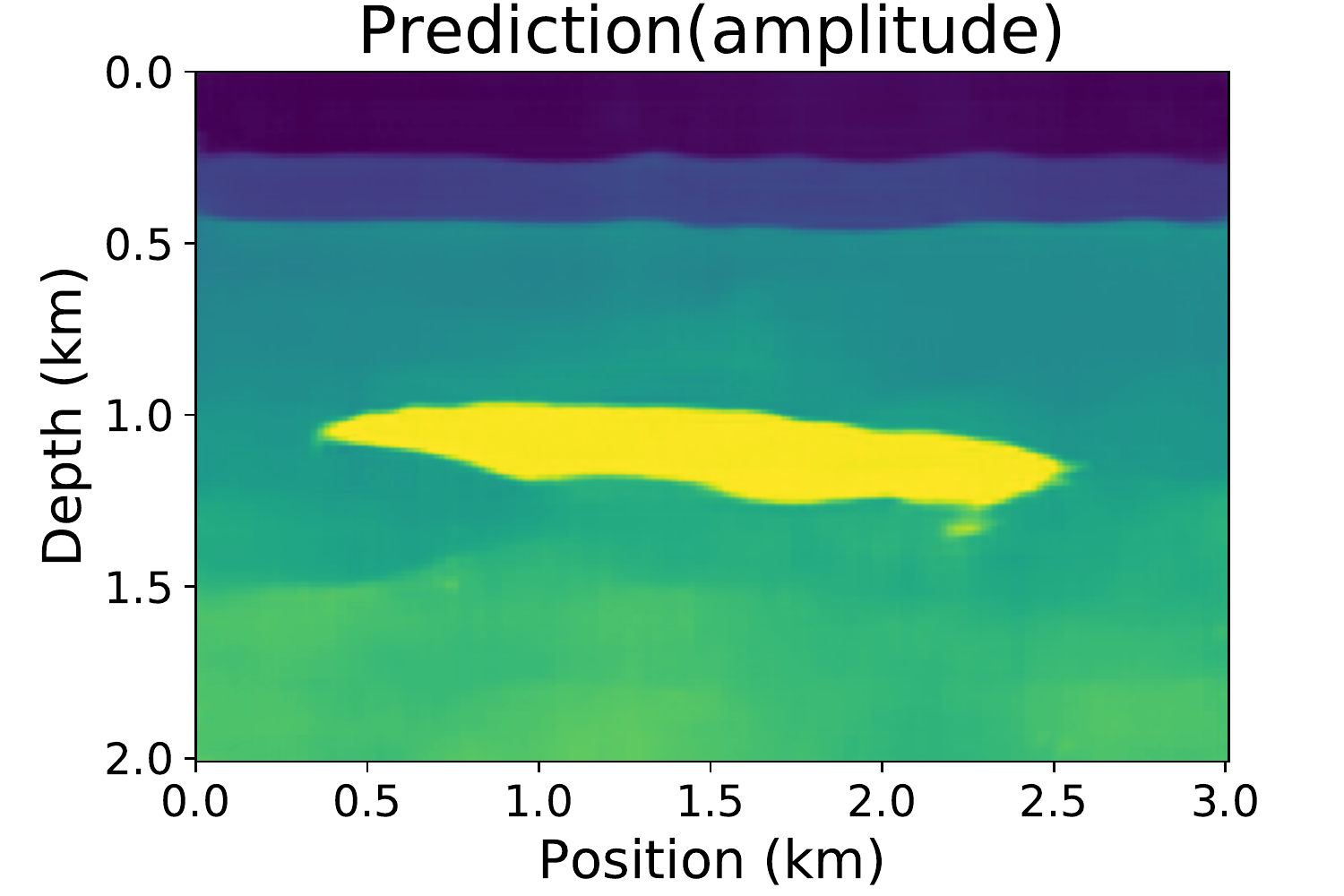}}
  \hspace{-0.7cm}
  \subfigure[]{\label{fig12-36}
  \includegraphics[width=0.35\textwidth]{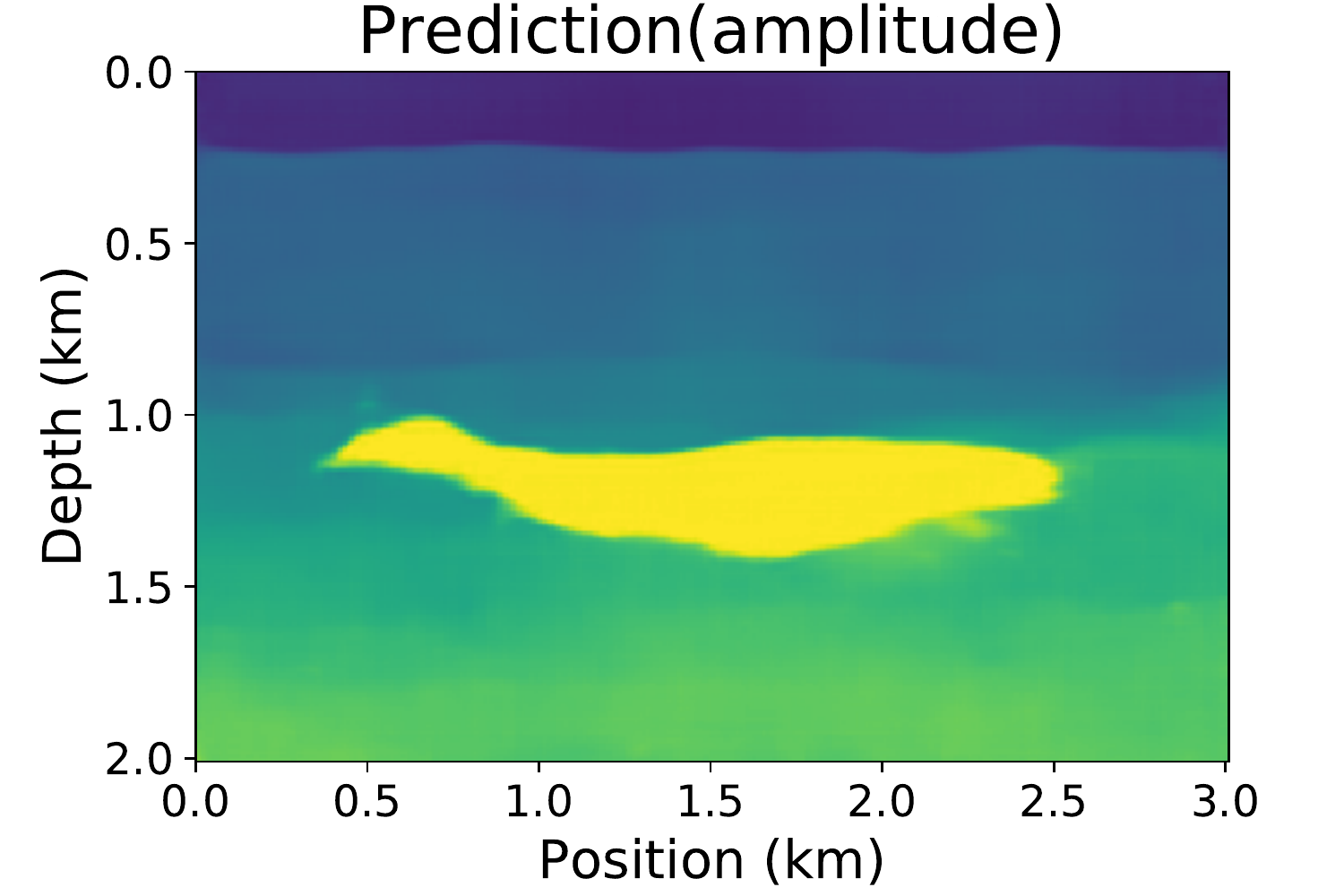}}
  \caption{Sensitivity of the proposed method to noise and amplitude (simulated models): (a)--(c) prediction results with the noisy seismic data; (d)--(f) prediction results with magnified seismic amplitude. Our method  showed acceptable results when the input data were perturbed.}
  \label{fig19}
\end{figure*}

\clearpage
\begin{figure*}
  \centering
  \includegraphics[scale=0.55]{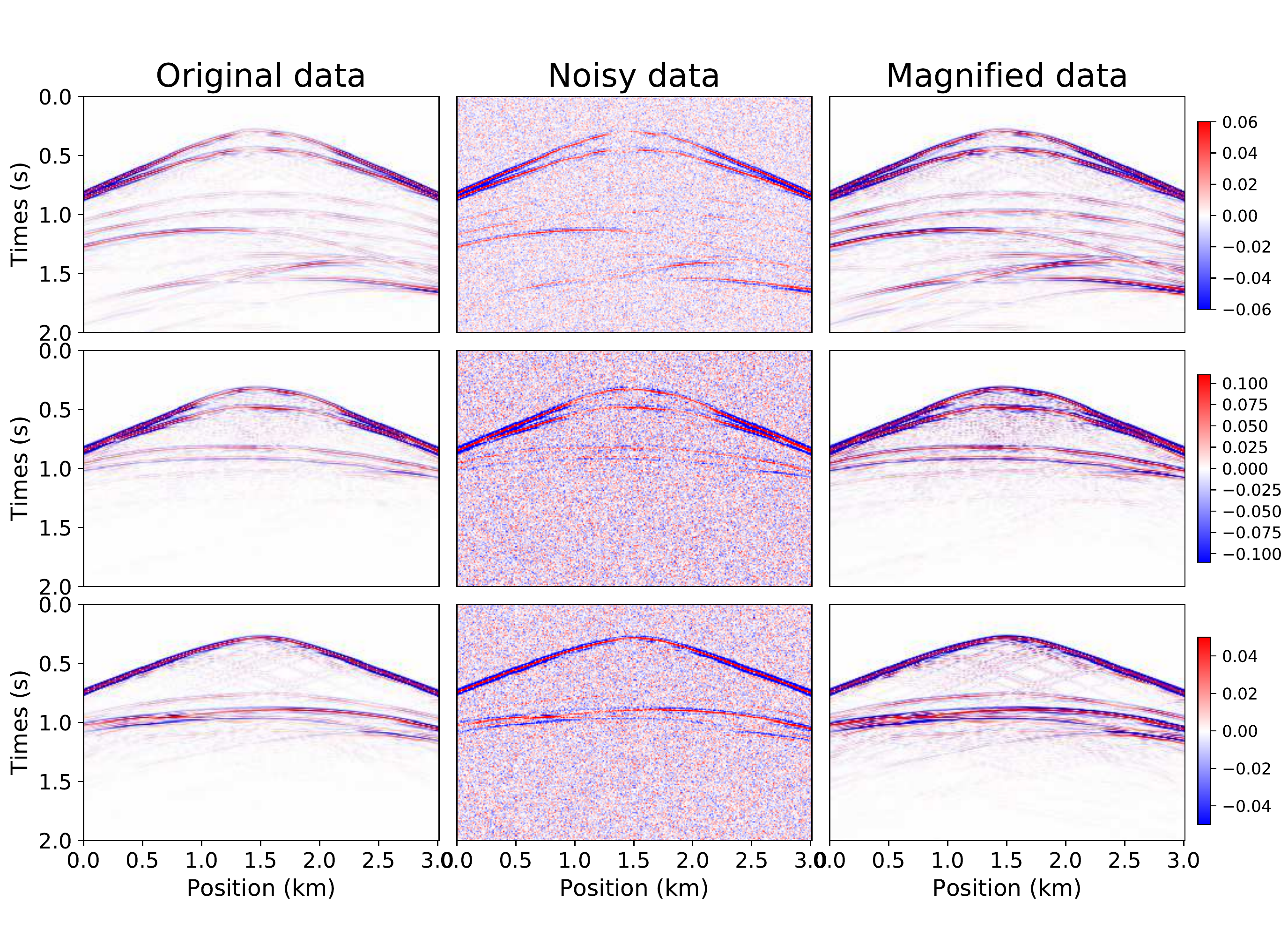}
  \caption{Comparison of records of simulated seismic data. Given in each row from left to right are original data, noisy data (with added Gaussian noise), and magnified data (to twice as large). The corresponding velocity models of each row  are the three models shown in Figure \ref{fig12-1}-Figure \ref{fig12-3}, respectively. }
  \label{fig20}
\end{figure*}

\clearpage
\begin{figure*}
\centering
  \hspace{-0.4cm}
  \subfigure[]{\label{fig13-1}
  \includegraphics[width=0.35\textwidth]{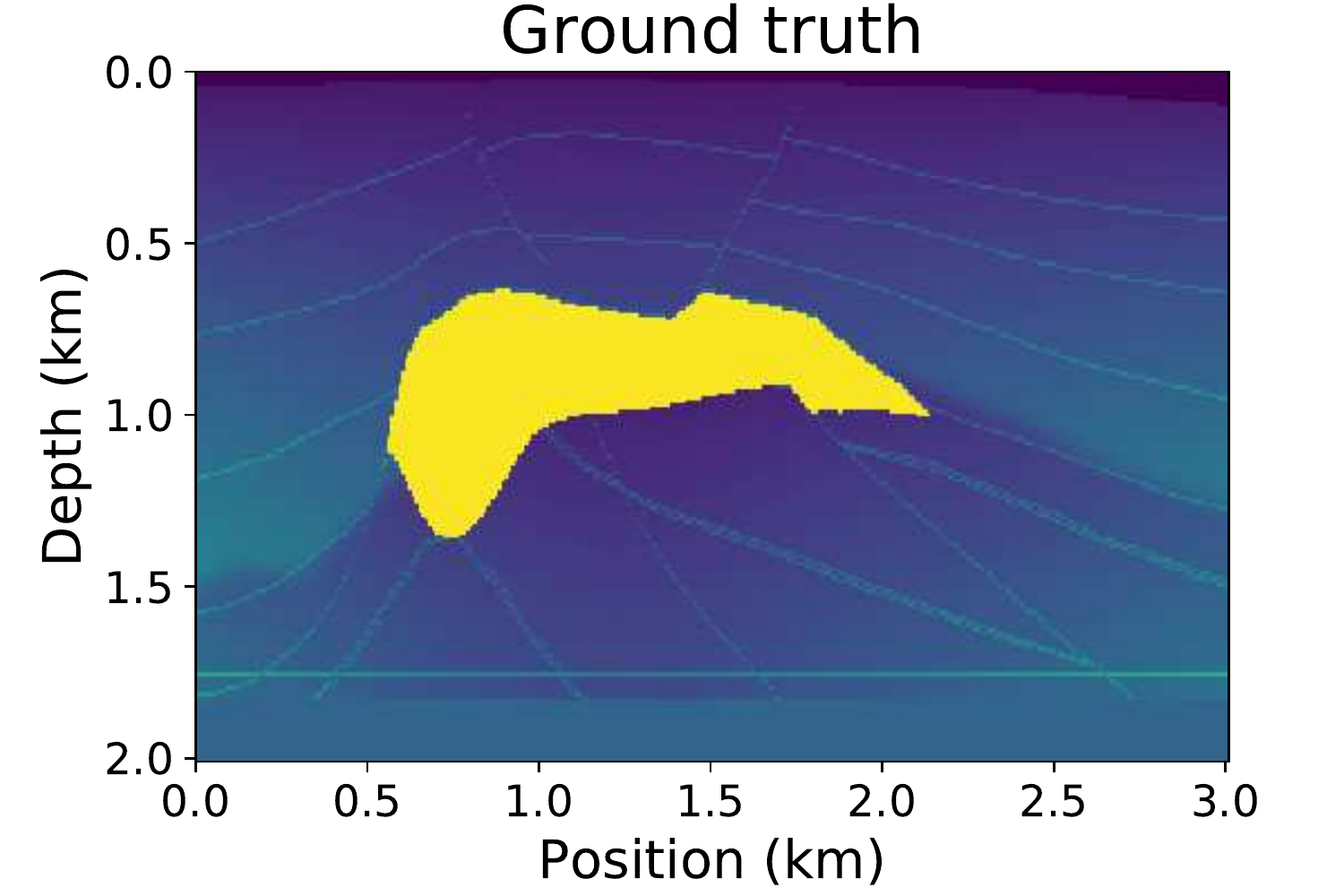}}
  \hspace{-0.7cm}
  \subfigure[]{\label{fig13-2}
  \includegraphics[width=0.35\textwidth]{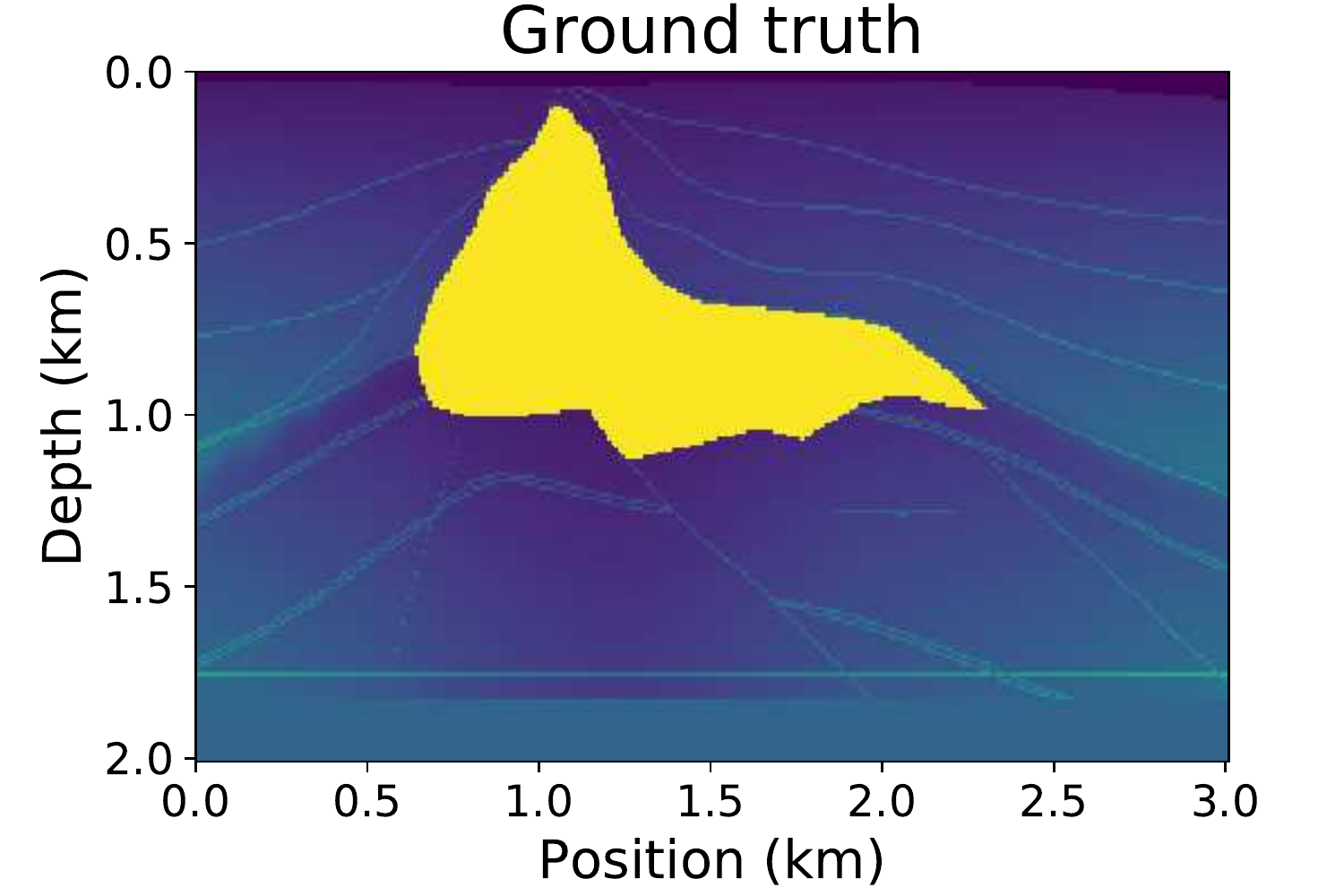}}
  \hspace{-0.7cm}
  \subfigure[]{\label{fig13-3}
  \includegraphics[width=0.35\textwidth]{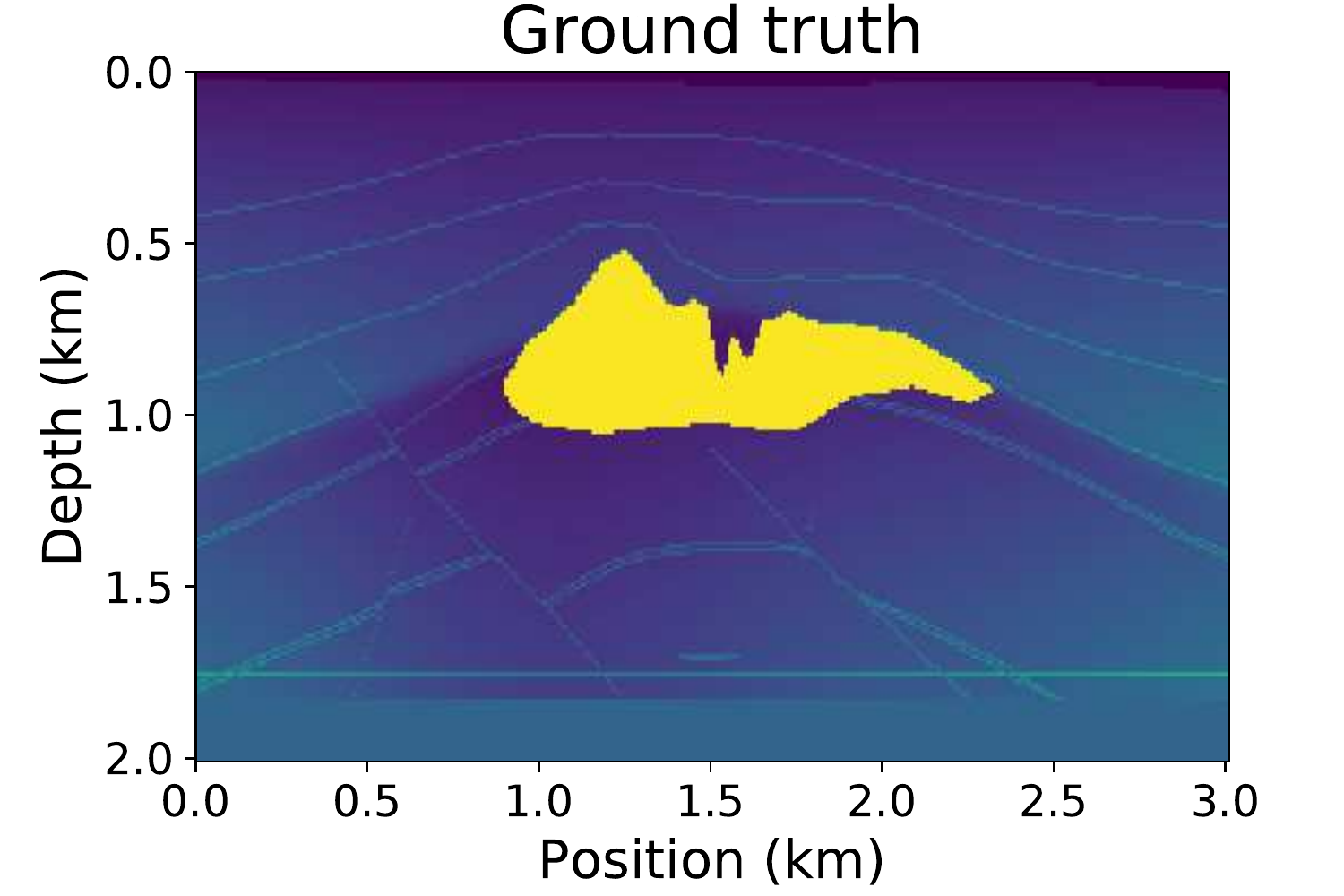}}

  \hspace{-0.4cm}
  \subfigure[]{\label{fig13-4}
  \includegraphics[width=0.35\textwidth]{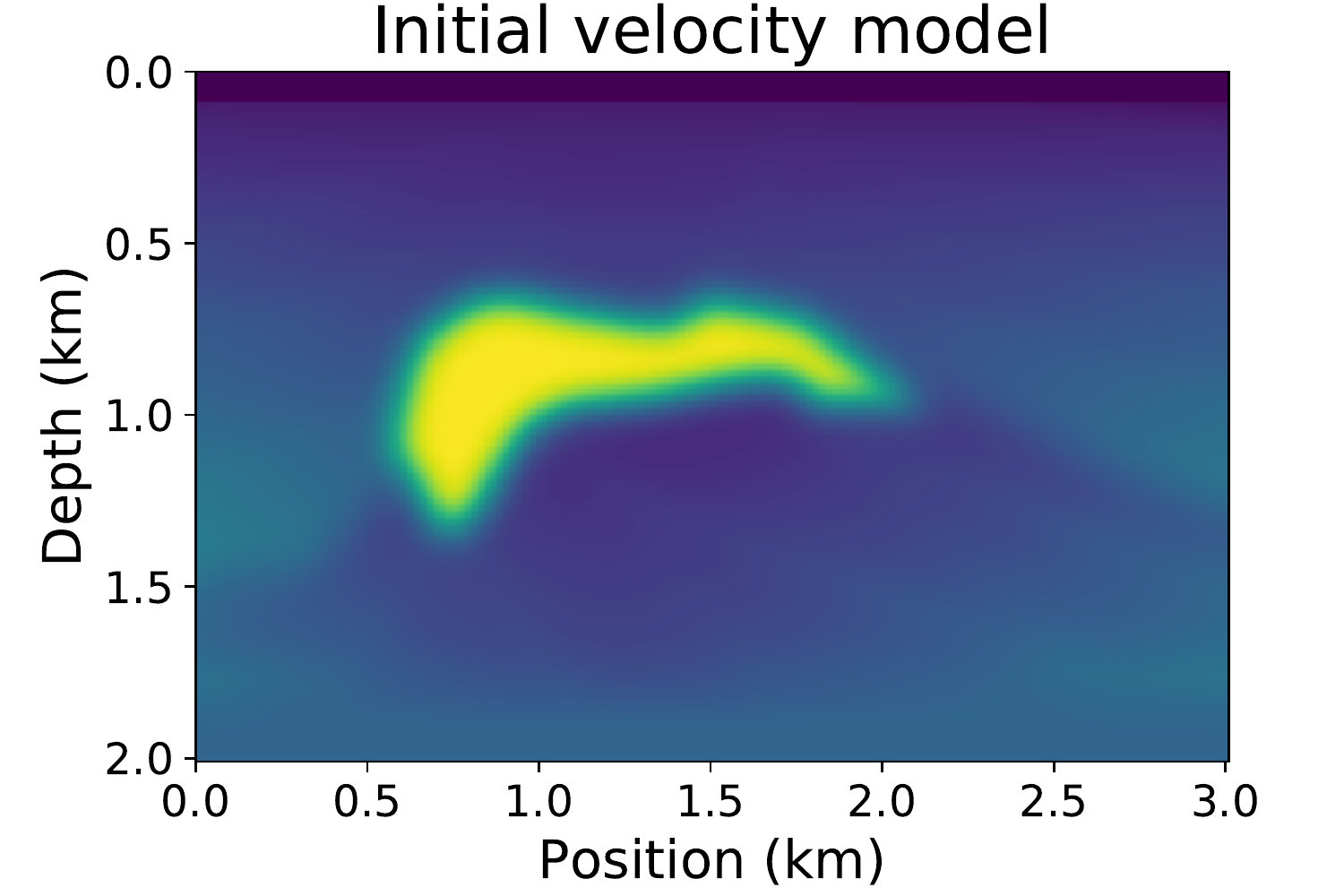}}
  \hspace{-0.7cm}
  \subfigure[]{\label{fig13-5}
  \includegraphics[width=0.35\textwidth]{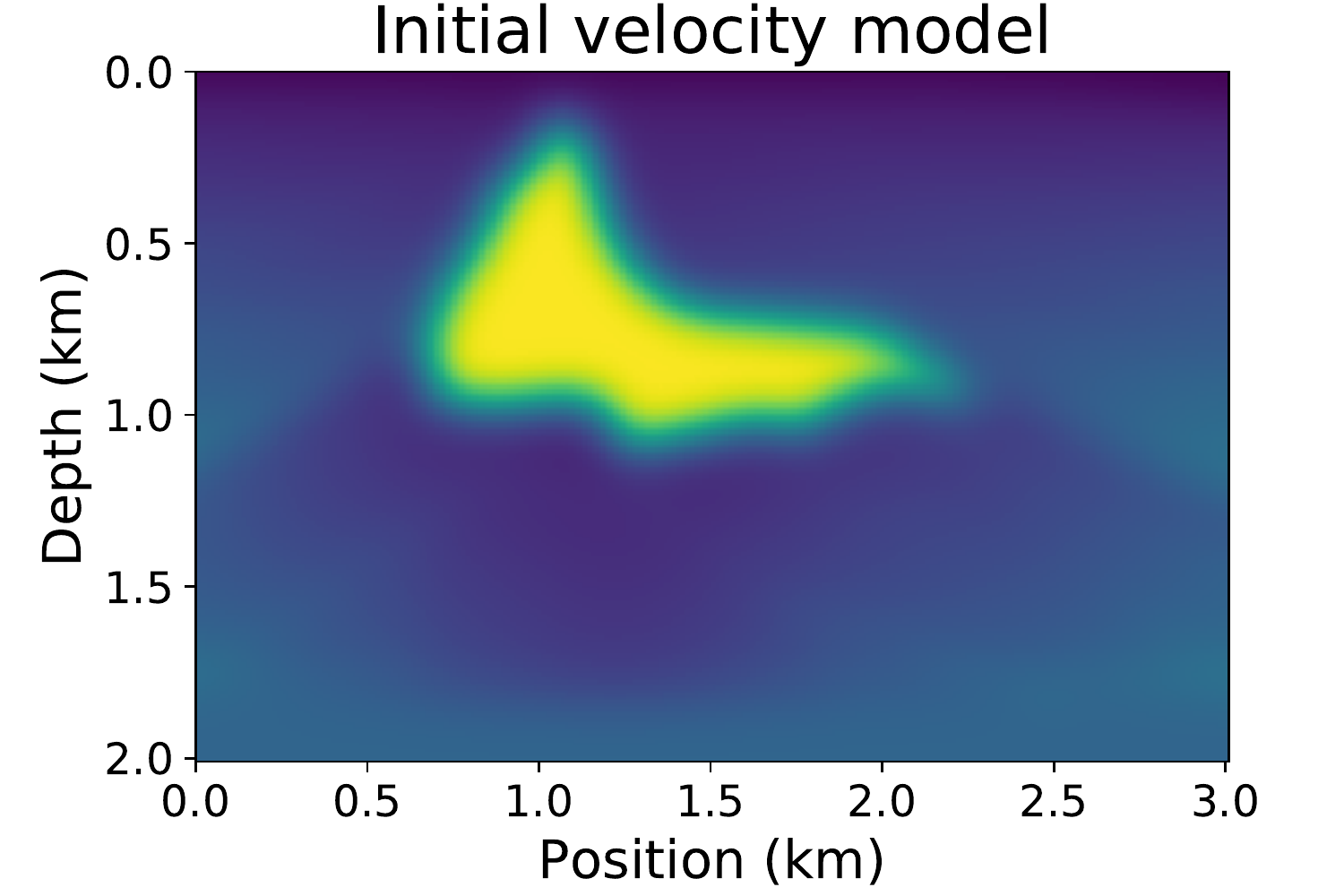}}
  \hspace{-0.7cm}
  \subfigure[]{\label{fig13-6}
  \includegraphics[width=0.35\textwidth]{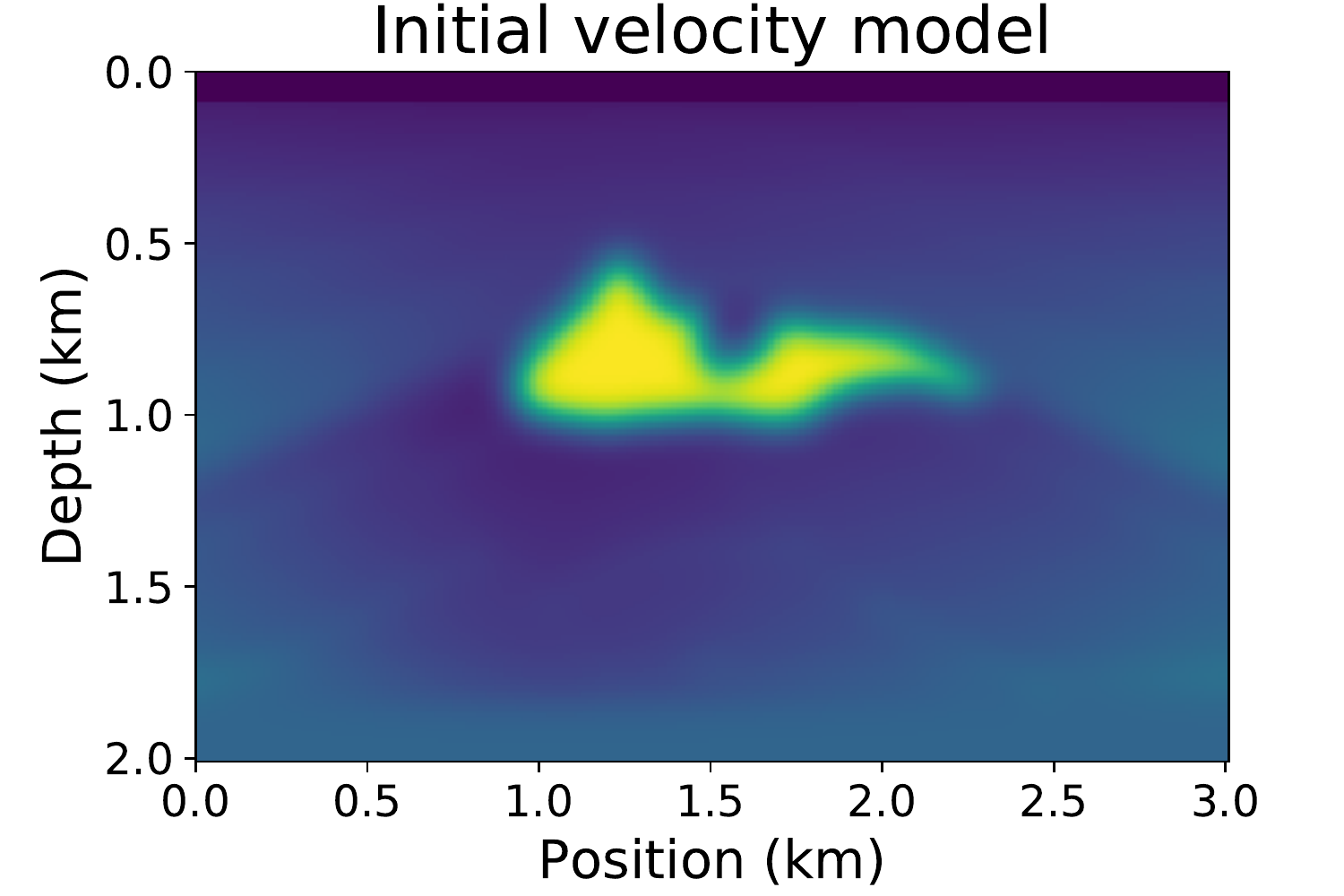}}

  \hspace{-0.4cm}
  \subfigure[]{\label{fig13-7}
  \includegraphics[width=0.35\textwidth]{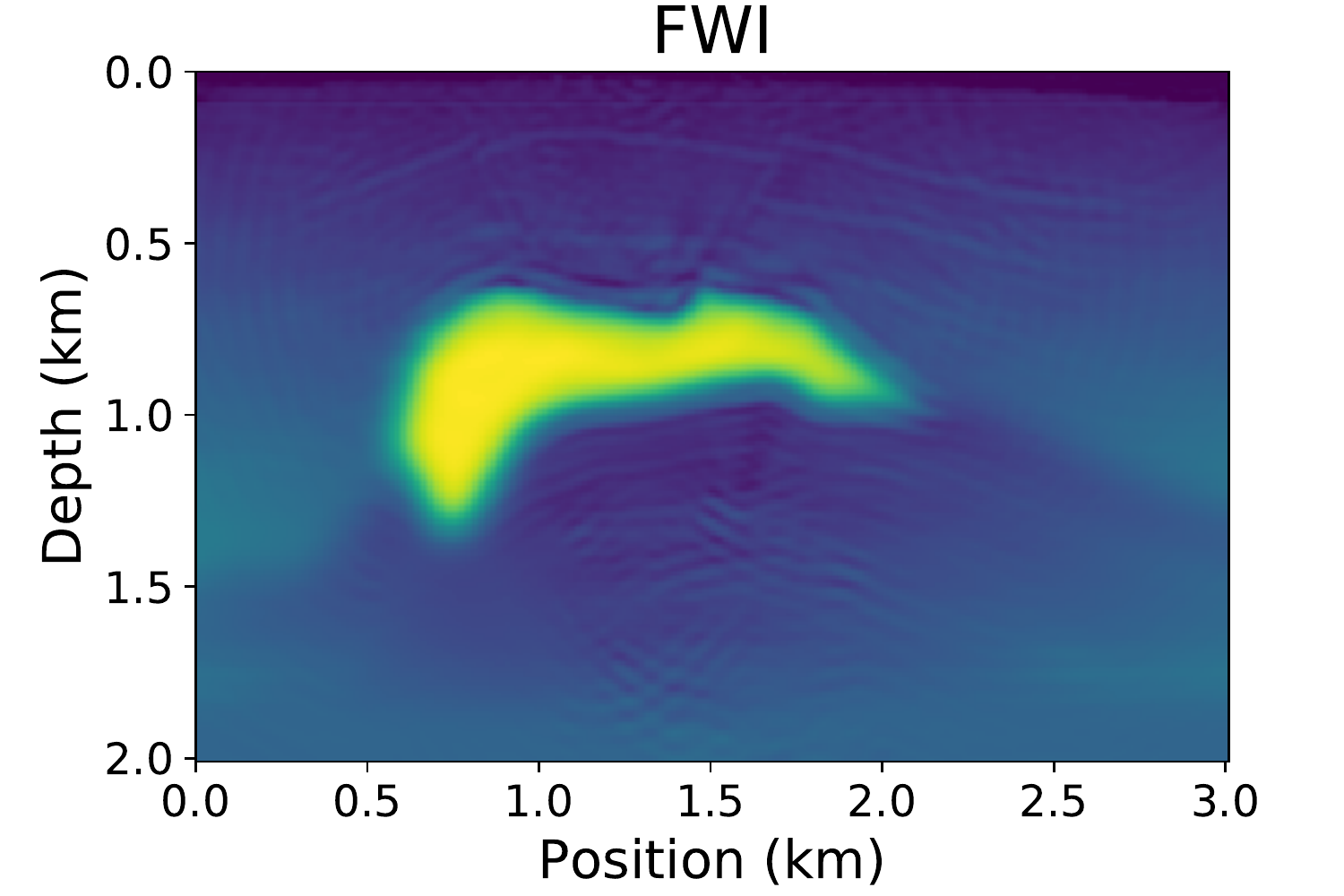}}
  \hspace{-0.7cm}
  \subfigure[]{\label{fig13-8}
  \includegraphics[width=0.35\textwidth]{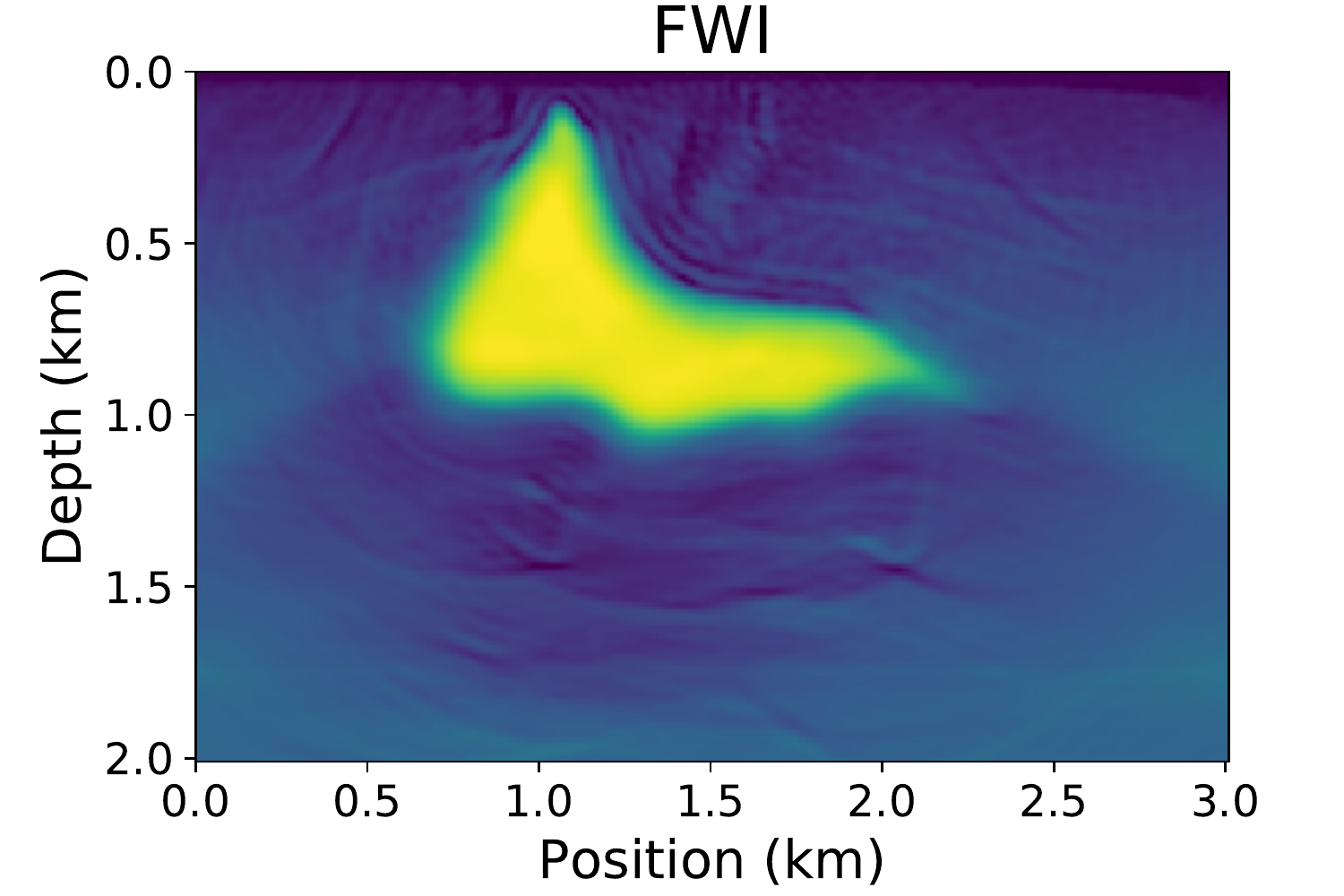}}
  \hspace{-0.7cm}
  \subfigure[]{\label{fig13-9}
  \includegraphics[width=0.34\textwidth]{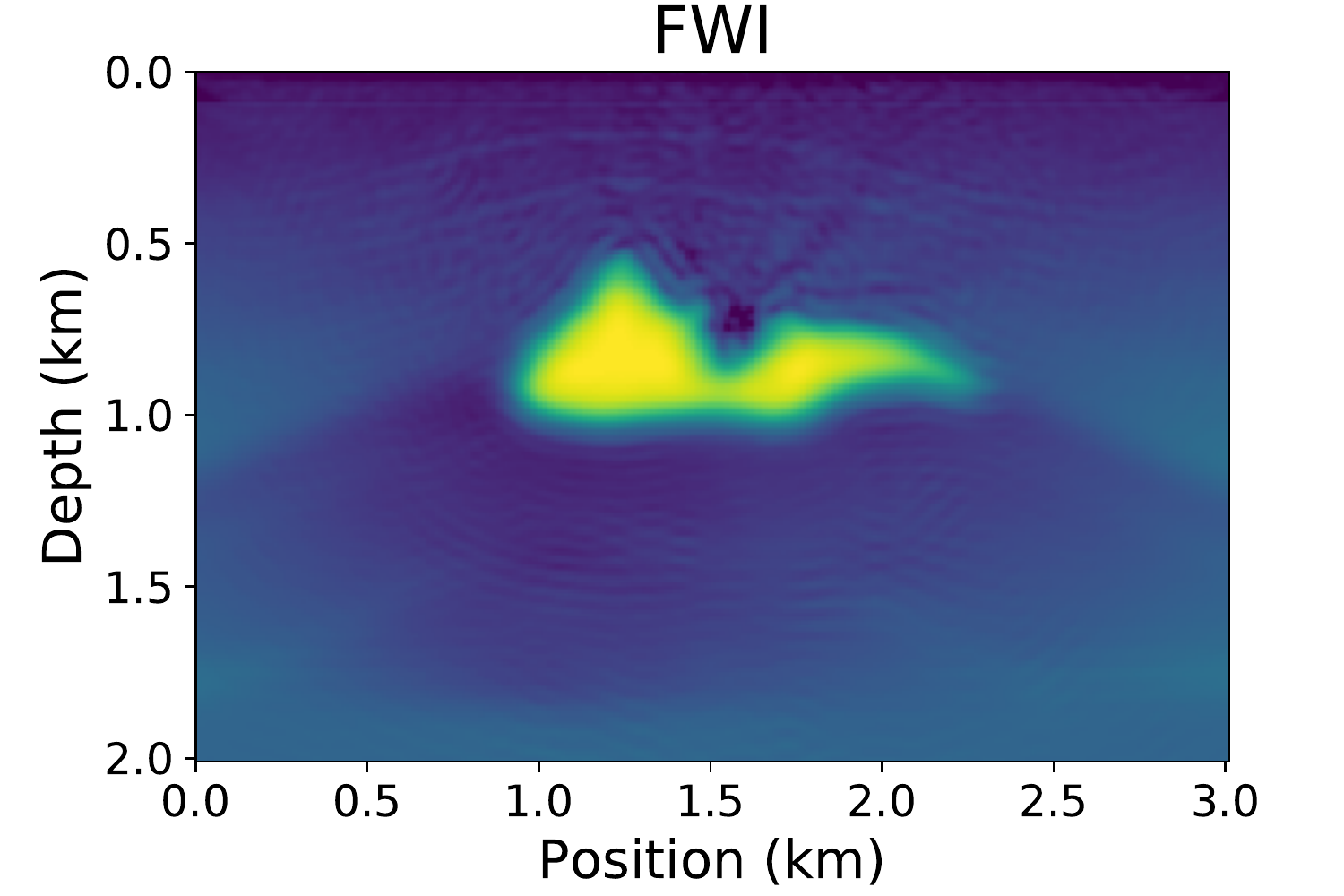}}

  \hspace{-0.4cm}
  \subfigure[]{\label{fig13-10}
  \includegraphics[width=0.34\textwidth]{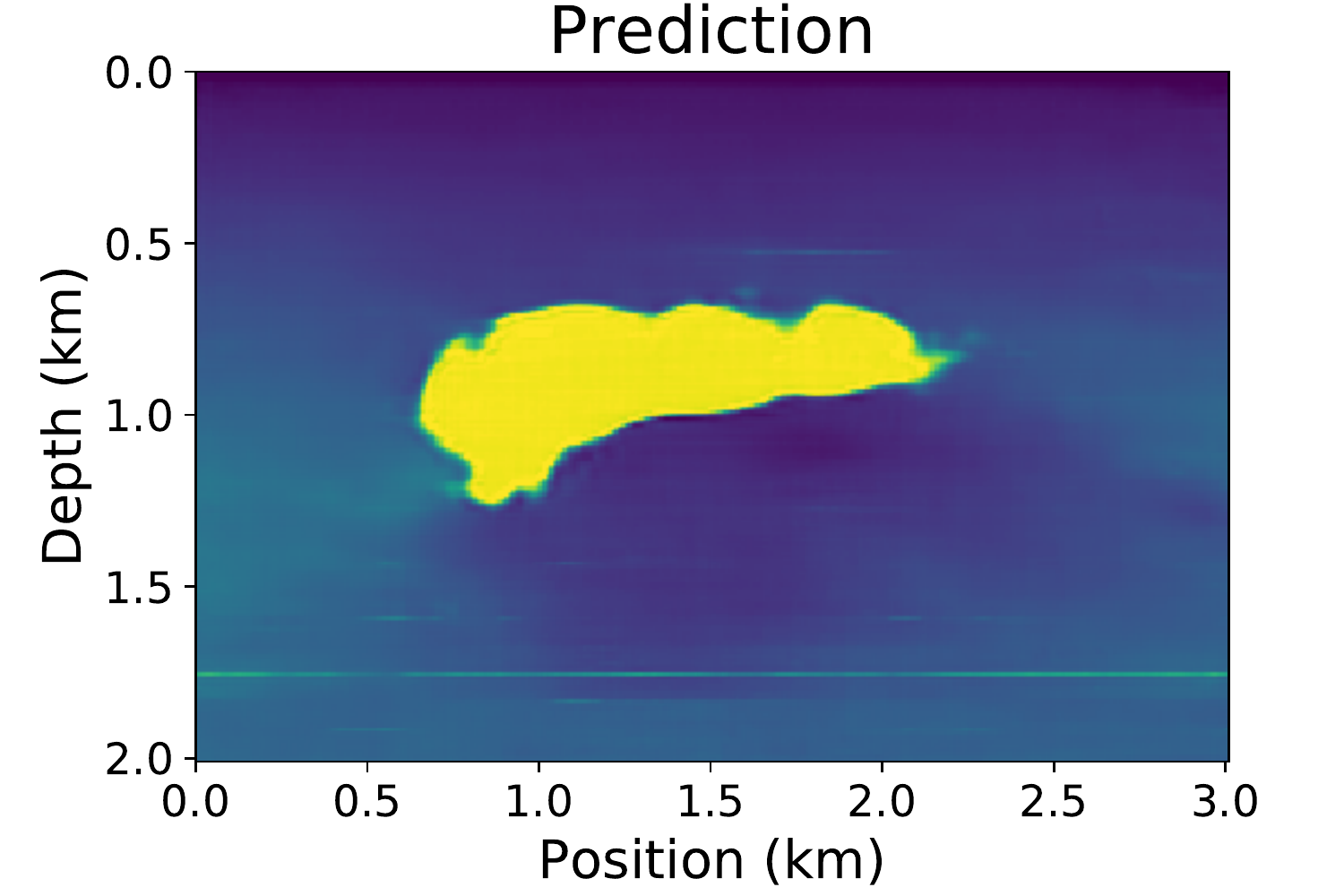}}
  \hspace{-0.7cm}
  \subfigure[]{\label{fig13-11}
  \includegraphics[width=0.34\textwidth]{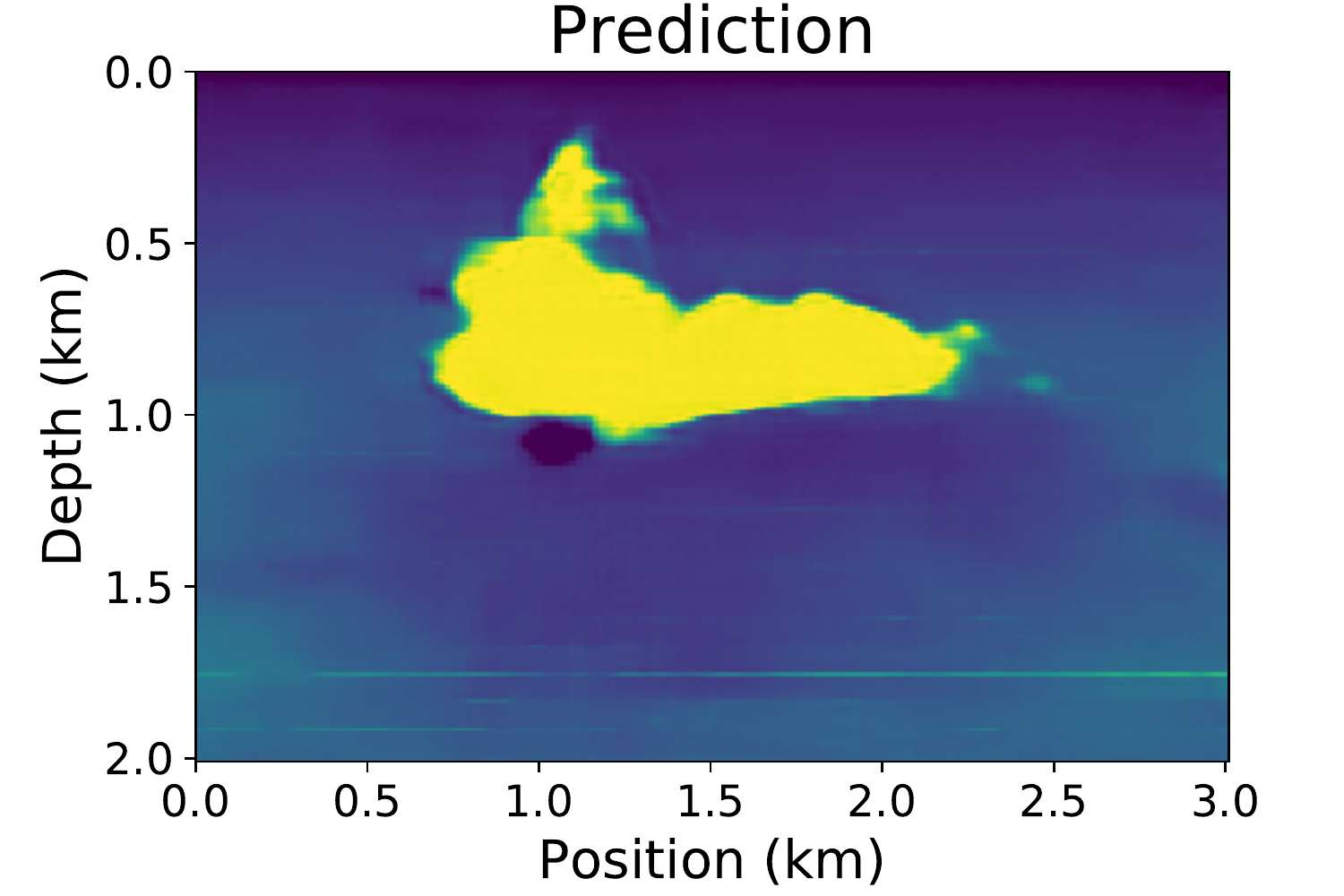}}
  \hspace{-0.7cm}
  \subfigure[]{\label{fig13-12}
  \includegraphics[width=0.34\textwidth]{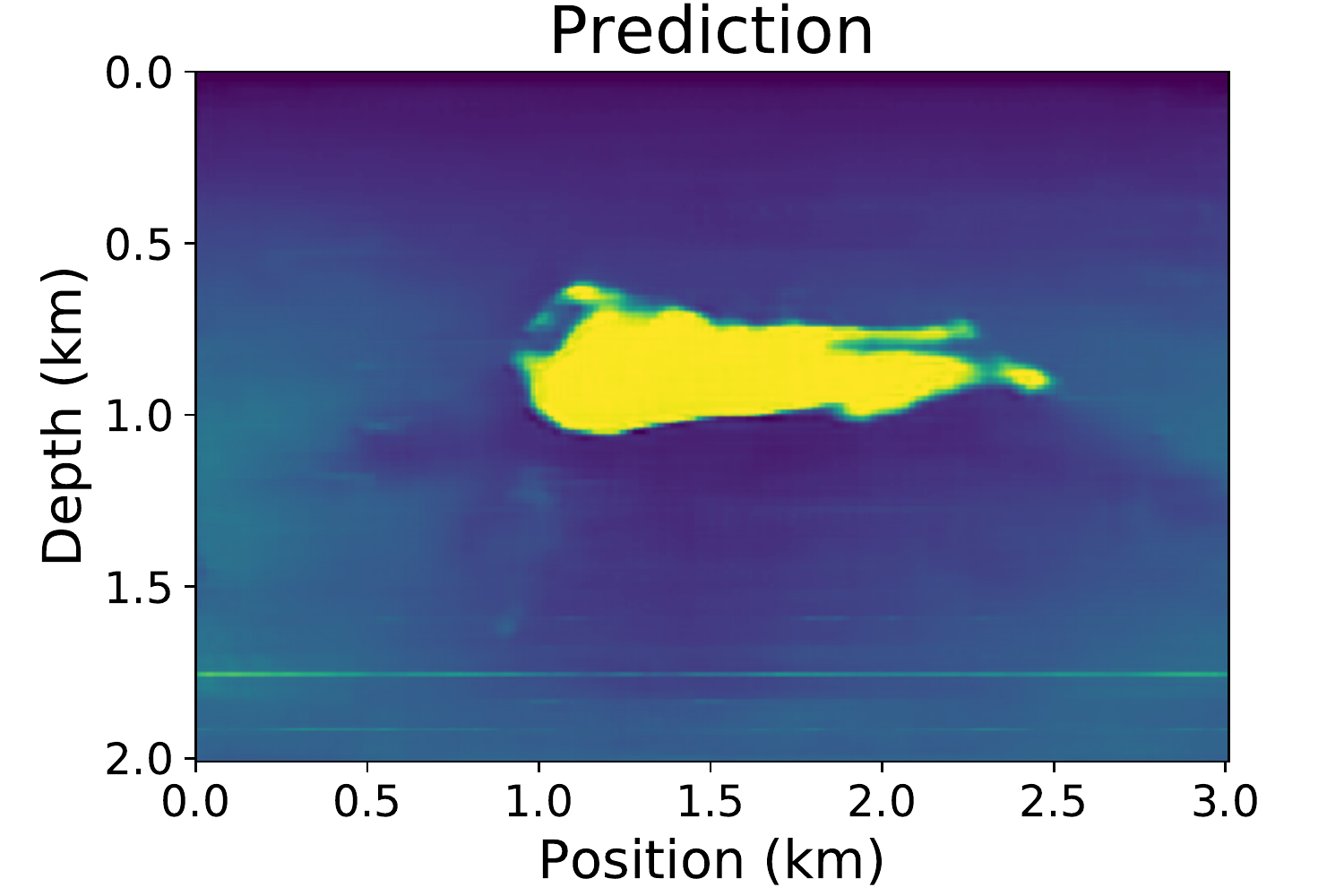}}
  \caption{Comparisons of the velocity inversion (SEG salt models): (a)--(c) ground truth; (d)--(f) initial velocity model of  FWI; (g)--(i) results of  FWI; (j)--(l) prediction of  our method.}
  \label{fig14}
\end{figure*}

\clearpage
\begin{figure*}
\centering
  \hspace{-0.4cm}
  \includegraphics[width=0.5\textwidth]{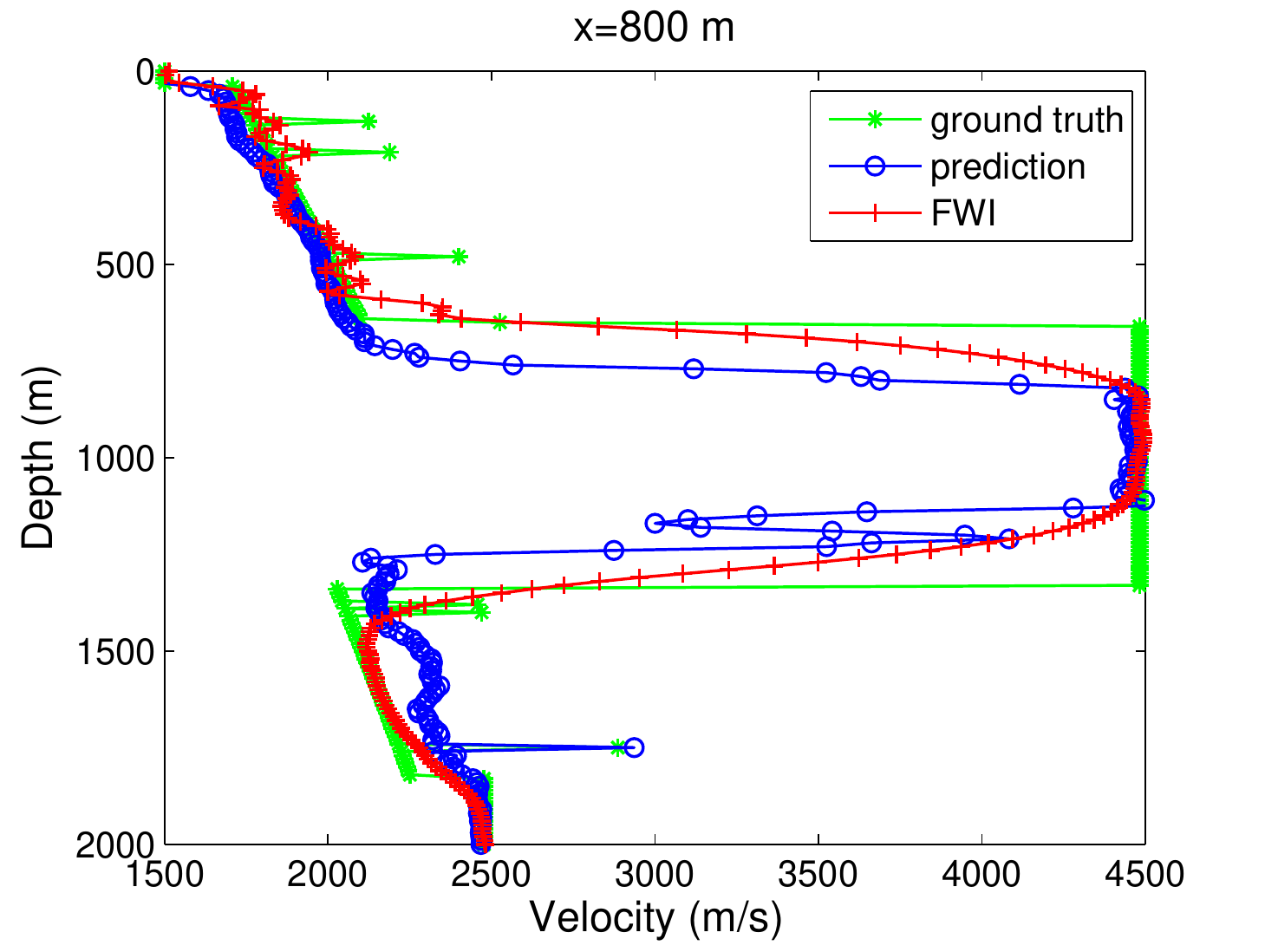}
  \hspace{-0.4cm}
  \includegraphics[width=0.5\textwidth]{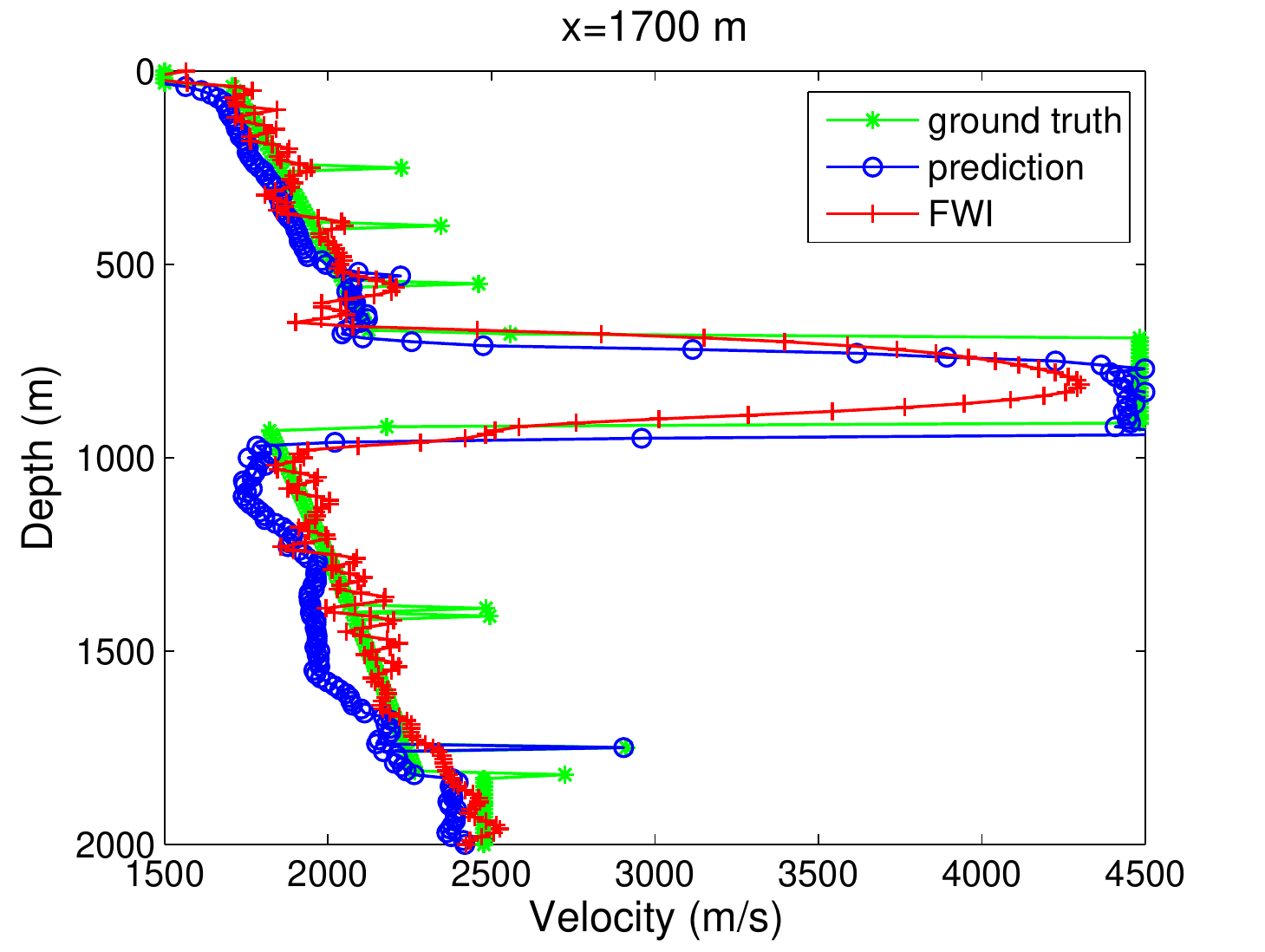}\\

  \hspace{-0.4cm}
  \includegraphics[width=0.5\textwidth]{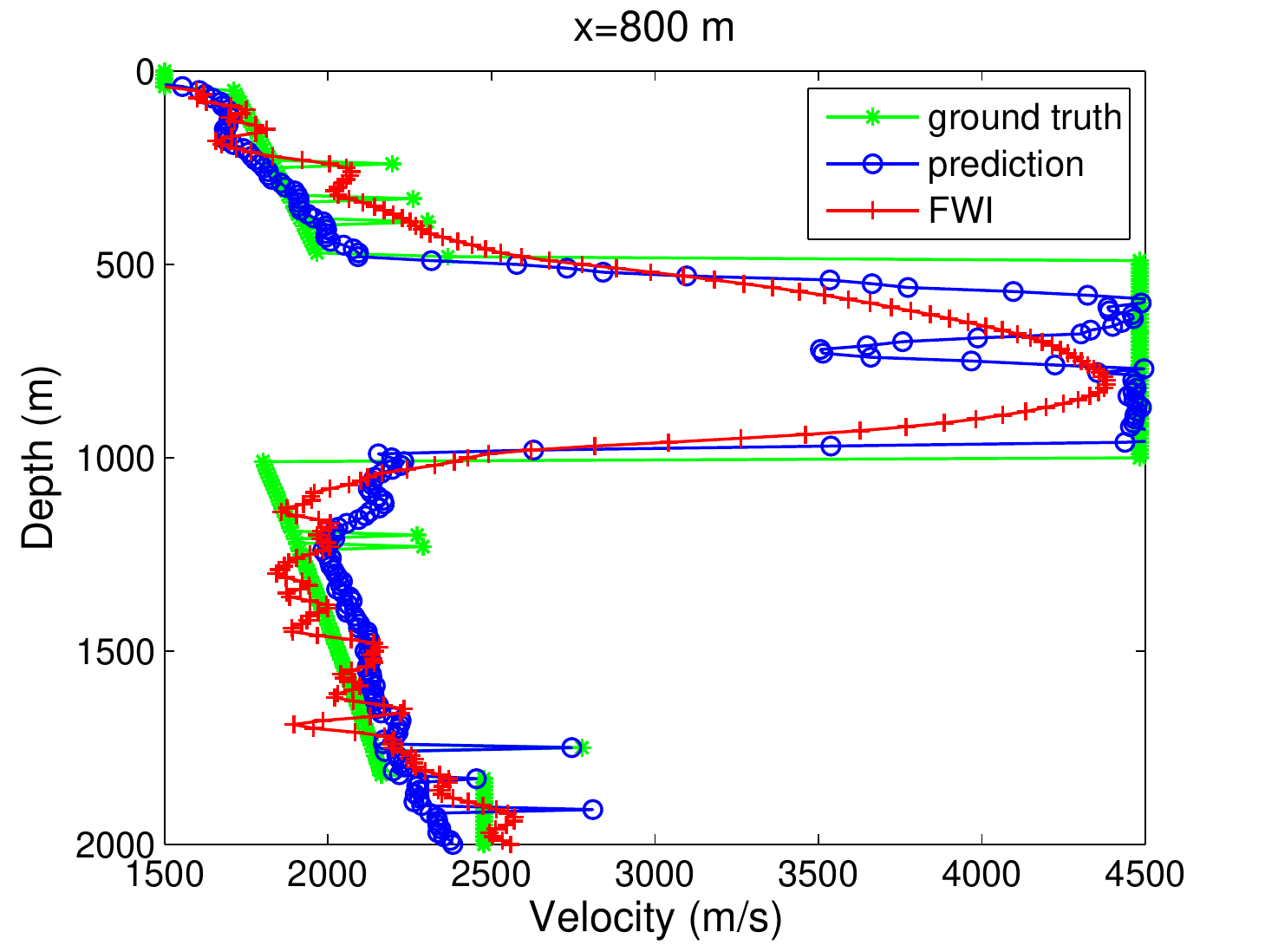}
  \hspace{-0.4cm}
  \includegraphics[width=0.5\textwidth]{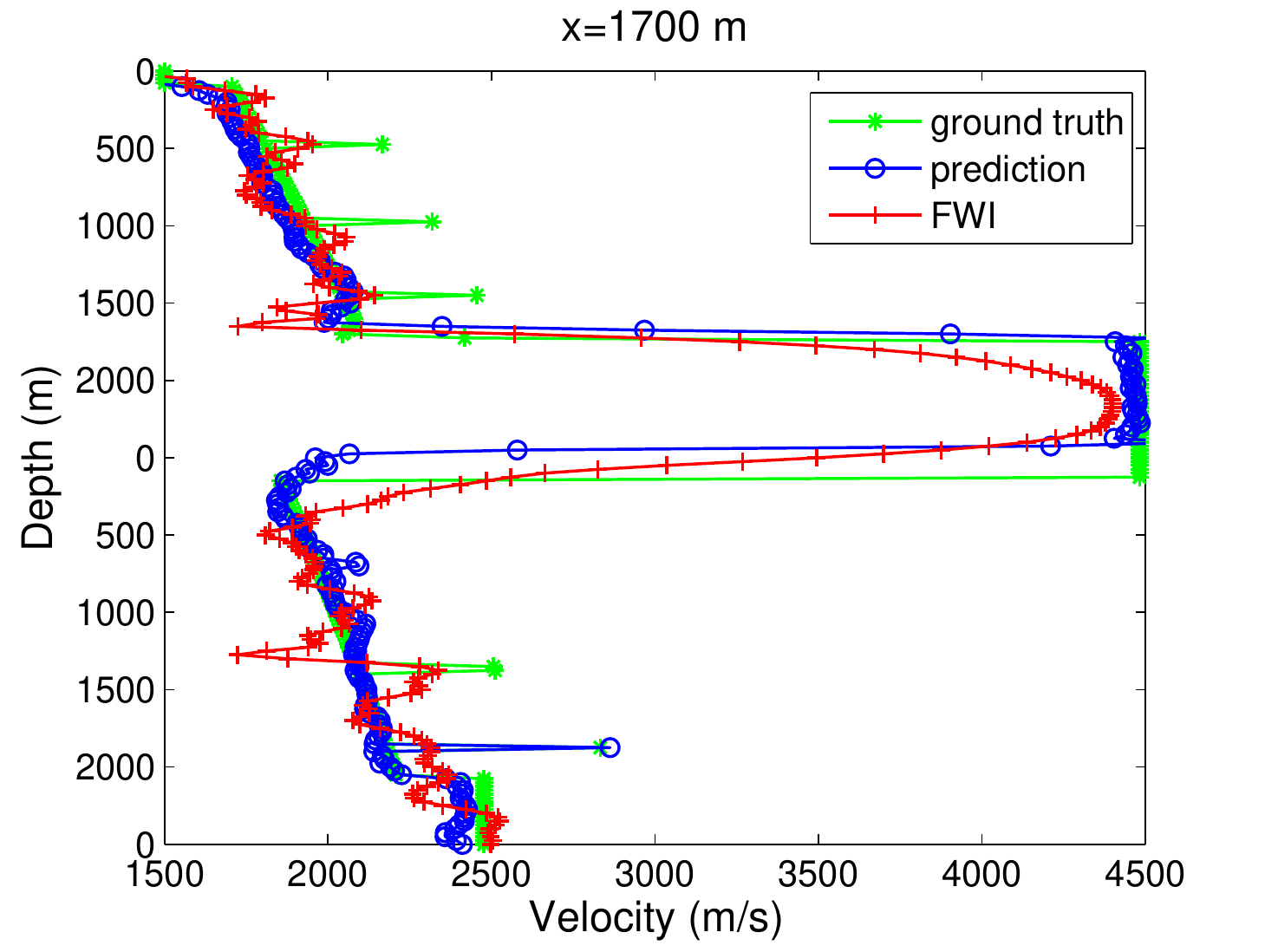}\\

  \hspace{-0.4cm}
  \includegraphics[width=0.5\textwidth]{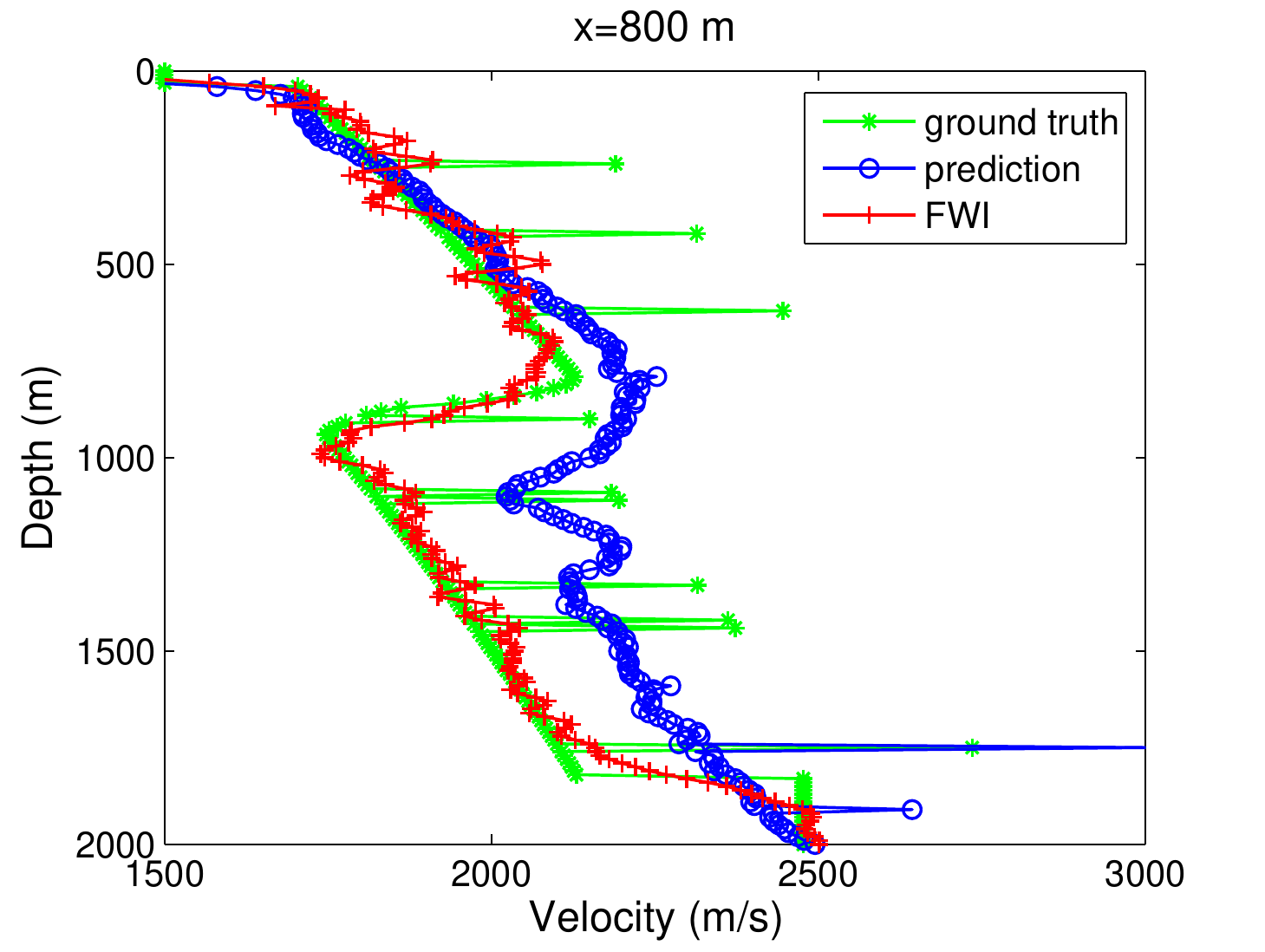}
  \hspace{-0.4cm}
  \includegraphics[width=0.5\textwidth]{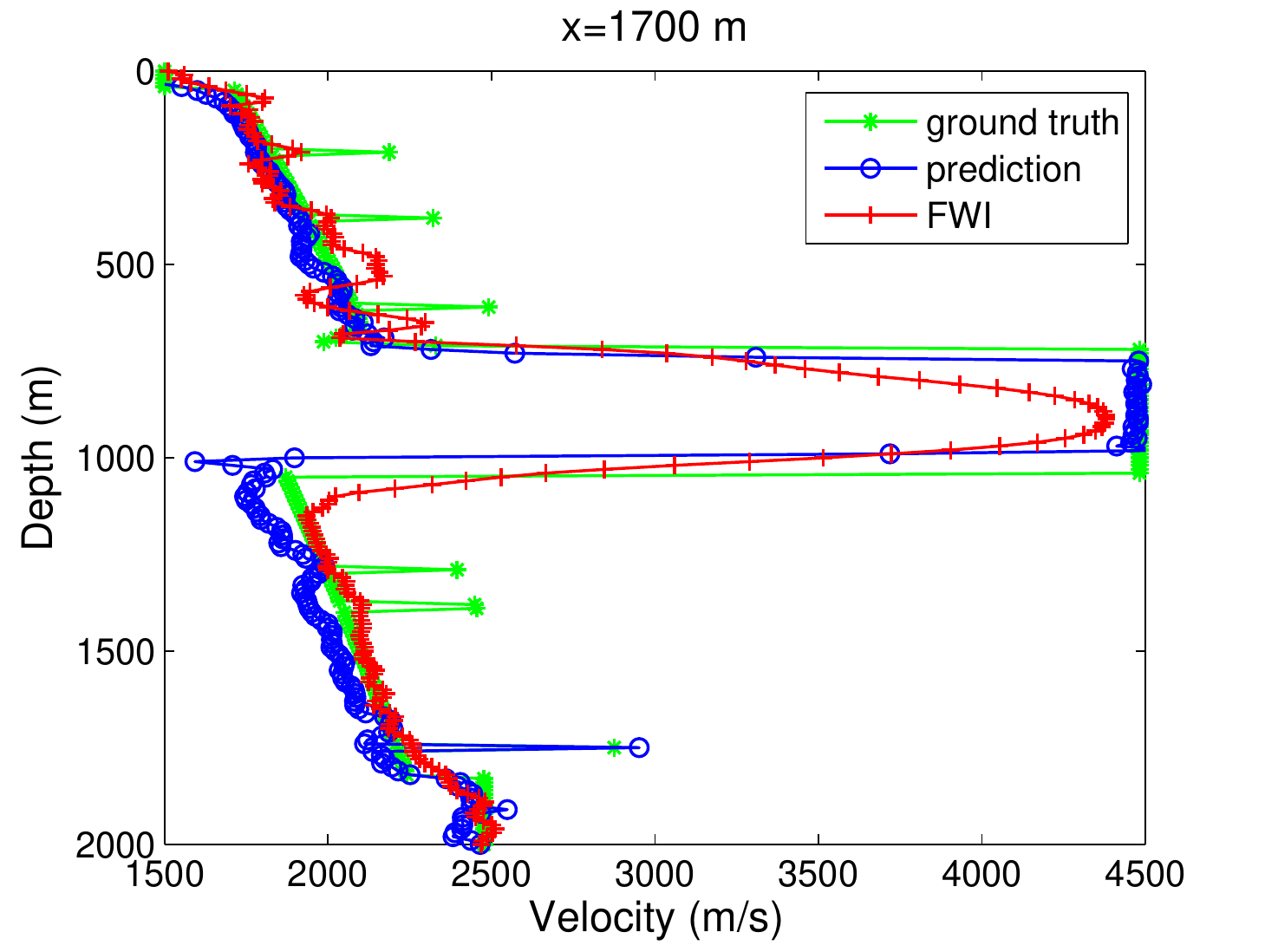}
  \caption{Vertical velocity profiles of our method and  FWI. The prediction, FWI, and the ground-truth velocities in the velocity versus depth profiles at two horizontal positions (x = 800 m, x = 1700 m) of the three test samples in Figure \ref{fig14} are shown in each row.}
  \label{fig15}
\end{figure*}

\clearpage
\begin{figure*}
\centering
  \hspace{-0.4cm}
  \subfigure[]{\label{fig13-31}
  \includegraphics[width=0.35\textwidth]{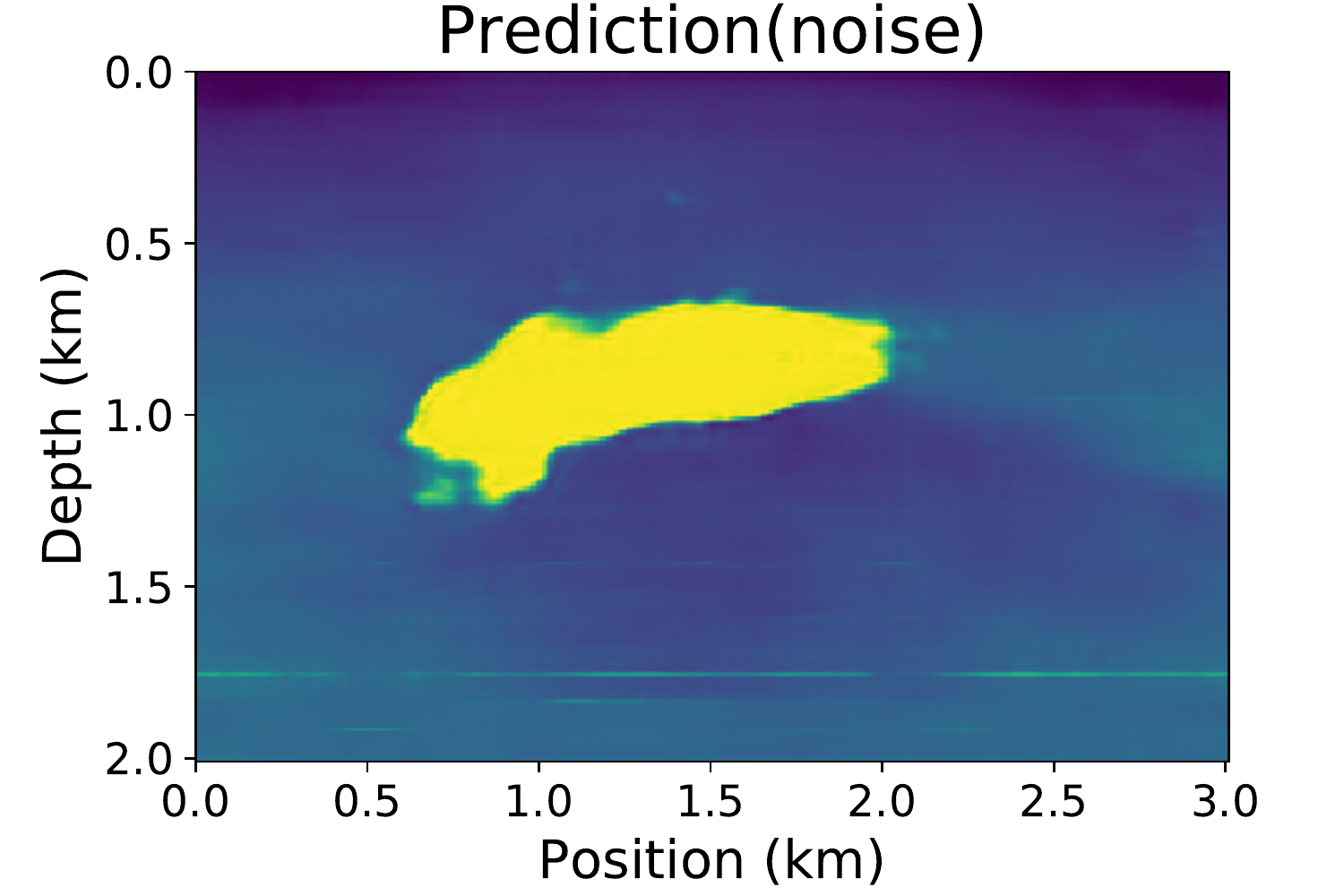}}
  \hspace{-0.7cm}
  \subfigure[]{\label{fig13-32}
  \includegraphics[width=0.35\textwidth]{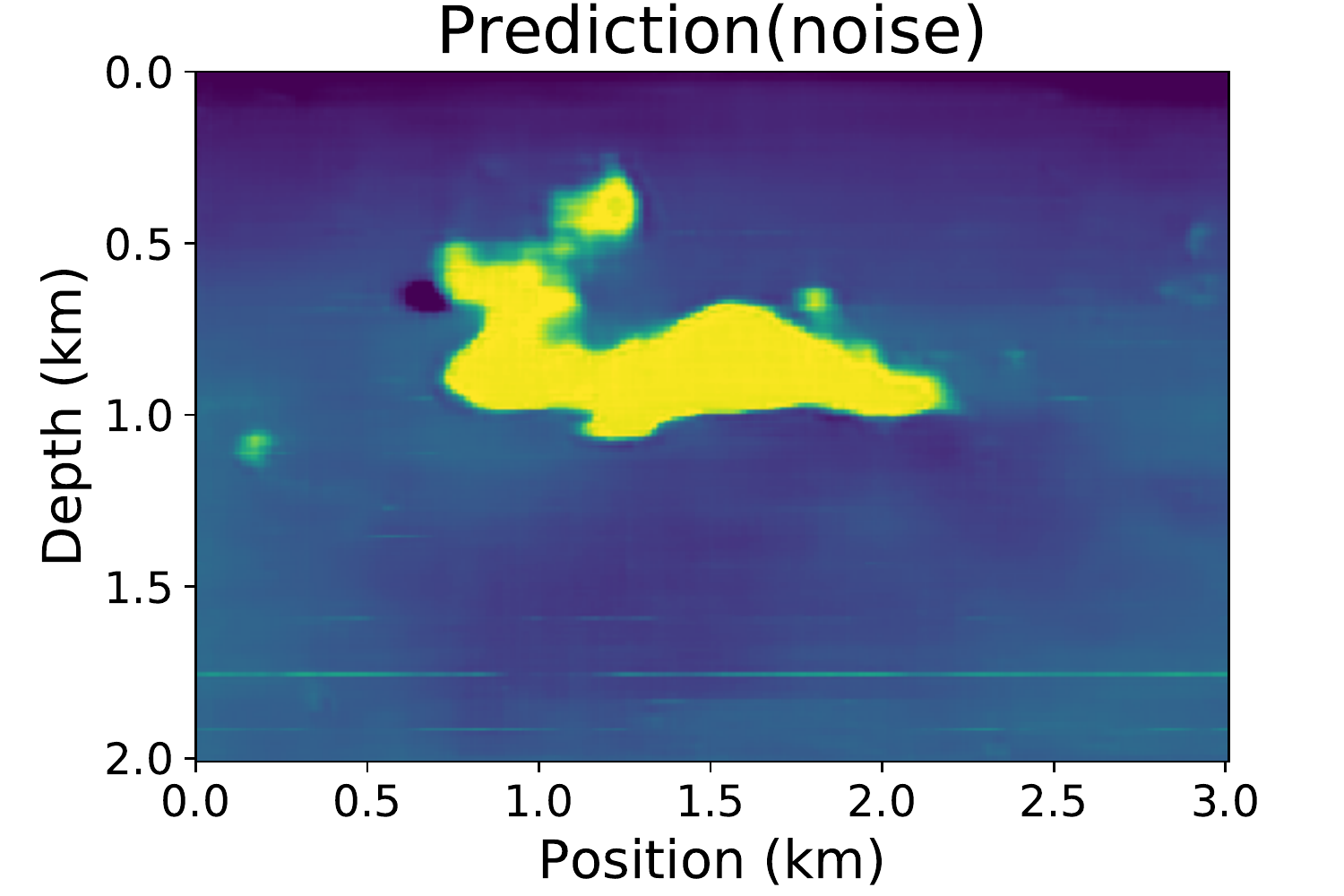}}
  \hspace{-0.7cm}
  \subfigure[]{\label{fig13-33}
  \includegraphics[width=0.35\textwidth]{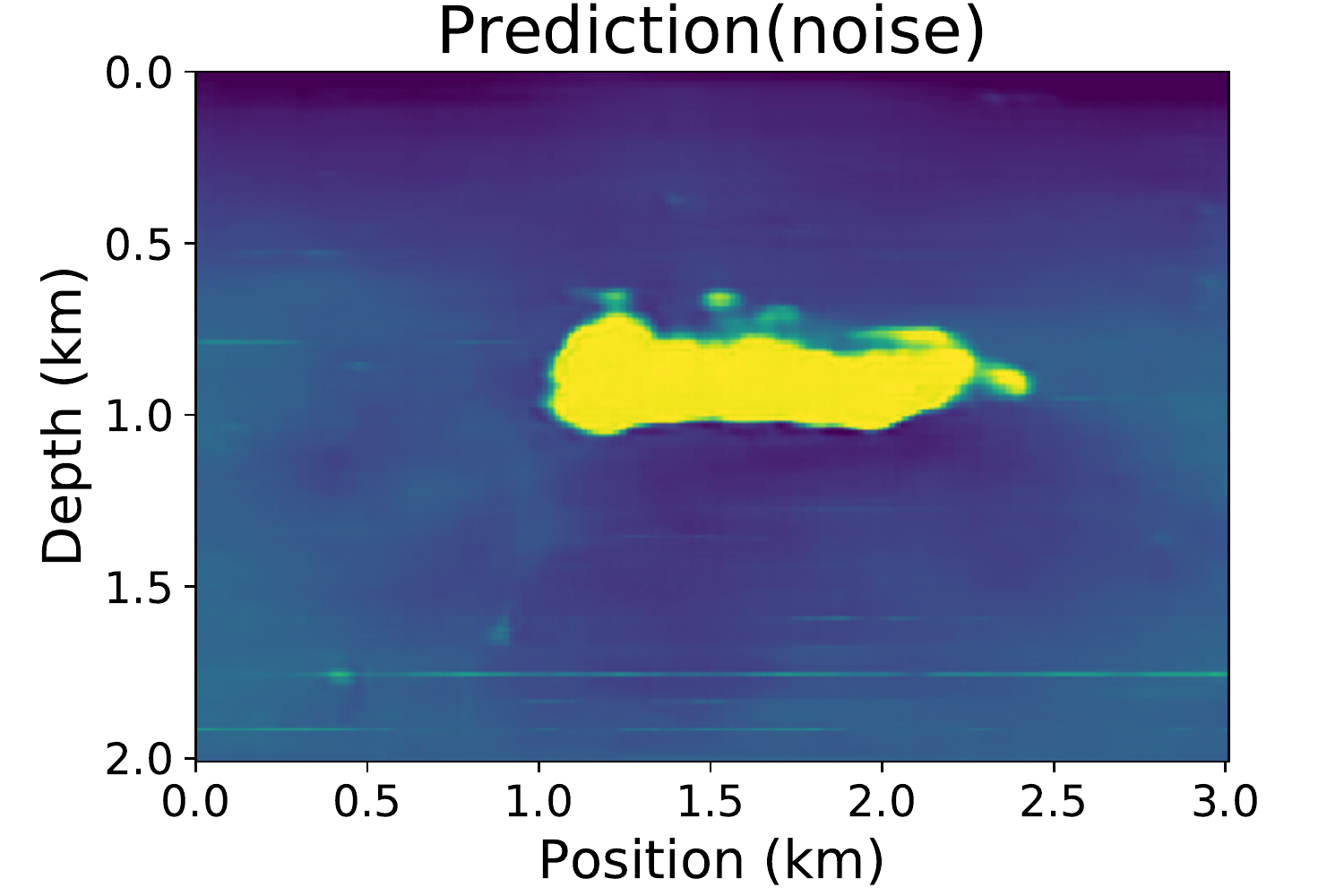}}

  \hspace{-0.4cm}
  \subfigure[]{\label{fig13-34}
  \includegraphics[width=0.35\textwidth]{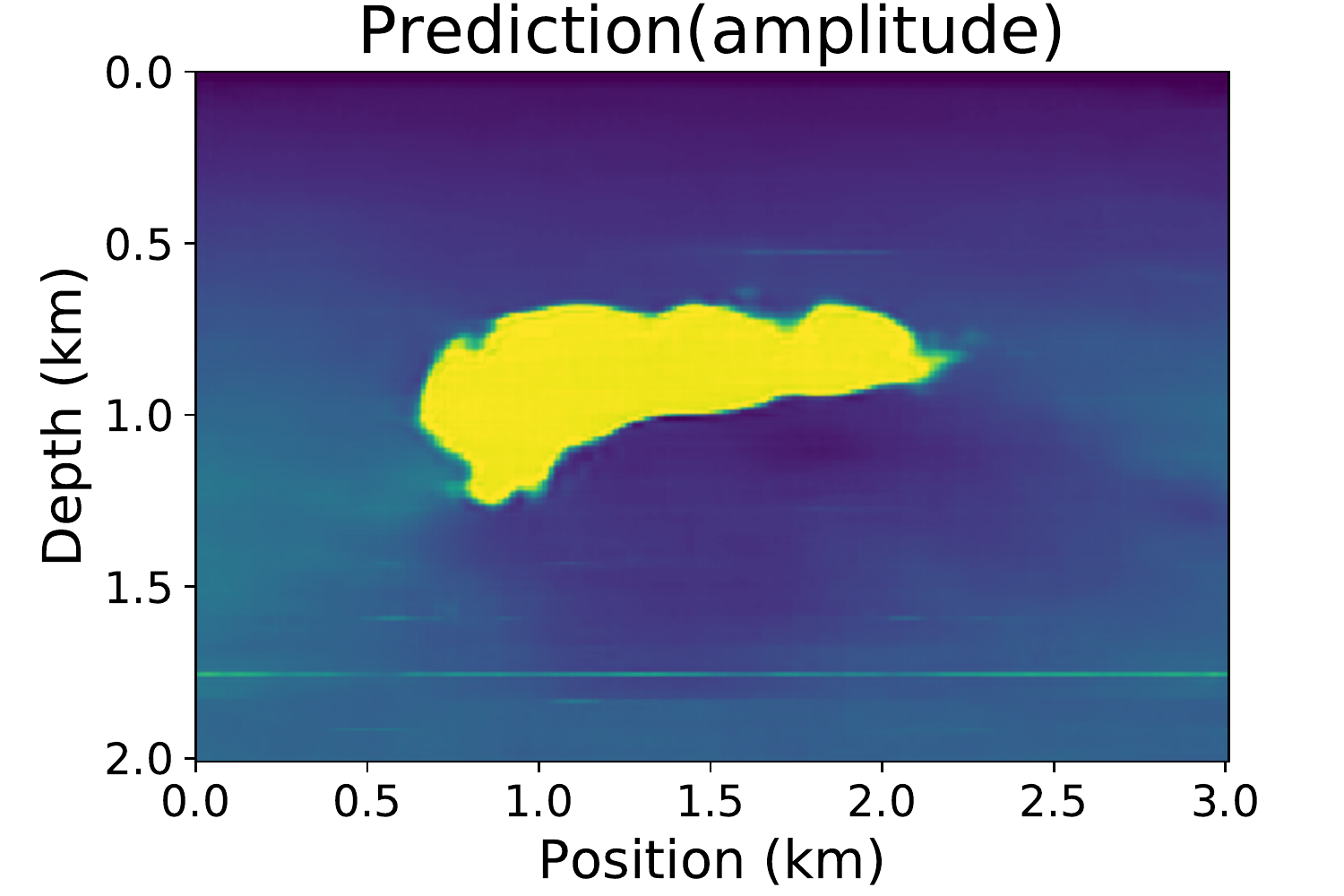}}
  \hspace{-0.7cm}
  \subfigure[]{\label{fig13-35}
  \includegraphics[width=0.35\textwidth]{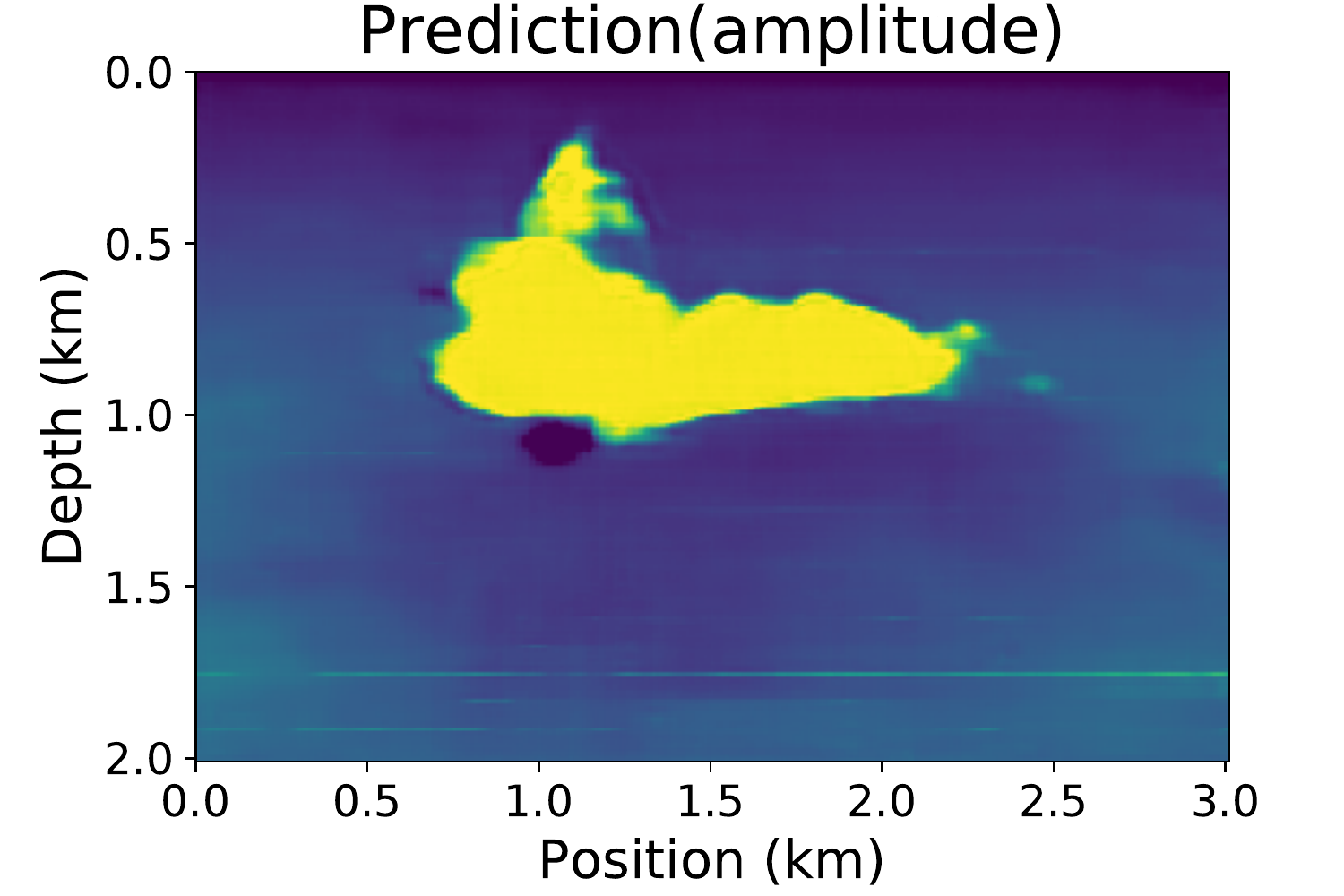}}
  \hspace{-0.7cm}
  \subfigure[]{\label{fig13-36}
  \includegraphics[width=0.35\textwidth]{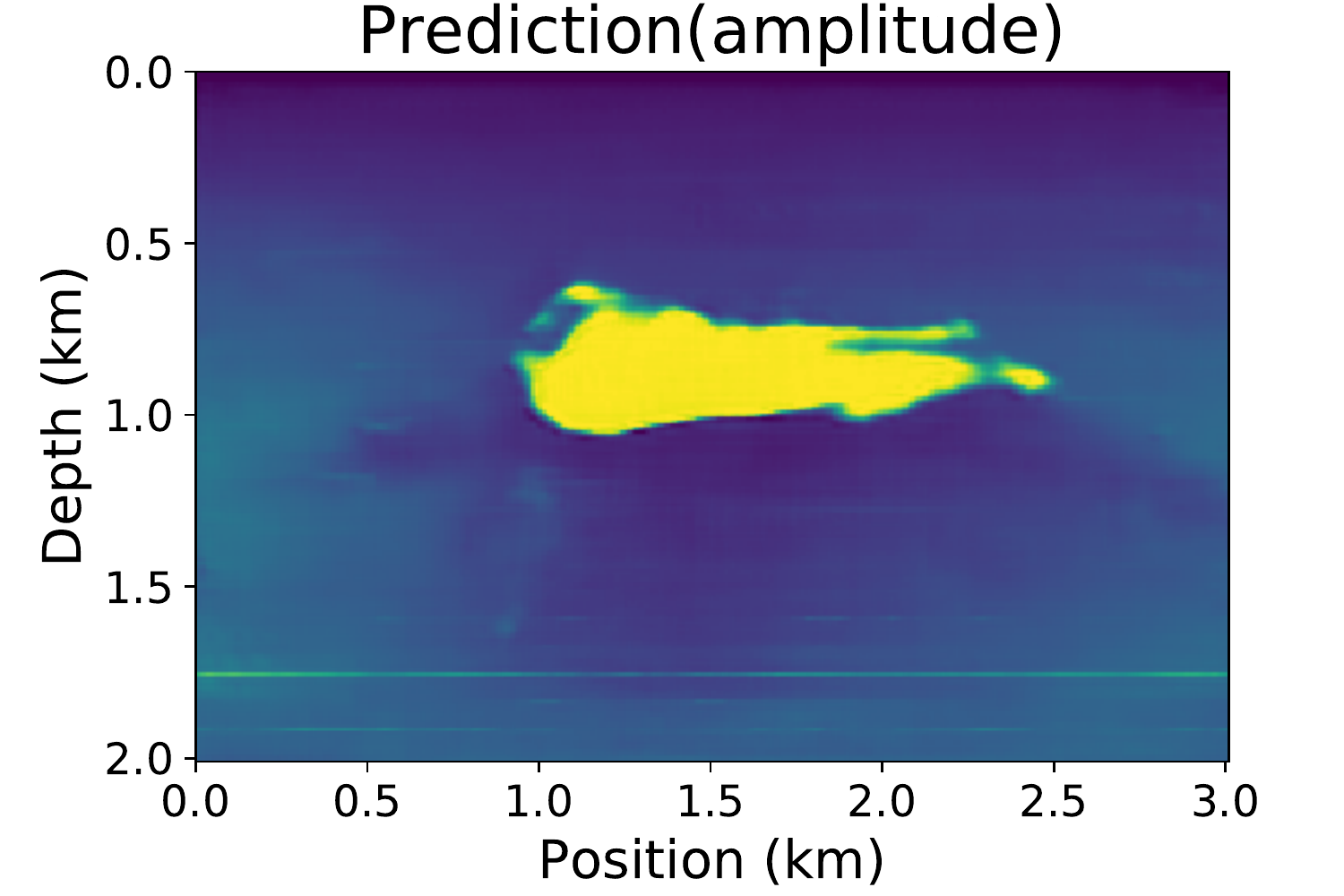}}
  \caption{Sensitivity of the proposed method to noise and amplitude (SEG models): (a)--(c) prediction results with the noisy seismic data; (d)--(f) prediction results with magnified seismic amplitude. For the  open  dataset,  our method  yielded acceptable predictions when the input data were perturbed.}
  \label{fig22}
\end{figure*}

\clearpage
\begin{figure*}
  \centering
  \includegraphics[scale=0.55]{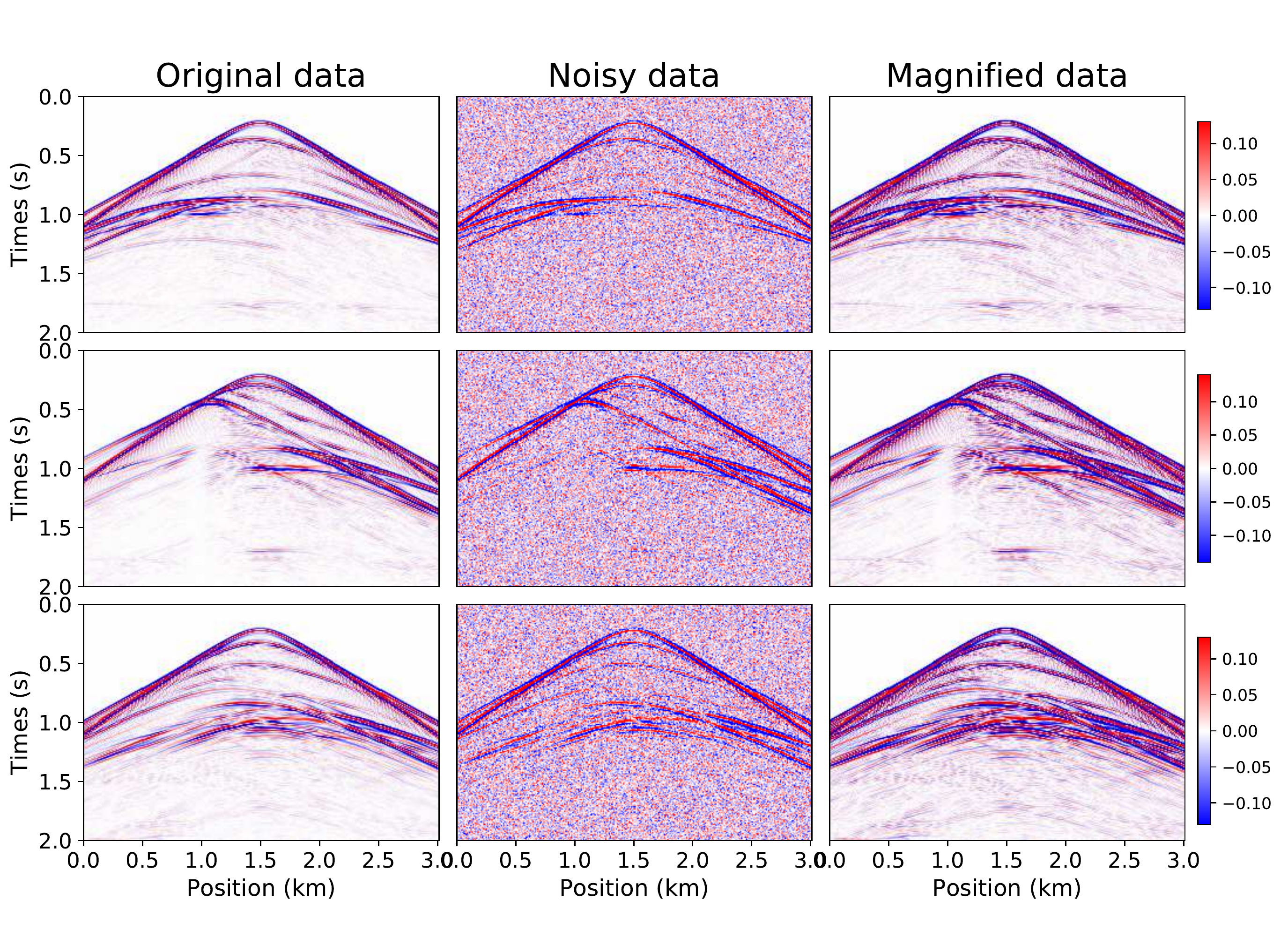}
  \caption{Comparison of records of SEG seismic data. Given in each row from left to right are original data, noisy data (with added Gaussian noise), and magnified data (to twice as larger).  The corresponding velocity models of each row are the three models shown in Figure \ref{fig13-1}--Figure \ref{fig13-3}, respectively.  }
  \label{fig21}
\end{figure*}

\clearpage
\begin{figure*}
\centering
  \hspace{-0.4cm}
  \subfigure[]{\label{fig16-1}
  \includegraphics[width=0.5\textwidth]{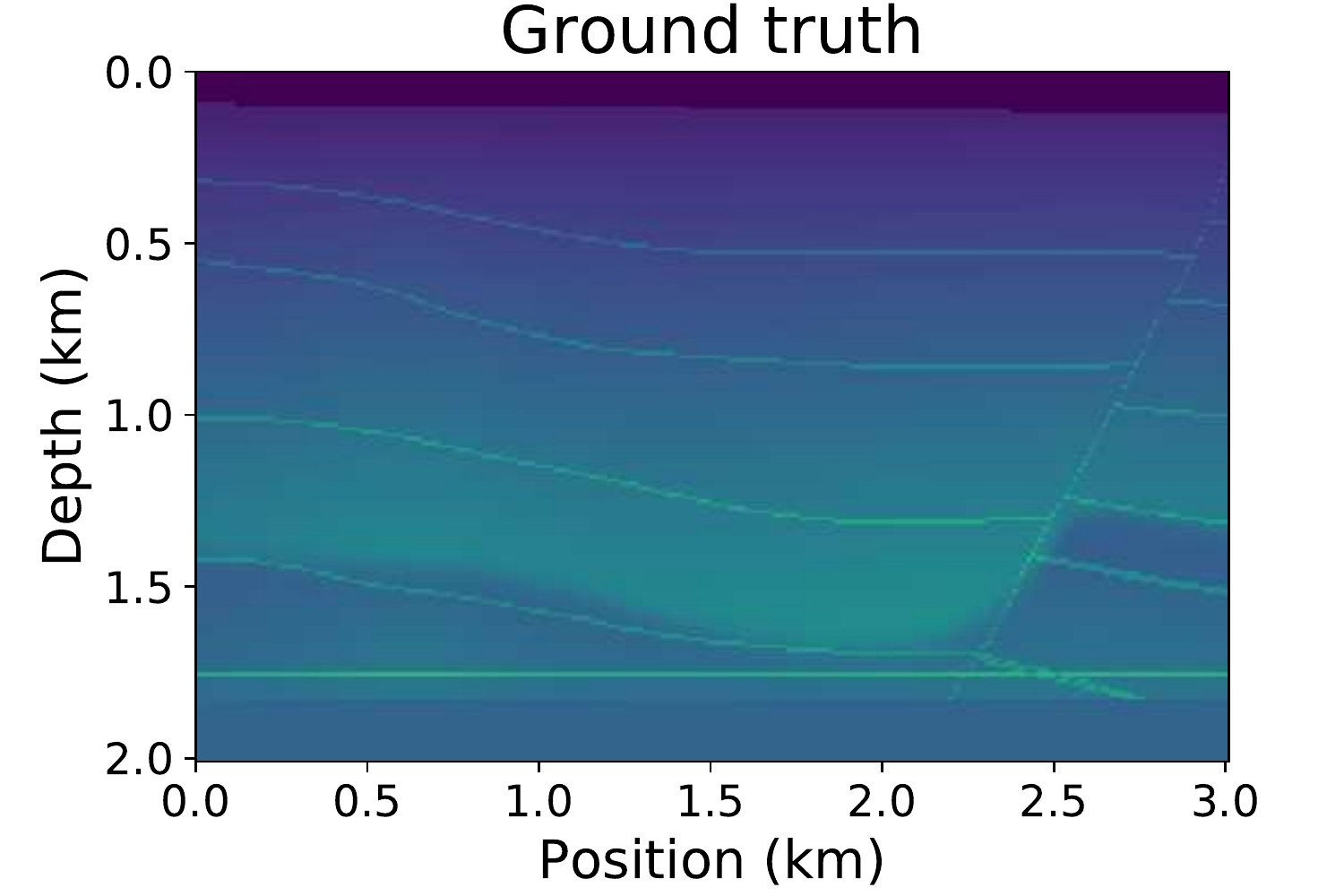}}
  \hspace{-0.4cm}
  \subfigure[]{\label{fig16-2}
  \includegraphics[width=0.5\textwidth]{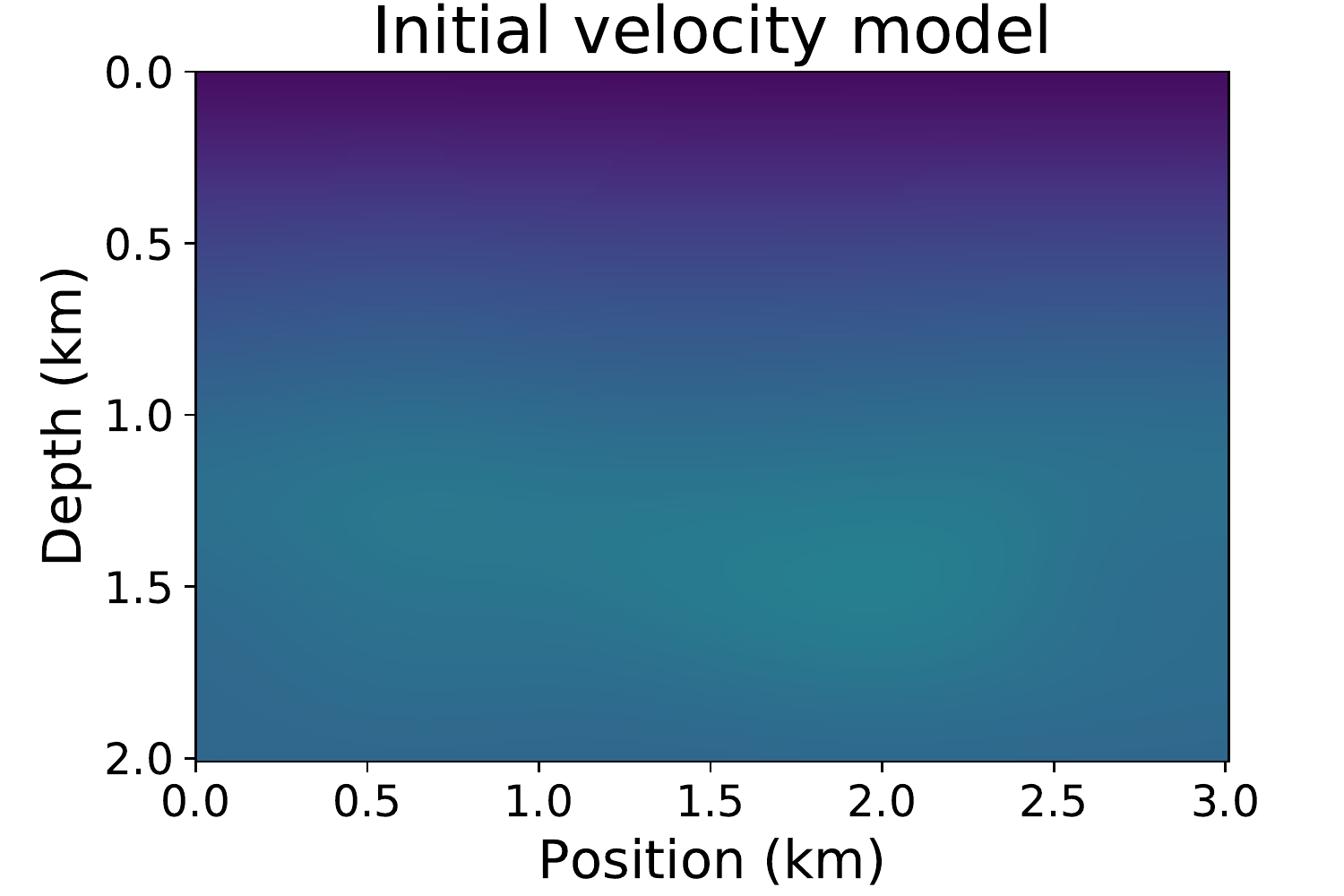}}

  \hspace{-0.4cm}
  \subfigure[]{\label{fig16-3}
  \includegraphics[width=0.5\textwidth]{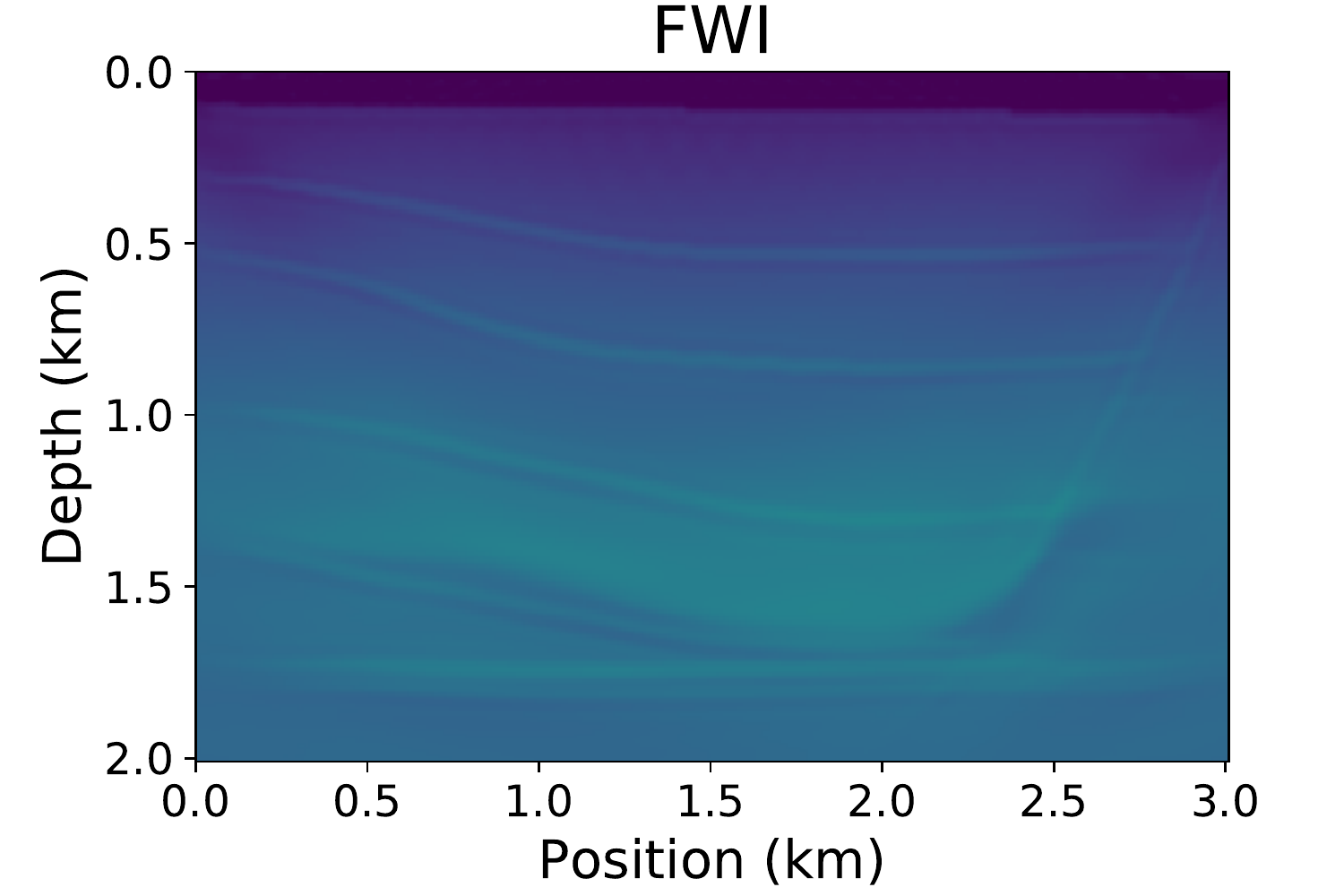}}
  \hspace{-0.4cm}
  \subfigure[]{\label{fig16-4}
  \includegraphics[width=0.5\textwidth]{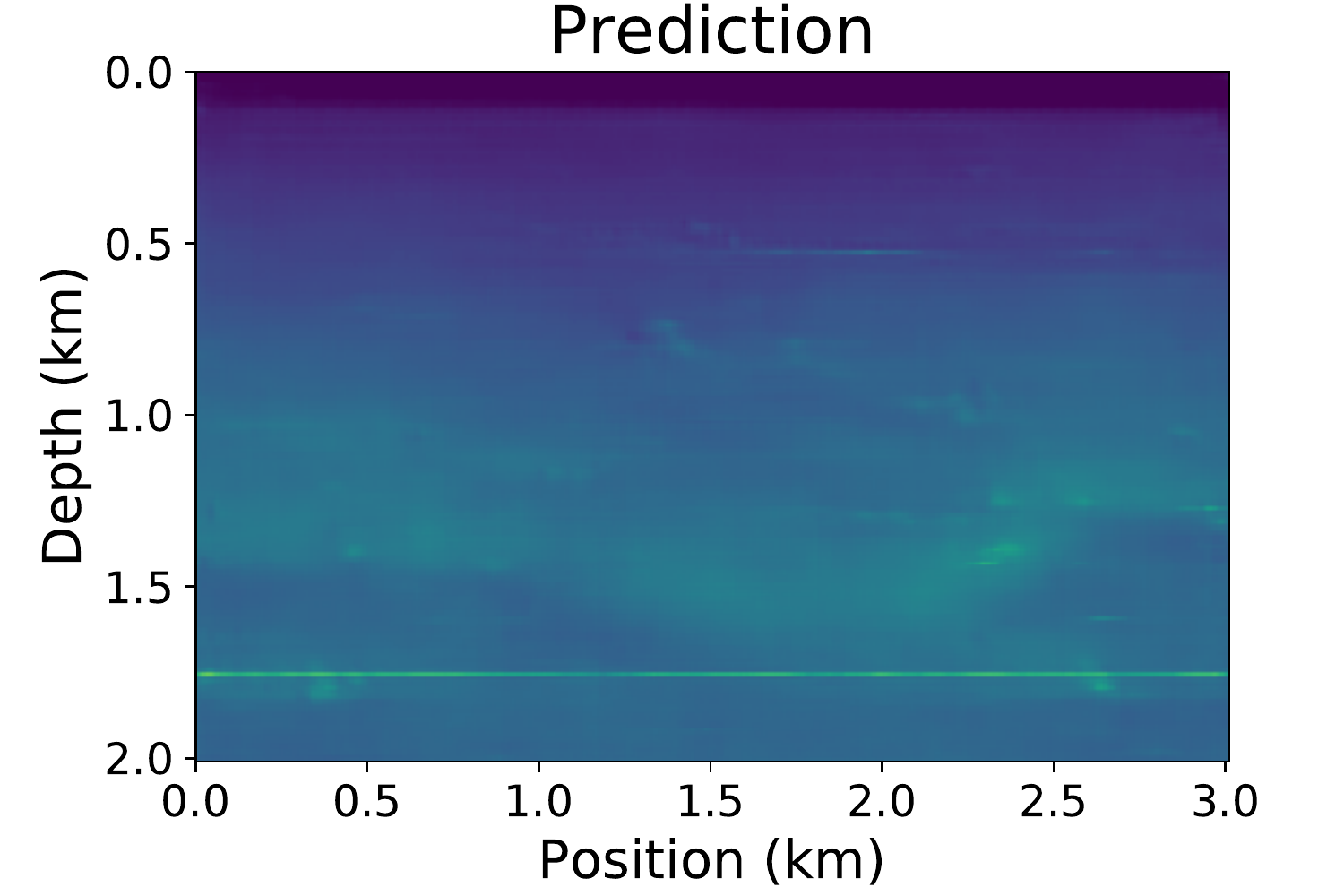}}

  \caption{Inversion of  velocity model without salt dome: (a) ground-truth velocity model; (b) initial velocity model of FWI; (c) result of FWI; (d) result of the proposed method. Our method showed a slightly lower performance than FW because only 10  training models without the salt body were utilized. More training data are required to obtain corresponding improvement. }
  \label{fig16}
\end{figure*}

\clearpage
\begin{figure*}
  \centering
  \hspace{-0.8cm}
  \subfigure[]{\label{fig11-1}
  \includegraphics[width=0.5\textwidth]{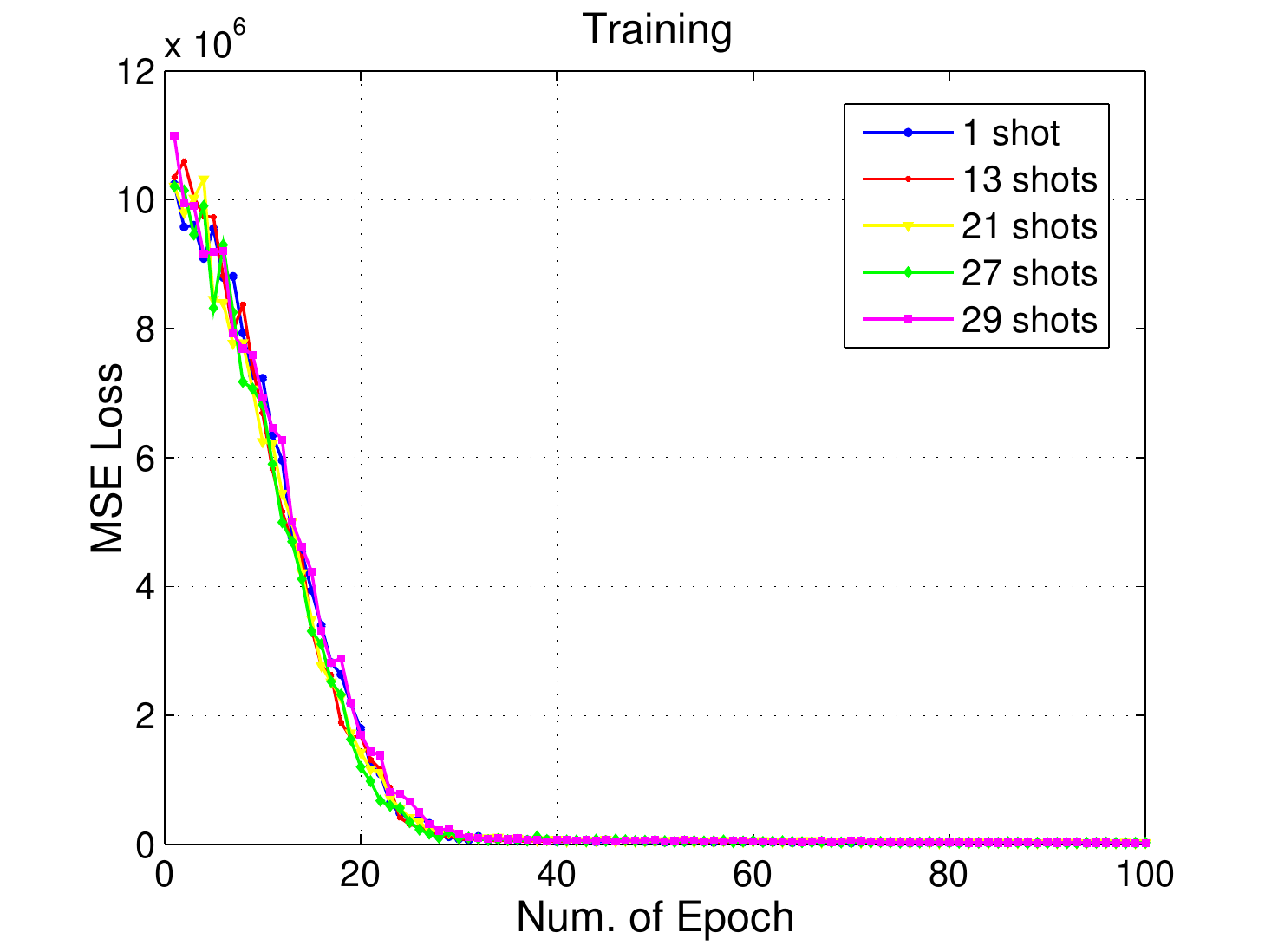}}
  \hspace{-0.8cm}
  \subfigure[]{\label{fig11-2}
  \includegraphics[width=0.5\textwidth]{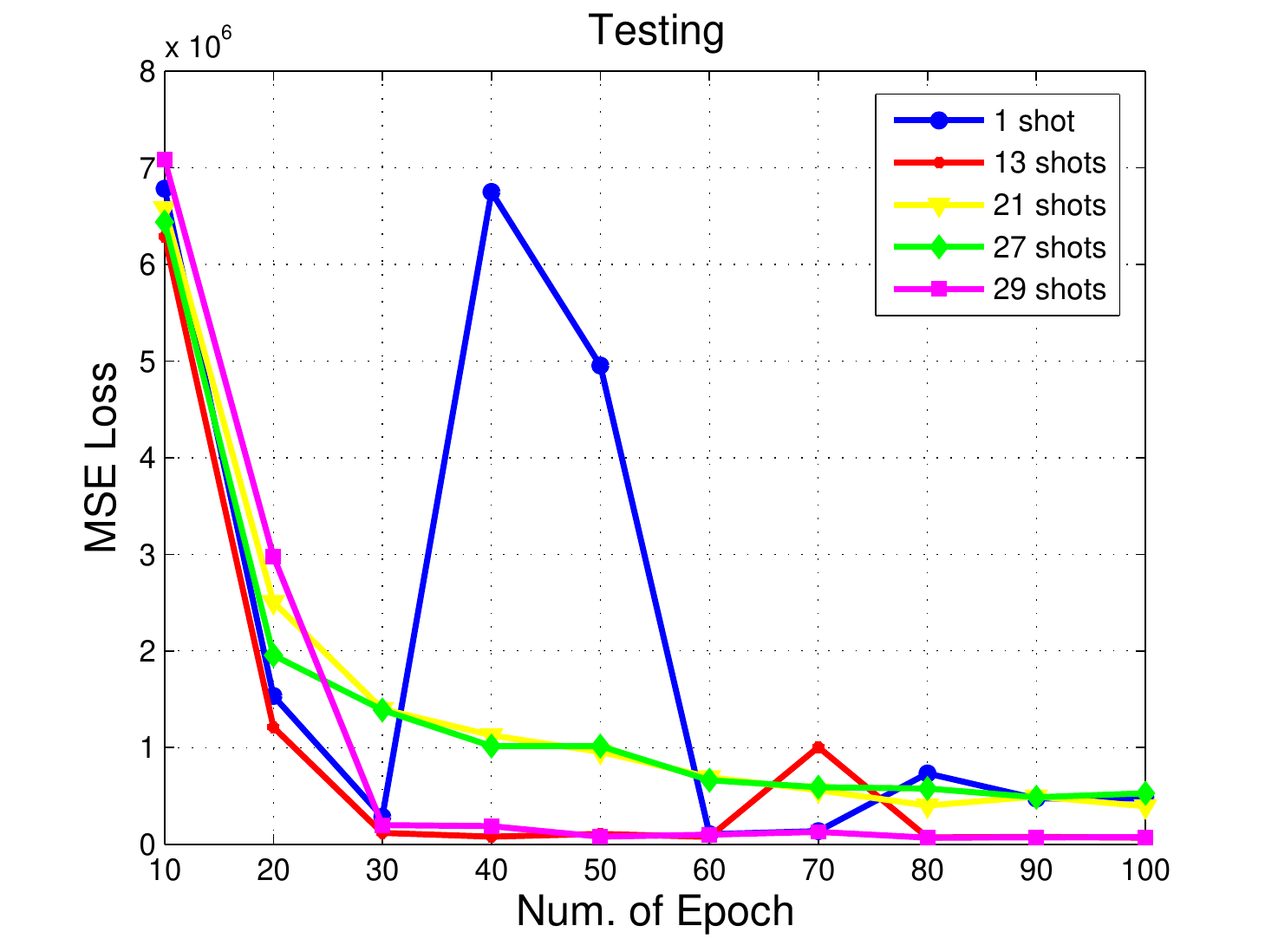}}

  \hspace{-0.8cm}
  \subfigure[]{\label{fig11-3}
  \includegraphics[width=0.5\textwidth]{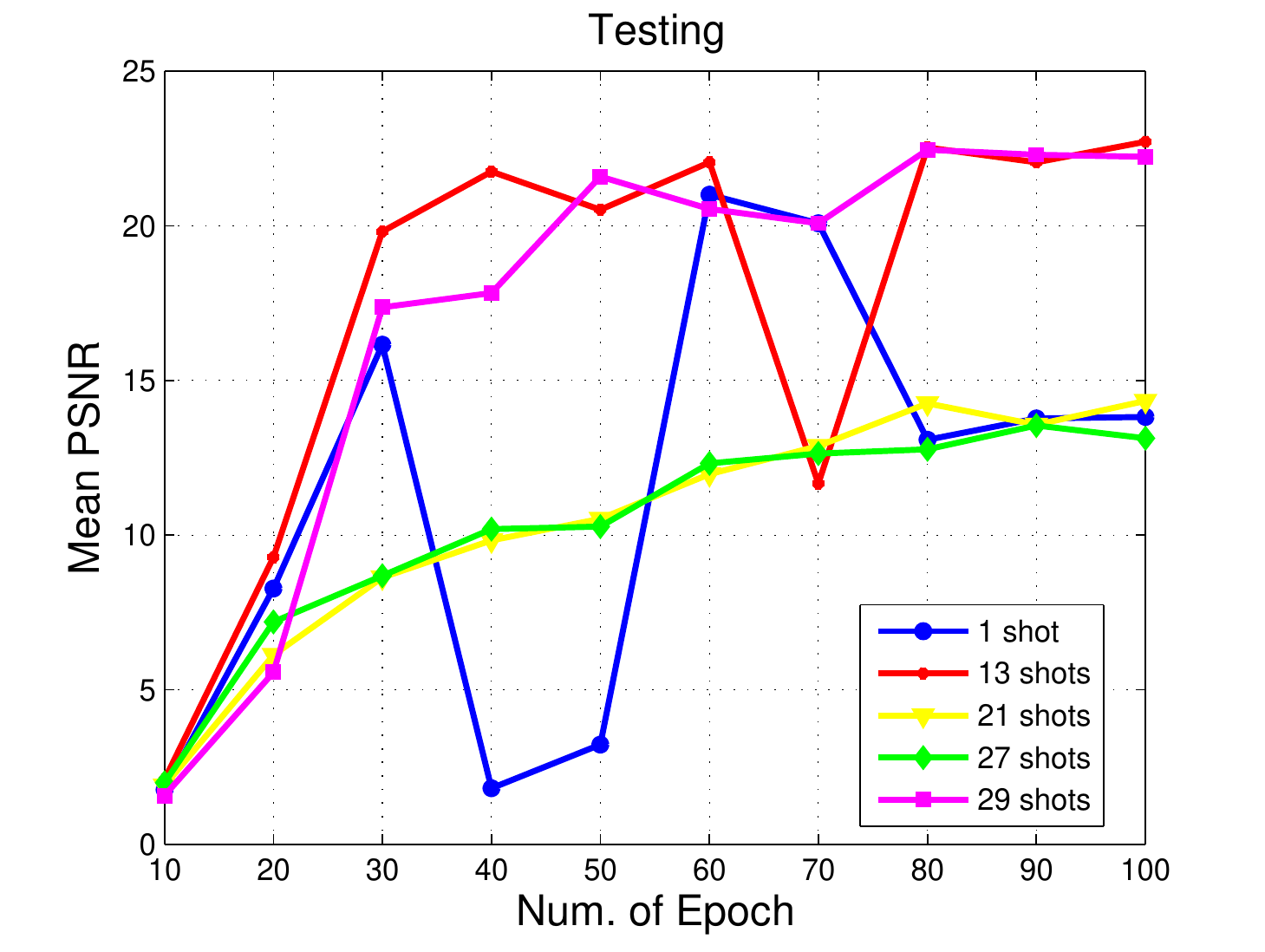}}
  \hspace{-0.8cm}
  \subfigure[]{\label{fig11-4}
  \includegraphics[width=0.5\textwidth]{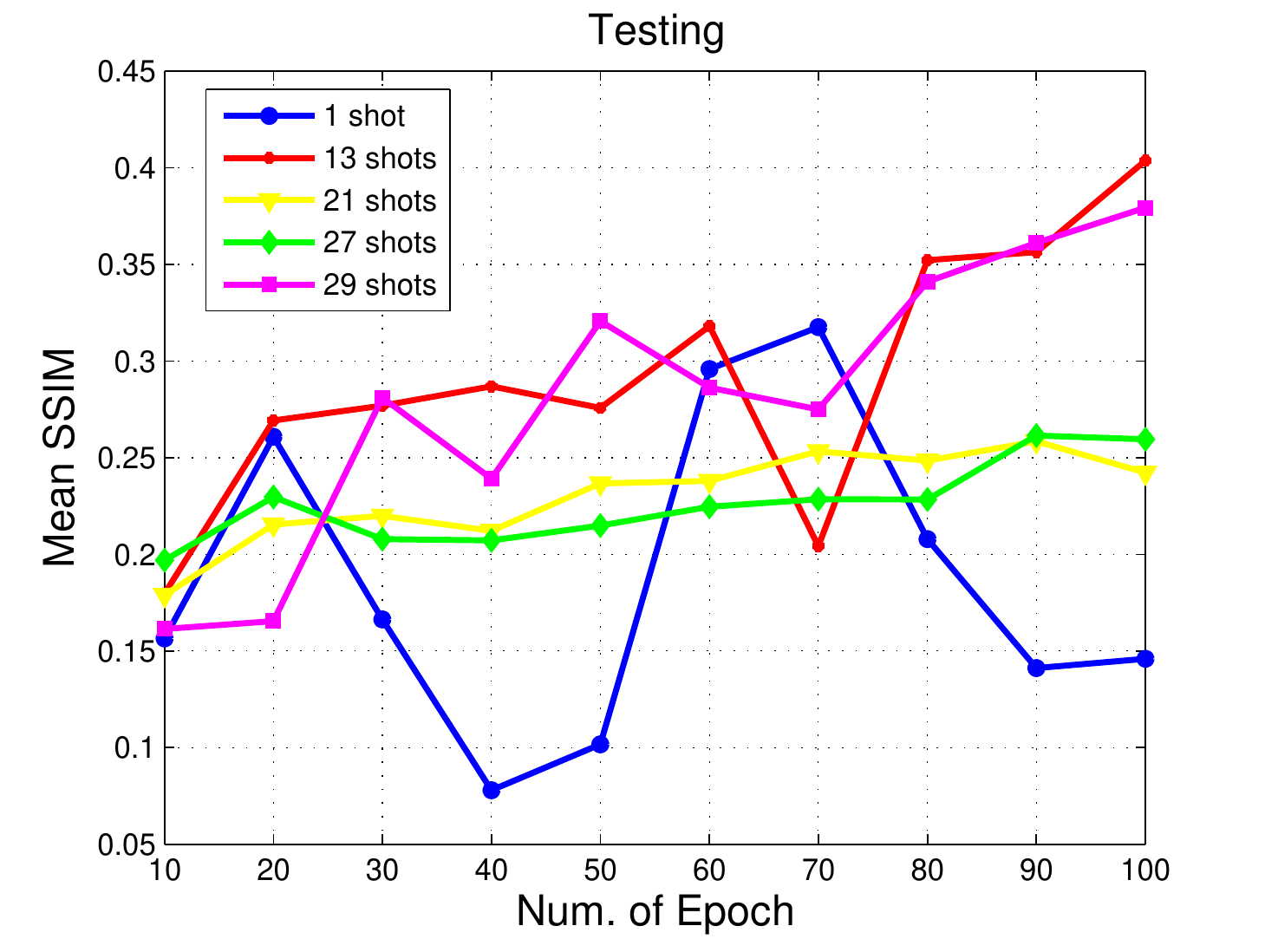}}

  \caption{{{Comparison of performance versus number of epoch between different numbers of training shots: (a) mean-square error during training stage; (b) mean square error during testing stage; (c) mean peak signal-to-noise ratio (PSNR) during the testing stage; (d) mean structural similarity (SSIM) during the testing stage. All testing evaluation was obtained for 100 testing models.}}}
  \label{fig11}
\end{figure*}

\clearpage
\begin{figure*}
\centering
  \hspace{-0.6cm}
  \subfigure[]{\label{fig17-1}
  \includegraphics[width=0.5\textwidth]{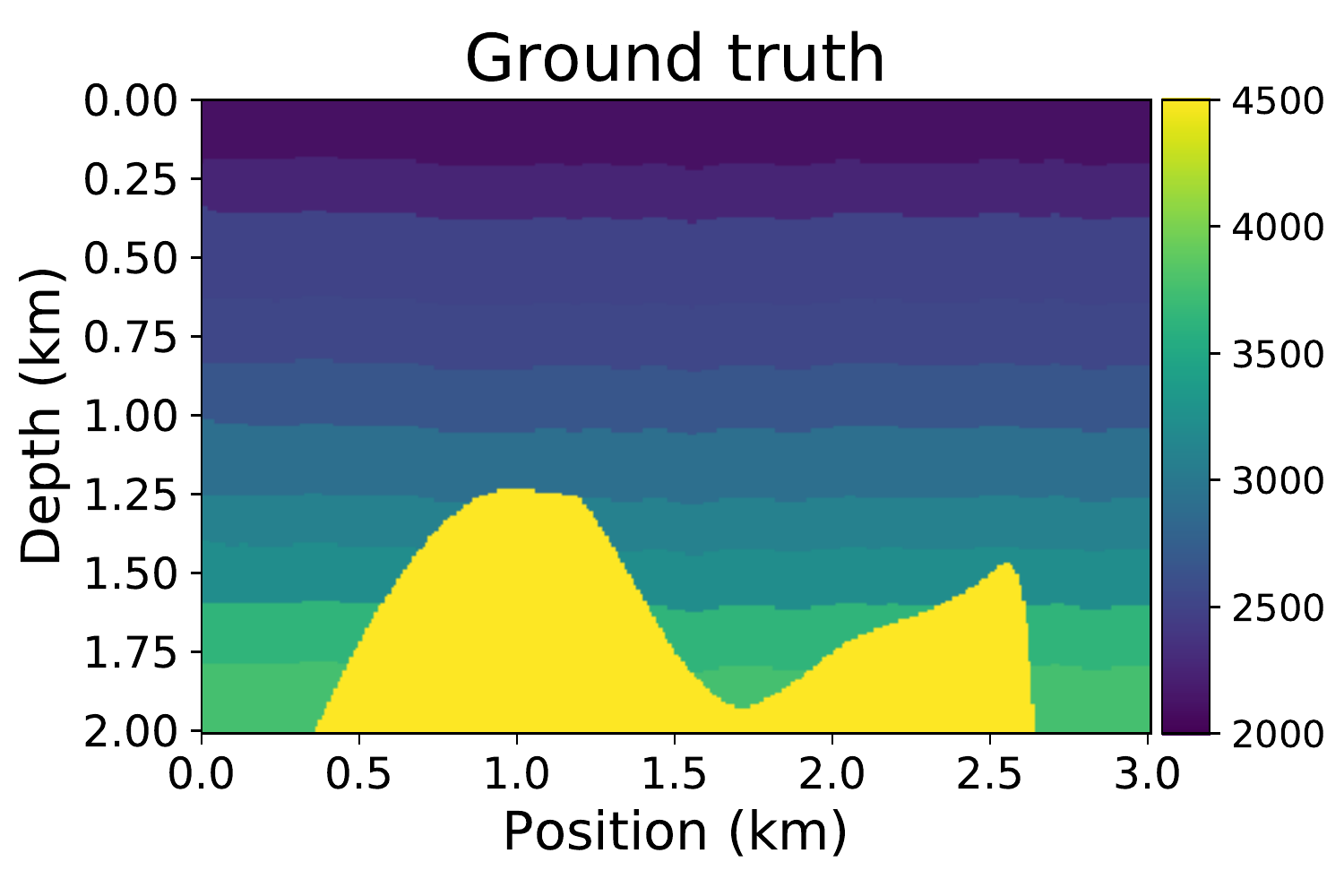}}
  \hspace{-0.2cm}
  \subfigure[]{\label{fig17-2}
  \includegraphics[width=0.5\textwidth]{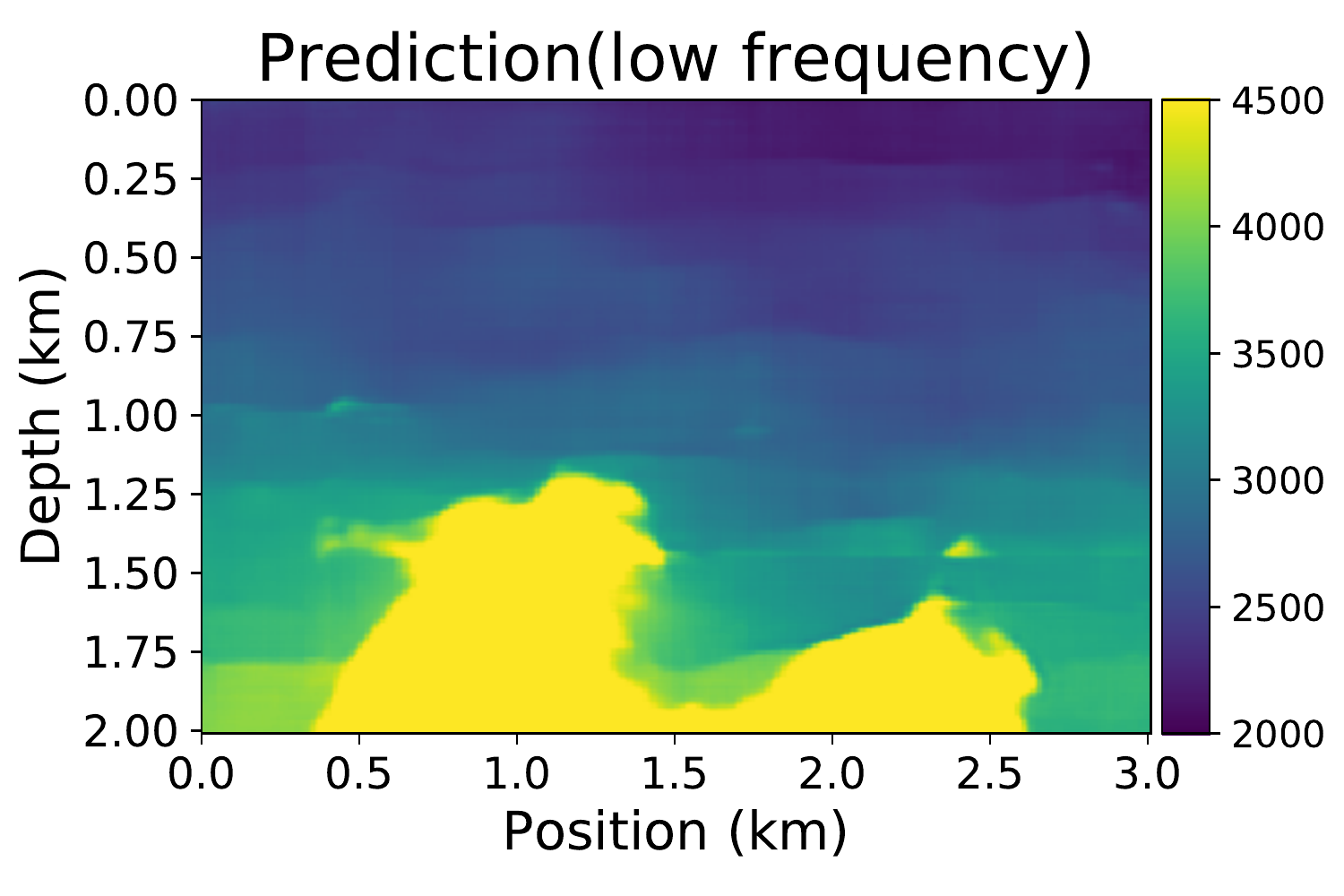}}

  \hspace{-0.4cm}
  \subfigure[]{\label{fig17-5}
  \includegraphics[width=0.5\textwidth]{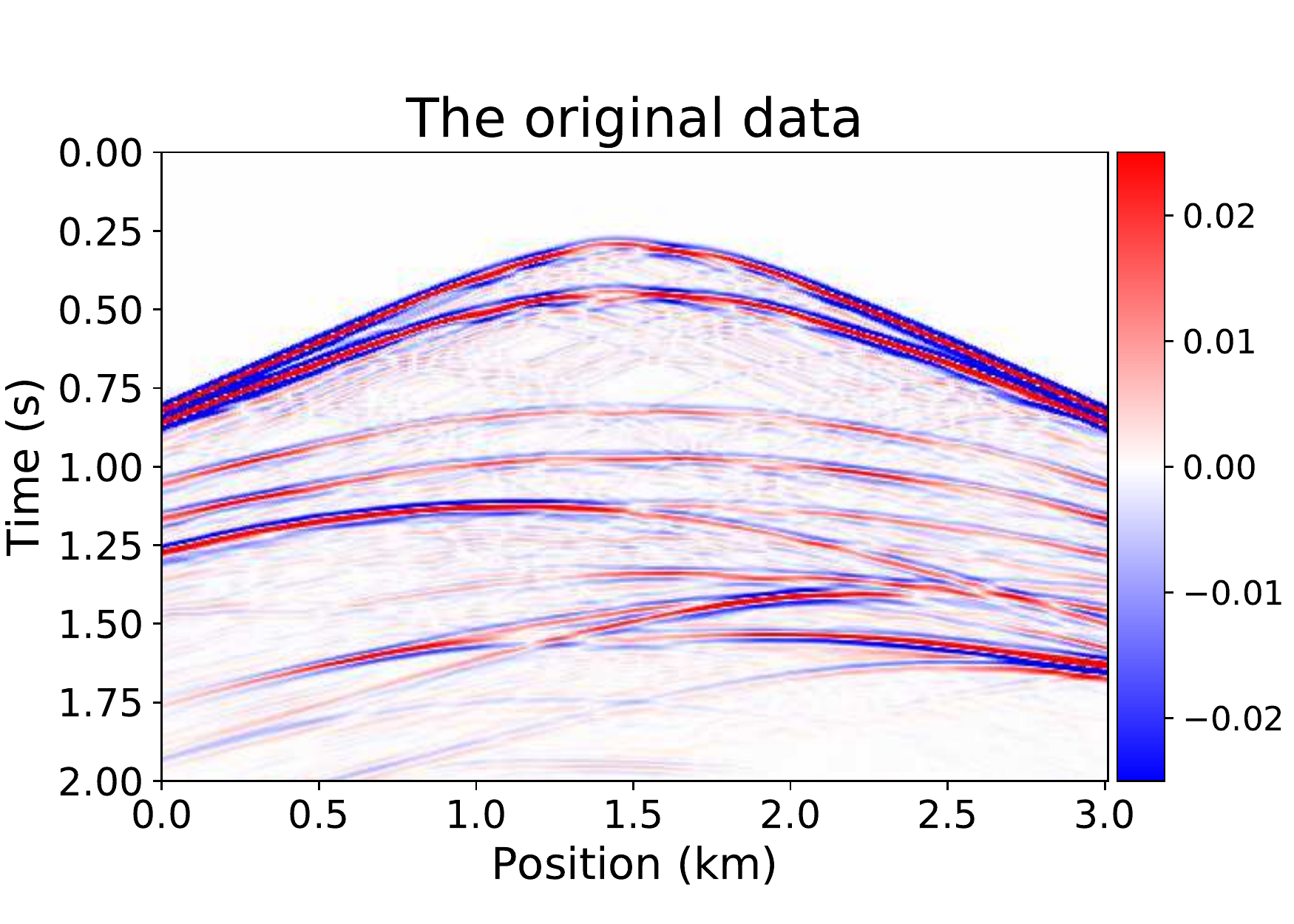}}
  \hspace{-0.4cm}
  \subfigure[]{\label{fig17-6}
  \includegraphics[width=0.5\textwidth]{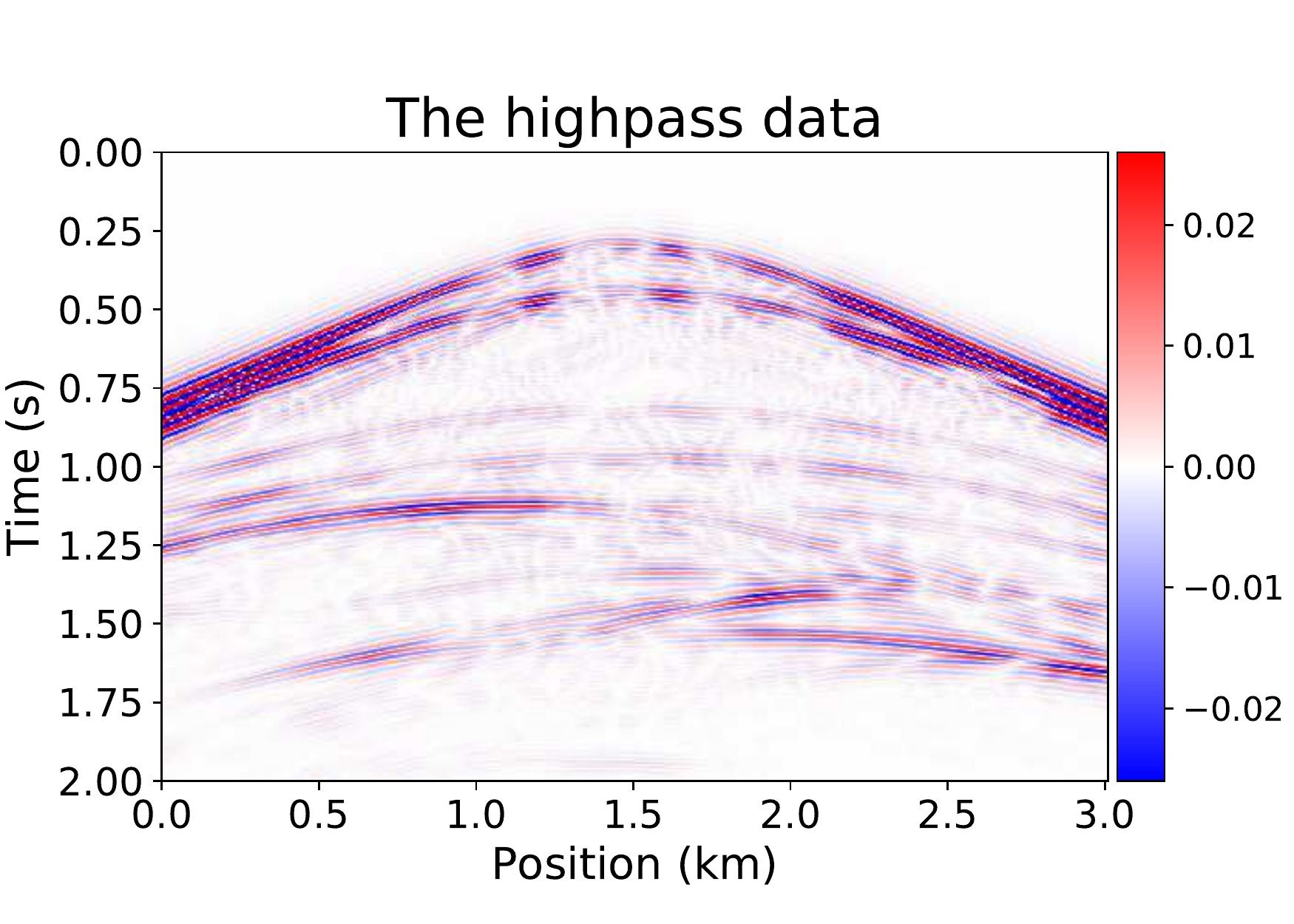}}
  \caption{Typical results obtained by using by our method when seismic data are lacking in low-frequency components: (a) ground truth; (b)  prediction with data lacking low frequencies; (c)  original seismic data (15th shot); (d) reconstructed data  lacking  1/10 normalized Fourier spectrum.  }
  \label{fig17}
\end{figure*}

\clearpage
\begin{figure*}
\centering
  \hspace{-0.4cm}
  \subfigure[]{\label{fig18-7}
  \includegraphics[width=0.35\textwidth]{f12-1-eps-converted-to}}
  \hspace{-0.7cm}
  \subfigure[]{\label{fig18-8}
  \includegraphics[width=0.35\textwidth]{f12-2-eps-converted-to}}
  \hspace{-0.7cm}
  \subfigure[]{\label{fig18-9}
  \includegraphics[width=0.35\textwidth]{f12-3-eps-converted-to}}

  \hspace{-0.4cm}
  \subfigure[]{\label{fig18-1}
  \includegraphics[width=0.35\textwidth]{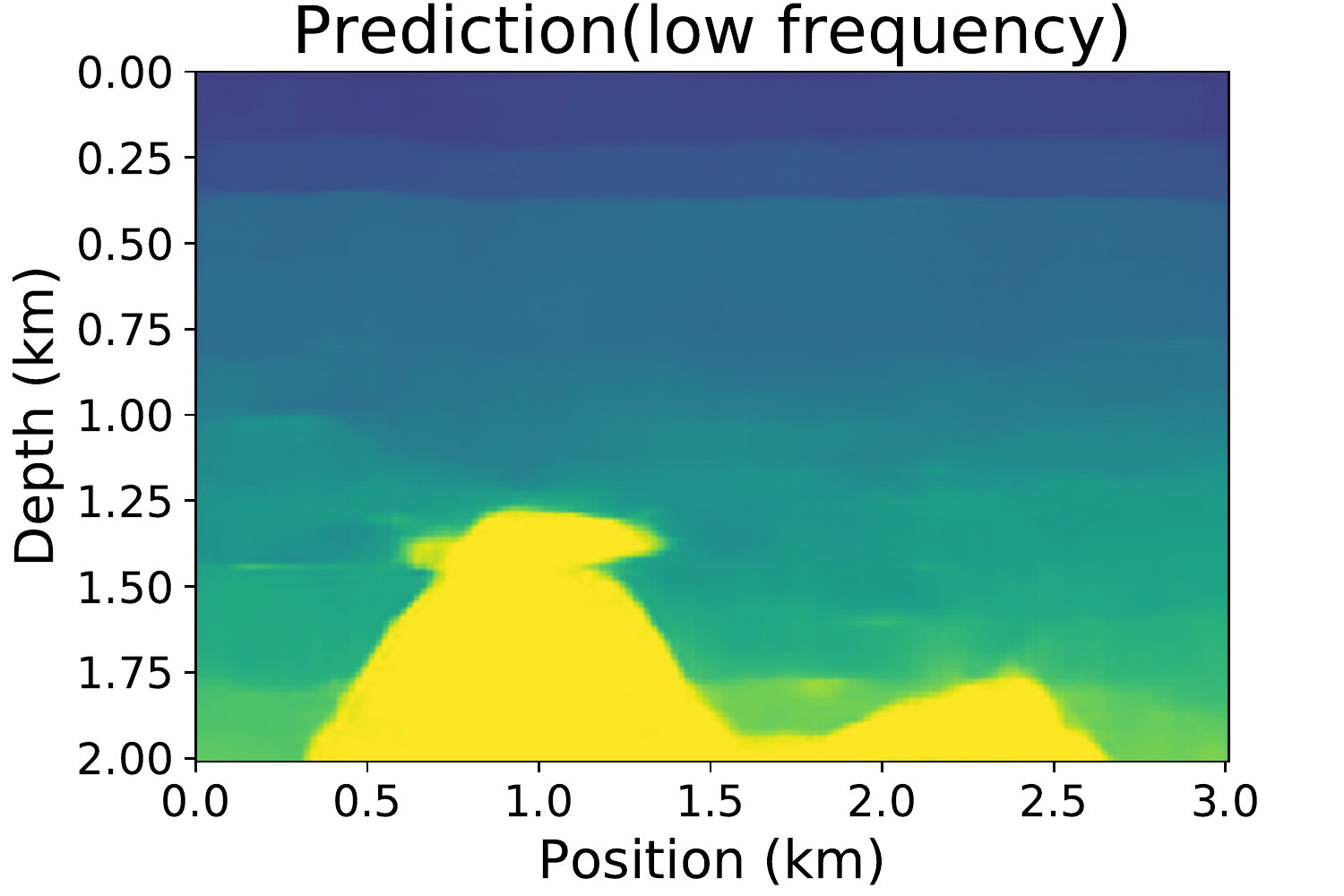}}
  \hspace{-0.7cm}
  \subfigure[]{\label{fig18-2}
  \includegraphics[width=0.35\textwidth]{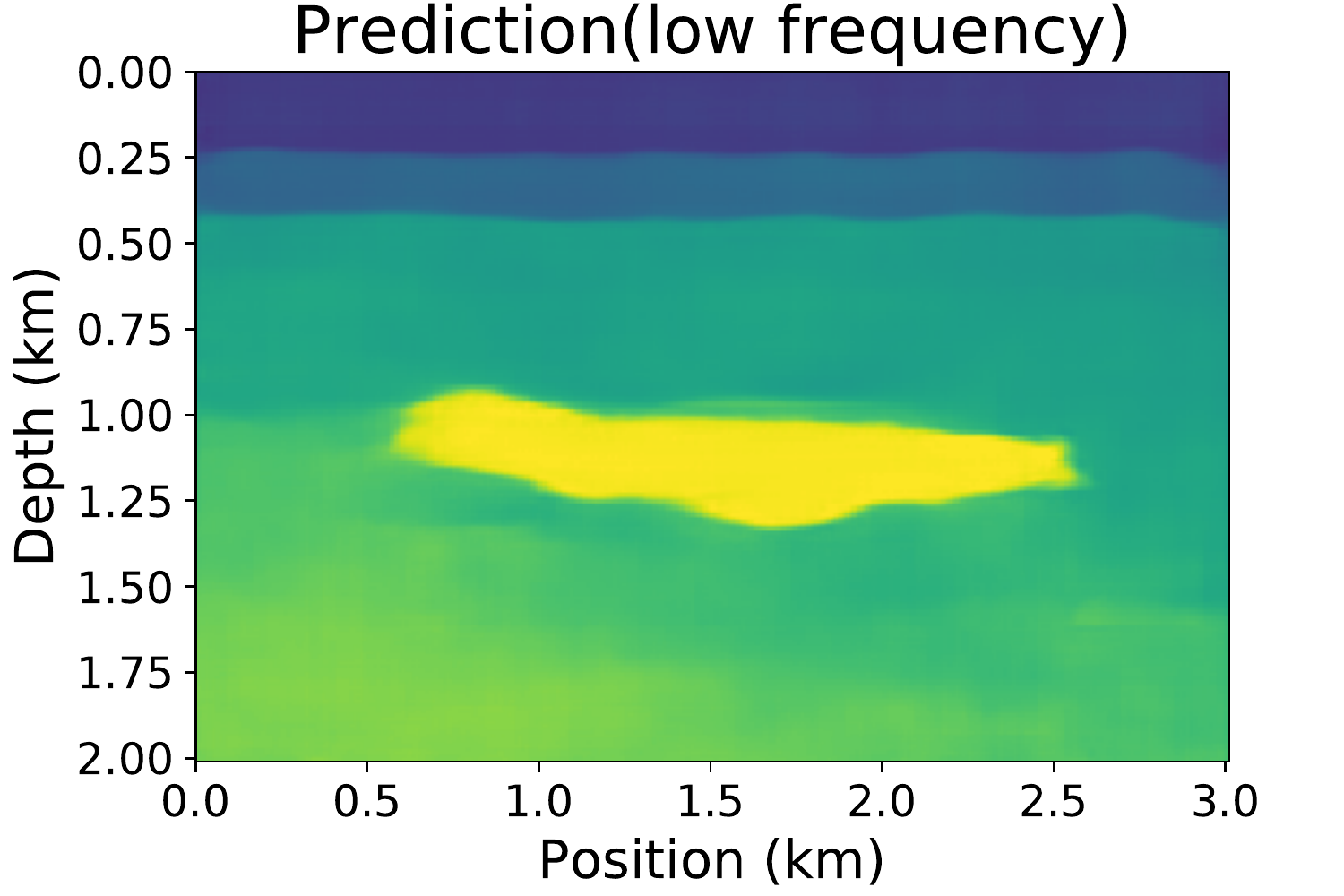}}
  \hspace{-0.7cm}
  \subfigure[]{\label{fig18-3}
  \includegraphics[width=0.35\textwidth]{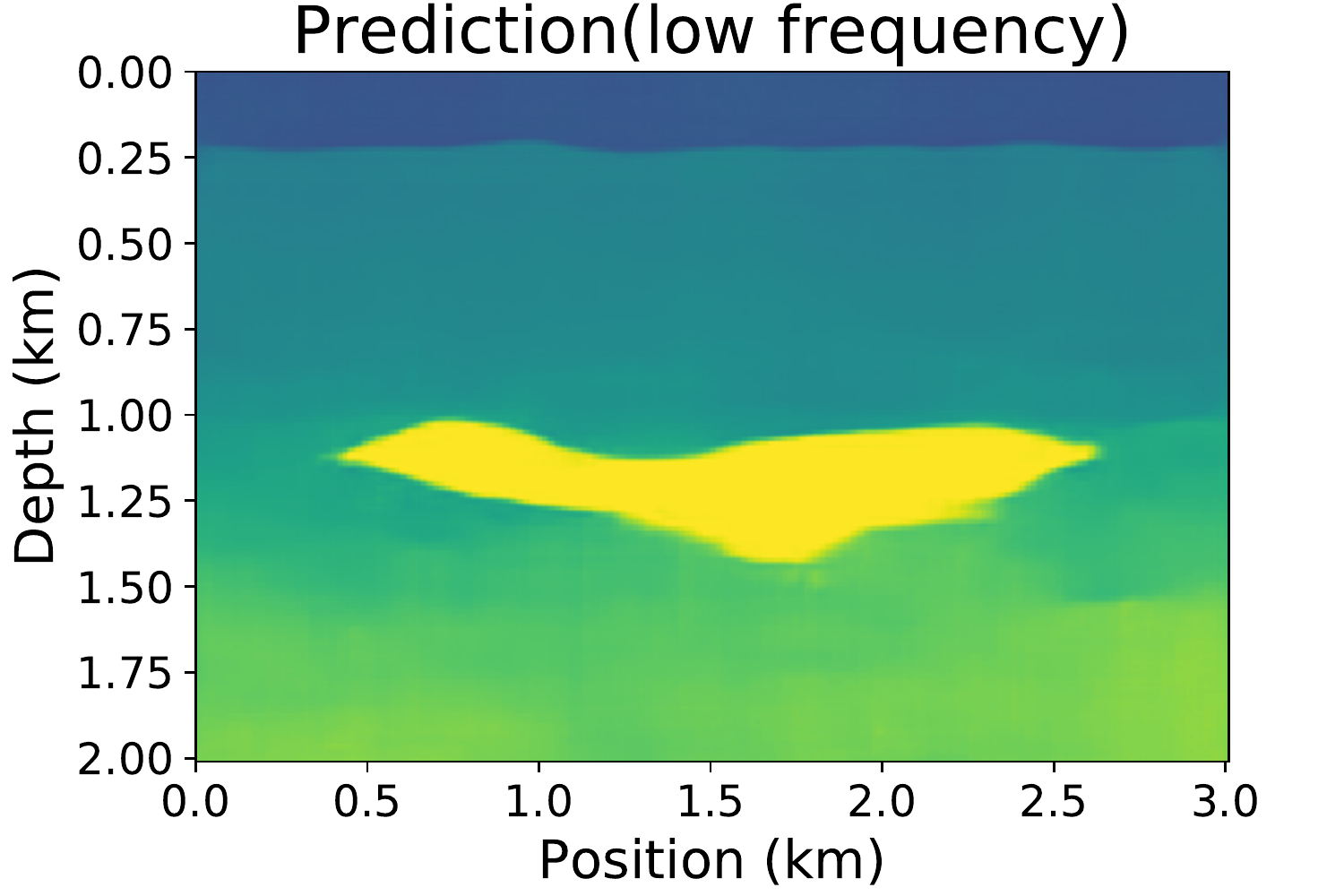}}

  \hspace{-0.4cm}
  \subfigure[]{\label{fig18-10}
  \includegraphics[width=0.35\textwidth]{f13-1-eps-converted-to}}
  \hspace{-0.7cm}
  \subfigure[]{\label{fig18-11}
  \includegraphics[width=0.35\textwidth]{f13-2-eps-converted-to}}
  \hspace{-0.7cm}
  \subfigure[]{\label{fig18-12}
  \includegraphics[width=0.35\textwidth]{f13-3-eps-converted-to}}

  \hspace{-0.4cm}
  \subfigure[]{\label{fig18-4}
  \includegraphics[width=0.35\textwidth]{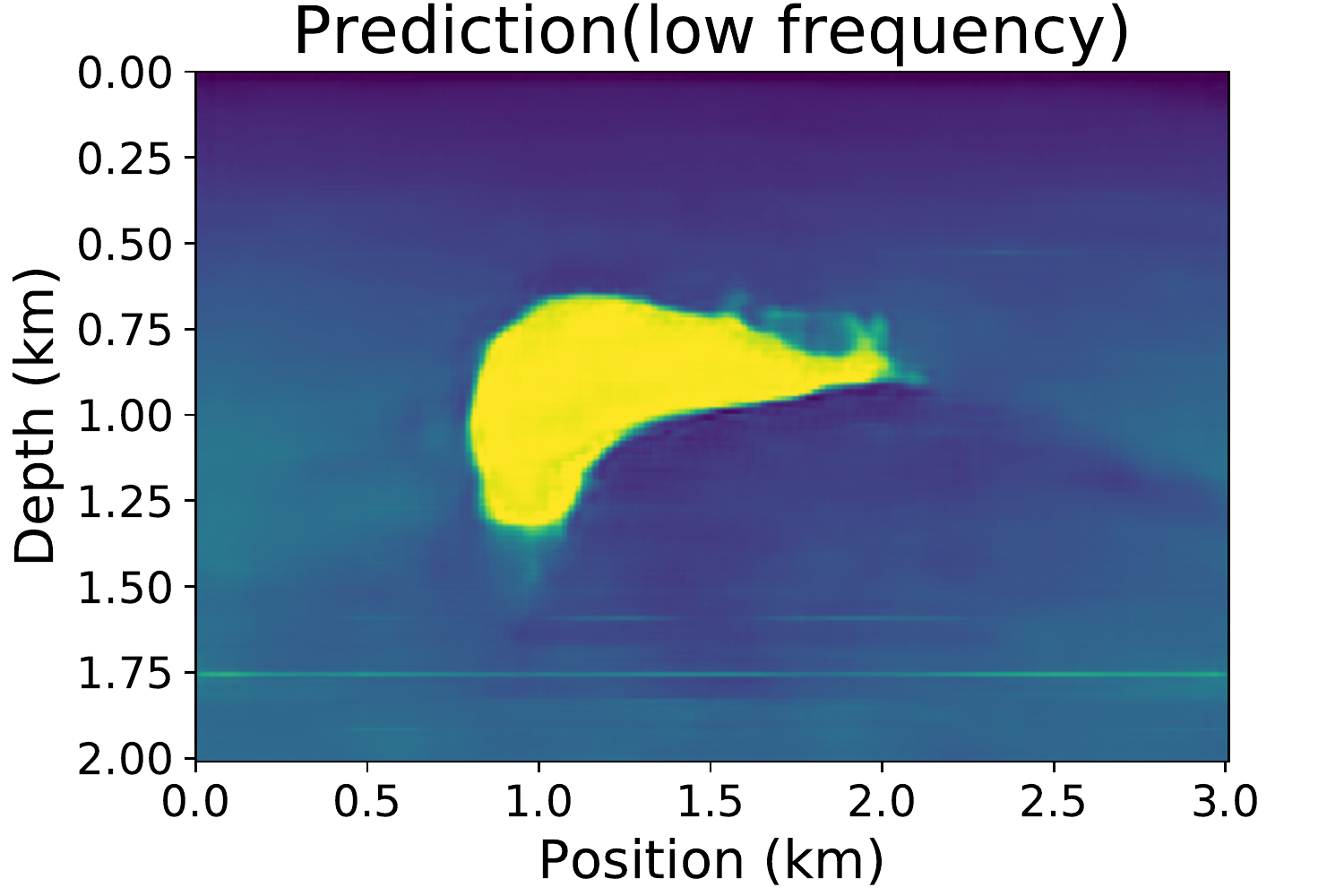}}
  \hspace{-0.7cm}
  \subfigure[]{\label{fig18-5}
  \includegraphics[width=0.35\textwidth]{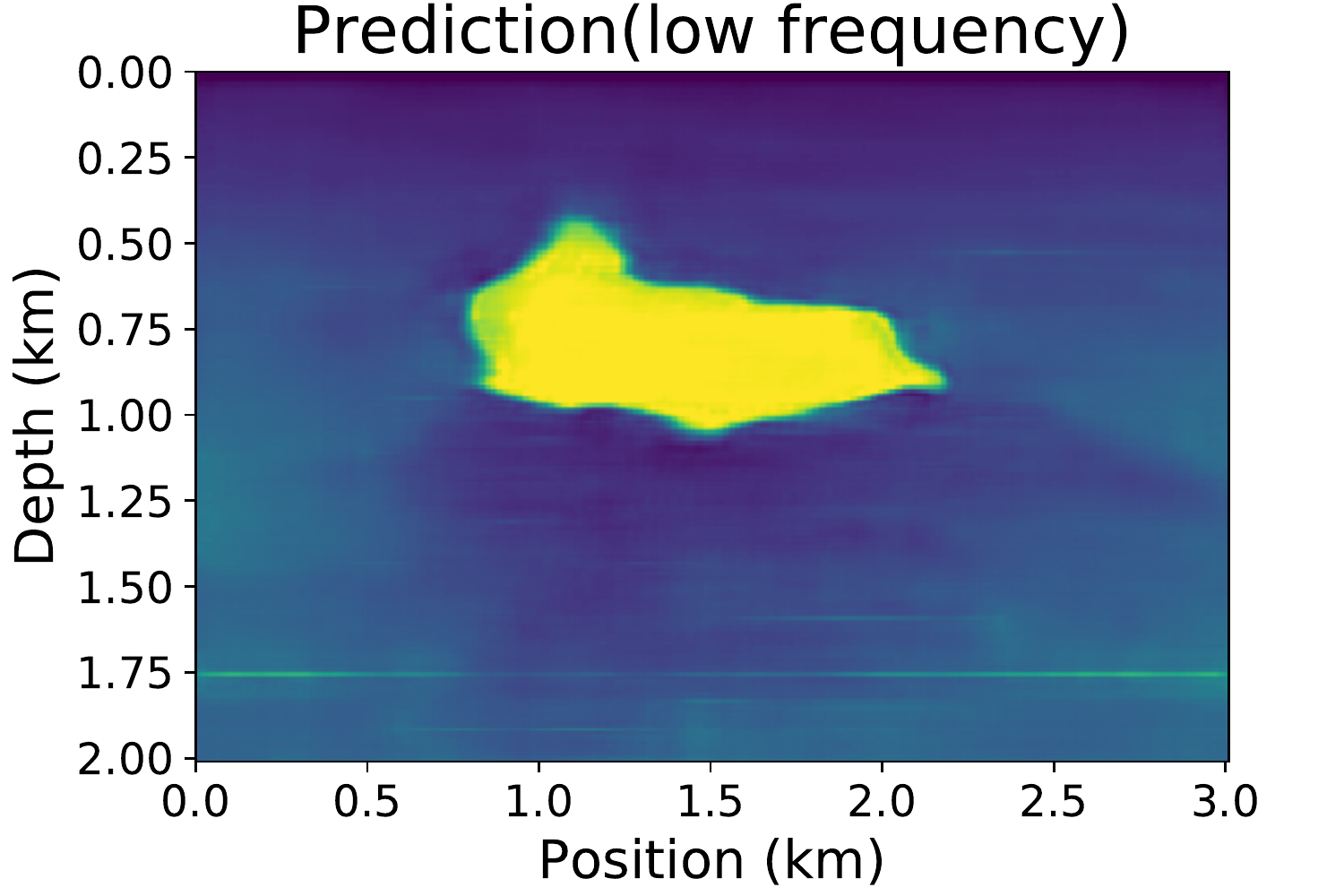}}
  \hspace{-0.7cm}
  \subfigure[]{\label{fig18-6}
  \includegraphics[width=0.35\textwidth]{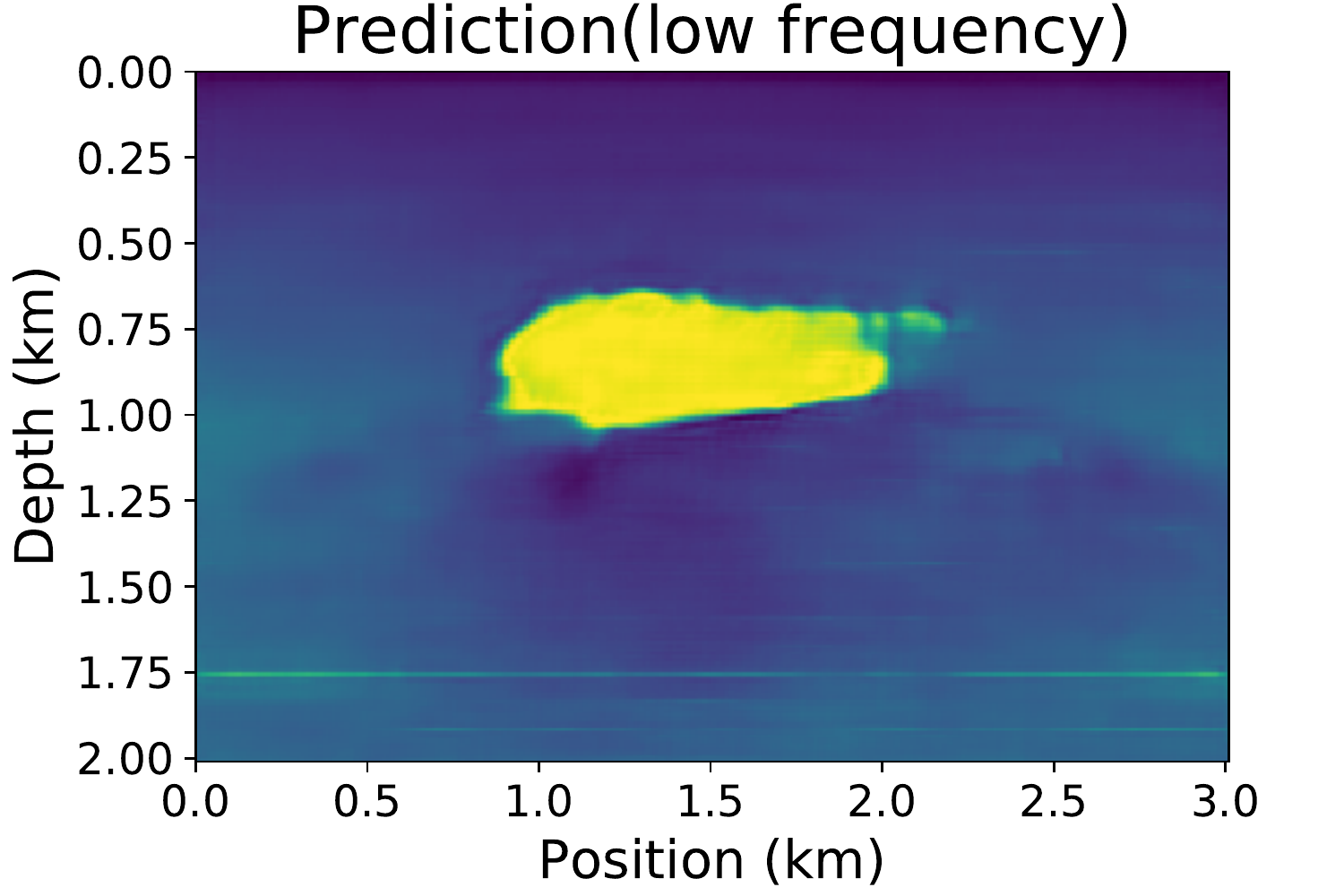}}
  \caption{Results of the velocity inversion obtained when the  training data lack low frequencies: (a)--(c) ground truth of simulated models; (d)--(f) prediction results; (g)--(i) ground truth of SEG salt models; (j)--(l): prediction results.}
  \label{fig18}
\end{figure*}

\clearpage
\begin{figure*}
\centering
  \hspace{-0.7cm}
  \includegraphics[width=0.7\textwidth]{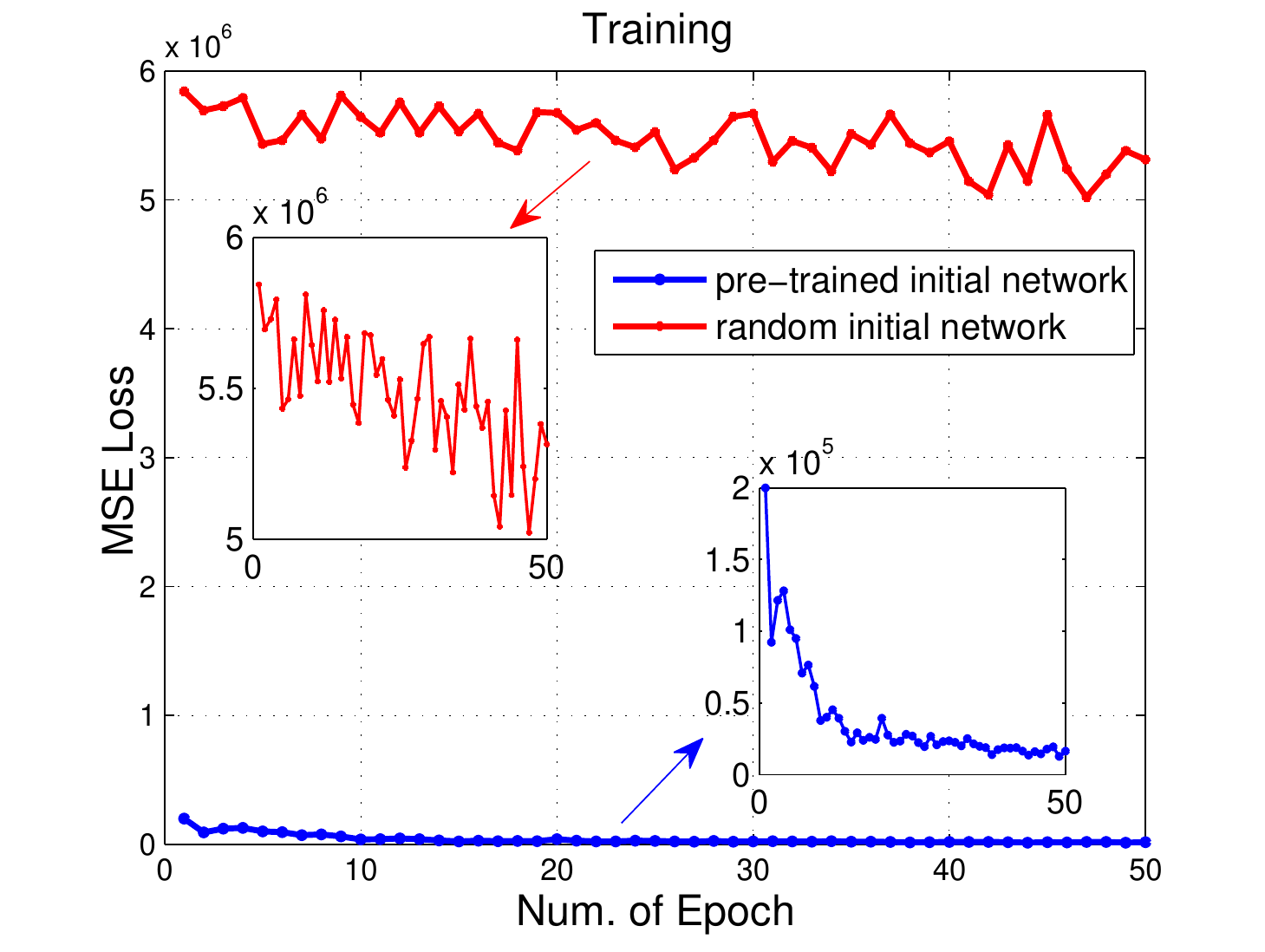}
  \caption{Comparison of training loss versus number of epochs. The red line denotes training with a random initial network. The blue line represents training with a pre-trained initial network (i.e., the trained network for the simulated dataset).}
  \label{fig24}
\end{figure*}


\clearpage
\begin{table}
  \centering
  \begin{tabular}{|c|c|}
  \hline
  Acronyms  &Corresponding definition \\
  \hline
  FCN & Fully convolutional neural network \\
  \hline
  DL &Deep learning \\
  \hline
  FWI &Full-waveform inversion \\
  \hline
  VMB &Velocity-model building \\
  \hline
  ML &Machine learning \\
  \hline
  DNN &Deep neural network \\
  \hline
  CNN &Convolutional neural network \\
  \hline
  SEG &Society of exploration geophysics \\
  \hline
  SGD &Stochastic gradient descent \\
  \hline
  \end{tabular}
  \caption{All acronyms used in this paper and their definitions.}
  \label{tab1}
\end{table}
\clearpage

\begin{table}
  \centering
  \begin{tabular}{|c|c|c|c|c|c|}
  \hline
  Task &Source &Spatial &Sampling time &Ricker &Maximum travel\\
   &num &interval &interval &wave &time\\
  \hline
  Velocity inversion &29 &10 m &0.001 s &25 Hz &2 s \\
  \hline
  \end{tabular}
  \caption{Parameters of forward modeling. }
  \label{tab4}
\end{table}

\clearpage
\begin{table}
\centering
\begin{tabular}{|c|c|}
\hline
Operation (Acronym) &Definition(2D) \\
\hline
Convolution (conv) &$output=K*input+b$ \\
\hline
Batch normalization (BN) &$out=\frac{input-mean[input]}{\sqrt{Var[input]+\varepsilon}}*\gamma+\beta$ \\
\hline
Rectified linear unit (Relu) &$out=max(0,input)$ \\
\hline
Max-pooling (max-pooling) &$out=max[input]_{w\times h}$ \\
\hline
Deconvolution / Transposed convolution (deconv) &$out=\overline{K}*input+b$ \\
\hline
Skip connection and concatenation &$out=[input,padding]_{channel}$ \\
\hline
\end{tabular}
\caption{Definitions of the different operations for our proposed  network.}
\label{tab2}
\end{table}

\clearpage
\begin{table}
\centering
\hspace{-0.7cm}
\begin{tabular}{|c|c|c|c|c|c|c|c|}
\hline
Task &Learning &Epoch &Batch &SGD &Number of &Number of\\
 &rate & &size &algorithm & training setd & testing setd\\
\hline
Inversion  & & & & & & \\
(simulated model) &1.0e-03 &100 &10 &Adam &1600 &100\\
\hline
Inversion  & & & & & &\\
(SEG salt model)  &1.0e-03 &50 &10 &Adam &130 &10\\
\hline
\end{tabular}
\caption{ Parameters of training process  in our proposed network.}
\label{tab3}
\end{table}

\clearpage
\begin{table}
\centering
\begin{tabular}{|l|c|c|c|c|}
\hline
\diagbox{Process}{Time}{Method} &\multicolumn{2}{c|}{ FCN-based method } &\multicolumn{2}{c|}{ FWI } \\
\hline
Training &1078 min &43 min &N/A &N/A  \\
\hline
Prediction &2 s &2 s  &37 min &25 min  \\
\hline
\end{tabular}
\caption{Time consumed for the training and testing processes. The three columns of each method from left to right indicate the GPU time for the simulated velocity-model inversion, and SEG salt-model inversion. The training time is the total time required for all training sets; the testing time is for only one model. N/A indicates that FWI had no training time.  }
\label{tab7}
\end{table}
\end{document}